%% file: phi3_Belle_Belle2.tex
\documentclass[a4paper,11pt]{article}
\pdfoutput=1 % if your are submitting a pdflatex (i.e. if you have
\usepackage{jheppub} % for details on the use of the package, please
\input{files/definitions}
\usepackage{siunitx}
\usepackage[T1]{fontenc} % if needed

\title{\boldmath Combined analysis of Belle and Belle~II data to determine the CKM angle $ \phi_{3} $ using $B^+ \to D(K_{S}^0 h^+ h^-) h^+$ decays}
\newcounter{AffiliationCounter}

\input{files/pub007}

\abstract{We present a measurement of the Cabibbo-Kobayashi-Maskawa unitarity triangle angle $ \phi_{3} $ (also known as~$\gamma$) using a model-independent Dalitz plot analysis of \linebreak $B^+\to D\left(K_{S}^{0}h^{+}h^{-}\right)h^+$, where $D$ is either a $D^0$ or $\bar{D}{}^0$ meson and $h$ is either a $\pi$ or $K$. This is the first measurement that simultaneously uses Belle and Belle~II data, combining samples corresponding to integrated luminosities of
\SI{711}{fb^{-1}}
and \SI{128}{fb^{-1}}, respectively. All data were accumulated from energy-asymmetric $e^+e^-$ collisions at a centre-of-mass energy corresponding to the mass of the $\Upsilon(4S)$ resonance. We measure 
$\phi_3 = \left(78.4 \pm 11.4 \pm 0.5 \pm 1.0 \right)^{\circ}$, where the first uncertainty is statistical, the second is the experimental systematic uncertainty and the third is from the uncertainties on external measurements of the $D$-decay strong-phase parameters.}

\begin{document} 

\begin{flushright}
Belle II Preprint 2021-003\\
KEK Preprint 2021-28\\
\end{flushright}

\maketitle
\flushbottom
%%%%%%%%%%%%%%%%%%%%%%%%%%%%%%%%%%%%%%%%%%%%%%%%%%%%%%%%%%%%%%%%%%%%%%%%%%%
\section{Introduction}
\label{sec:intro}
The Cabibbo-Kobayashi-Maskawa (CKM) unitarity triangle describes {\it CP} violation within the Standard Model (SM) of particle physics. The sides and angles of the unitarity triangle are related to the elements of the CKM matrix \cite{cabibbo,ckm}. The angle $\phi_3$, also known as $\gamma$, is defined as $\phi_3\equiv\arg{\left(-V_{ud}V^{\ast}_{ub}/V_{cd}V_{cb}^{\ast}\right)}$, where $V_{qq^{\prime}}$ are CKM matrix elements. The angle is measurable via the interference of the tree-level quark transitions $\bar{b}\to \bar{c}u\bar{s}$ and $\bar{b}\to \bar{u}c\bar{s}$,\footnote{Throughout this paper charge-conjugate processes are included unless explicitly stated otherwise.} which involve the emission of a $W^+$ boson from the $\bar{b}$ quark. The tree-level nature of these transitions results in negligible theoretical uncertainties when interpreting the measured observables in terms of $\phi_3$ \cite{zupan}. Therefore, assuming the absence of new physics at tree level, the measurement of $\phi_3$ provides a test of the SM when compared to indirect determinations. The latter are derived from independent measurements of the sides and other angles of the unitarity triangle, which can be  influenced by beyond-the-SM particles via loop amplitudes \cite{blankeburas}. The world-average value of the direct measurements of $\phi_3$ is $\left(66.2^{+3.4}_{-3.2}\right)^{\circ}$ \cite{hflav}. The indirect determination of $\phi_3$ is $\left(63.4\pm 0.9\right)^{\circ}$  \cite{king}. Therefore, improvement in the direct determination of $\phi_3$ is required to better constrain possible beyond-the-SM contributions to {\it CP} violation. 

The world average is dominated by a result reported by the  LHCb collaboration \cite{lhcb_gamma} that uses measurements of direct {\it CP} violation in the decays $B^+\to D\left( K^{0}_{\rm S}h^+h^-\right)h^+$, where $D$ is either a $D^0$ or $\bar{D}{}^0$ and $h$ is a pion or kaon \cite{bpggsz1,bpggsz2,bpggsz3}. The interference between $\bar{b}\to \bar{c}u\bar{s}$ and $\bar{b}\to \bar{u}c\bar{s}$ arises because the $D^0$ or $\bar{D}{}^0$ decay to the same $K_{\rm S}^0h^+h^-$ final state. The \linebreak $D\to K^0_{\rm S}h^+h^-$ decays proceed via several intermediate resonances, which results in a variation of the {\it CP} asymmetry over phase space such that $\phi_3$, as well as other parameters related to the $B$-decay amplitude, can be determined from a single decay. This property is in contrast to measurements of $\phi_3$ that use $B^+\to DK^+$ decays in which the $D$ decays to a two-body final state, such as the {\it CP}-eigenstate $K^{0}_{\rm S}\pi^0$ \cite{glw1,glw2} or $K^{\pm}\pi^{\mp}$ \cite{ads1,ads2}. 

Two approaches are used to determine $\phi_3$ from $B^+\to D\left( K^{0}_{\rm S}h^+h^-\right)h^+$ decays, which are either dependent \cite{bpggsz1,bpggsz2,bpggsz3} or independent \cite{bpggsz1} of modelling the $D^0\to K_{\rm S}^0 h^+ h^-$ decay amplitude. The model-dependent method relies upon a detailed description of the intermediate-resonance structure of the $D$-decay amplitude. The model-independent method uses {\it CP}-asymmetry measurements in disjoint regions (bins) of the $D$-decay phase space that can then be related to $\phi_3$ using measurements of $D$-decay strong-phase parameters determined by the CLEO  \cite{libby} and BESIII \cite{kspipi_cisi,kskk_cisi} collaborations. The loss of information that results from binning data in the model-independent method is compensated by the removal of potentially large and difficult-to-determine systematic uncertainties from the assumptions of the amplitude model, which are unavoidable in the model-dependent method. These model-dependent uncertainties have been estimated to be between $3^\circ$ \cite{babarmoddep} to $9^\circ$ \cite{bellemoddep}. Due to this systematic limitation, the world-leading measurement \cite{lhcb_gamma} and most other recent measurements, use the model-independent method.

In this paper we report a model-independent measurement of $\phi_3$ using \linebreak $B^+\to D\left( K^{0}_{\rm S}h^+h^-\right)h^+$ decays produced in asymmetric $e^+e^-$ collisions at a centre-of-mass (c.m.)\ energy corresponding to the mass of the $\Upsilon(4S)$ resonance. The data sample used corresponds to integrated luminosities of \SI{711}{fb^{-1}}
and \SI{128}{fb^{-1}} accumulated by the Belle and Belle~II experiments, respectively. This result is the first obtained from a combined analysis of Belle and Belle~II data. The results in this paper supersede a previous model-independent measurement using the full Belle data set \cite{anton}. The measurement presented here includes significant improvements compared to the earlier Belle measurement, which are related to background reduction, $K^0_{\rm S}$ meson selection, reduced systematic uncertainties, and the addition of $D\to K_{\rm S}^0 K^+K^-$ decays.  

The organisation of this paper is as follows. Section~\ref{sec:bpggsz} presents an outline of the analysis, which includes the formalism related to the model-independent method. Section~\ref{sec:detector} describes the Belle and Belle~II experiments. Section~\ref{sec:data} introduces the data sets and simulation samples used to perform the measurements. Sections~\ref{sec:selection}~and~\ref{sec:xy_parameters} contain descriptions of the $B^+\to D\left( K^{0}_{\rm S}h^+h^-\right)h^+$ candidate selection and {\it CP}-violating observable extraction, respectively. Section~\ref{sec:systematics} describes the estimation of systematic uncertainties. Sections~\ref{sec:phi3} and \ref{sec:conclusion} present the results and conclusions, respectively.     
%%%%%%%%%%%%%%%%%%%%%%%%%%%%%%%%%%%%%%%%%%%%%%%%%%%%%%%%%%%%%%%%%%%%%%%%%%
\section{Analysis overview and formalism}
\label{sec:bpggsz}
The analysis proceeds by selecting samples of $B^+\to Dh^+$ decays in the Belle and Belle~II data sets. The $B^+\to D\pi^+$ decay is selected along with the $B^+\to DK^+$ decay because it is a more abundant and topologically identical control sample. An unbinned maximum likelihood fit to the combined sample of $B^+\to Dh^+$ candidates is used to determine {\it CP}-violating observables.  In addition, nuisance parameters are determined, which limits dependence upon simulated data. The fit is to data categorised in several ways, including by the bin in which they are reconstructed in the $D$-decay Dalitz plot. To be model-independent, the fit relies upon the strong-phase parameters of the $D$ decay measured in each of these bins, which are taken as external inputs. The {\it CP}-violating observables are then interpreted as constraints on $\phi_3$ and hadronic parameters related to the $B$-decay amplitude. To avoid experimental bias  analysis procedures are devised and validated on simulated samples before being applied to data. The remainder of this section describes the model-independent formalism used.

The two interfering decays sensitive to $\phi_3$ are $B^{+}\to \bar{D}{}^0 K^+$ and $B^{+}\to D^0 K^+$, where the latter is both CKM- and colour-suppressed compared to the former. Thus, we write the total $B^+\to D\left(K^{0}_{\rm S}h^+ h^-\right)K^+$, amplitude as 
\begin{equation}
 A_{B^+}\left(m^2_{-},m^2_{+}\right) \propto A_{\bar{D}}\left(m^2_{-},m^2_{+}\right) + r_B^{DK}e^{i\left(\delta_B^{DK}-\phi_3\right)}A_{D}\left(m^2_{-},m^2_{+}\right)\; ,
 \label{eq:bplusamp}
\end{equation}
where $A_{\bar{D}}\left(m^2_{-},m^2_{+}\right)$ $\left[A_{D}\left(m^2_{-},m^2_{+}\right)\right]$ is the $\bar{D}{}^0\to K^0_{\rm S}h^{+}h^{-}$ $\left[D^0\to K^0_{\rm S}h^{+}h^{-}\right]$ decay amplitude at a point in the Dalitz plot described by $m_-^2$ and $m_+^2$, which are the squared invariant masses of the $K^{0}_{\rm S}h^-$ and $K^{0}_{\rm S}h^+$ particle combinations, respectively. Here $r_{B}^{DK}$ and $\delta_{B}^{DK}$ are the ratio of the magnitudes of the suppressed to favoured $B^+\to DK^+$ amplitudes and the relative strong-phase difference between them, respectively. The world-average value of $r_B^{DK}$ is $0.0996\pm 0.0026$ \cite{hflav}, which means that the direct {\it CP}-violating effects are of the order 10\%. The expression for the $B^-\to D\left(K^{0}_{\rm S}h^+ h^-\right)K^-$ amplitude $A_{B^-}$ is obtained from 
Eq.~(\ref{eq:bplusamp}) by substituting $\phi_3\to -\phi_3$ and $A_{D}\left(m_-^2,m_+^2\right)\longleftrightarrow A_{\bar{D}}\left(m_-^2,m_+^2\right)$. In this paper {\it CP} violation in $D$ decays is considered to be negligible, such that $A_{\bar{D}}\left(m_-^2,m_+^2\right)=A_{D}\left(m_+^2,m_-^2\right)$.  

In the model-independent method, the $D$-decay Dalitz plot is divided into $2\times \mathcal{N}$ bins that are indexed from $i=-\mathcal{N}$ to $i=\mathcal{N}$, with $i=0$ excluded. The bins are defined symmetrically about the line $m_+^2=m_-^2$ such that if the point $\left(m^2_-,m^2_+\right)$ lies within bin $i$ then point $\left(m^2_+,m^2_-\right)$ lies within bin $-i$; bins in which $m_-^2 > m_+^2$ are labelled with $i>0$. The strong phase of the $D^0$-decay amplitude at a point $\left(m_-^2,m_+^2\right)$ is written as $\delta_D\left(m_-^2,m_+^2\right)$, from which the $D$-amplitude-weighted average of the cosine of the strong-phase difference between $D^0$ and $\bar{D}{}^0$ decays within bin $i$ is defined as \cite{bpggsz1}
\begin{equation}
    c_i = \frac{\int_i\:dm_-^2\:dm_+^2\left|A_D\left(m_-^2,m_+^2\right)\right|\left|A_D\left(m_+^2,m_-^2\right)\right|\cos\left[\delta_D\left(m_-^2,m_+^2\right)-\delta_D\left(m_+^2,m_-^2\right)\right]}{\sqrt{\int_i\:dm_-^2\:dm_+^2\left|A_D\left(m_-^2,m_+^2\right)\right|^2\int_i\:dm_-^2\:dm_+^2\left|A_D\left(m_+^2,m_-^2\right)\right|^2}} \;,
\end{equation}
where the integral is over the $i$-th bin. The $D$-amplitude-weighted average of the sine of the strong-phase difference within a bin $s_i$ is defined in an analogous manner. These definitions result in the conditions $c_i=c_{-i}$ and $s_i=-s_{-i}$. Further, the $A_{B^+}$ $(A_{B^-})$ amplitudes can be squared and integrated over each bin to give the expectation for the $B^+$ $\left(B^-\right)$ yields in each bin $N^+_i$ $\left(N^-_i\right)$,
\begin{eqnarray}
 N_{i}^+ & = & h_{B^+}\left[F_{-i}+ \left\{\left(x_+^{DK}\right)^2+\left(y_+^{DK}\right)^2\right\}F_{i\phantom{-}}+2\sqrt{F_iF_{-i}}\left(x_{+}^{DK}c_i-y_{+}^{DK}s_i\right)\right]\;,\nonumber\\
 N_{-i}^+ & = & h_{B^+}\left[F_{i\phantom{-}}+ \left\{\left(x_+^{DK}\right)^2+\left(y_+^{DK}\right)^2\right\}F_{-i}+2\sqrt{F_iF_{-i}}\left(x_{+}^{DK}c_i+y_{+}^{DK}s_i\right)\right]\;,\nonumber\\
 N_{i}^- & = & h_{B^-}\left[F_{i\phantom{-}}+ \left\{\left(x_-^{DK}\right)^2+\left(y_-^{DK}\right)^2\right\}F_{-i}+2\sqrt{F_iF_{-i}}\left(x_{-}^{DK}c_i+y_{-}^{DK}s_i\right)\right]\;, \nonumber\\
N_{-i}^- & = & h_{B^-}\left[F_{-i}+ \left\{\left(x_-^{DK}\right)^2+\left(y_-^{DK}\right)^2\right\}F_{i\phantom{-}}+2\sqrt{F_iF_{-i}}\left(x_{-}^{DK}c_i-y_{-}^{DK}s_i\right)\right]\;,
\label{eq:byields}
\end{eqnarray}
where $h_{B^{\pm}}$ are independent normalisation constants, $x_\pm^{DK} = r_B^{DK}\cos\left(\delta_B^{DK}\pm\phi_3\right)$, and $y_{\pm}^{DK}=r_B^{DK}\sin\left(\delta_B^{DK}\pm\phi_3\right)$.\footnote{{\it CP} violation in the total decay rate is negligible, i.e., when integrated over the full Dalitz plot \cite{lhcb_gamma}. Therefore, to avoid any bias due to detector asymmetry, independent normalisation constants are used for $B^{+}$  and $B^-$ decays.} Here $F_i$ is the fractional yield in each bin for a pure sample of $D^0$ decays accounting for any experiment-specific efficiency variation over the Dalitz plot \cite{binflip}:
\begin{equation}
    F_{i} = \frac{\int_i\: dm_-^2\:dm_+^2 \left|A_D\left(m_-^2,m_+^2\right)\right|^2 \eta\left(m_-^2,m_+^2\right)}
  {\sum_j\int_j\: dm_-^2\:dm_+^2 \left|A_D\left(m_-^2,m_+^2\right)\right|^2 \eta\left(m_-^2,m_+^2\right)}, 
 \end{equation}
where the sum in the denominator is over all $2\mathcal{N}$ bins and $\eta\left(m_-^2,m_+^2\right)$ is the acceptance profile over the Dalitz plot, which depends on both laboratory-frame decay kinematics and the experimental setup. 
%Neglecting {\it CP} violation in $D$ decay, the fractional yield in each bin for a pure sample of $\bar{D}{}^0$ decays is $\bar{F}_{i}=F_{-i}$.

The $4\mathcal{N}$ observables defined in Eqs.~(\ref{eq:byields}) depend upon $4\mathcal{N}+4$ parameters: $\phi_3$, $r_B^{DK}$, $\delta_B^{DK}$, $c_i$, $s_i$, $F_i$ and $h_{B^{\pm}}$. Therefore, independent measurements of the $2\mathcal{N}$ strong-phase parameters $c_i$ and $s_i$ are used to determine the other parameters from these yields. Furthermore, the $2\mathcal{N}-1$ fractional yields $F_{i}$ can be constrained using the simultaneous analysis of $B^+\to D\pi^+$ decays \cite{lhcb_gamma}, which have a branching fraction an order of magnitude larger than $B^+\to DK^+$ \cite{pdg}. An analogous set of yields for $B^+\to D\pi^+$ exist as those defined in Eq.~(\ref{eq:byields}), which depend upon $x_\pm^{D\pi} = r_B^{D\pi}\cos\left(\delta_B^{D\pi}\pm\phi_3\right)$ and $y_{\pm}^{D\pi}=r_B^{D\pi}\sin\left(\delta_B^{D\pi}\pm\phi_3\right)$, where $r_B^{D\pi}$ and $\delta_B^{D\pi}$ are the magnitude ratio and strong-phase difference between the Cabibbo- and colour-suppressed $B^+\to D^0\pi^+$ amplitude and the favoured $B^+\to \bar{D}{}^0\pi^+$ amplitude. The value of $r_B^{D\pi}$ is approximately 20 times smaller\footnote{This factor is  $\tan^2{\theta_{\rm C}}$, where $\theta_{\rm C}$ is the Cabibbo angle, which is the relative Cabibbo suppression of $r_B^{D\pi}$ with respect to $r_B^{DK}$.} than $r_B^{DK}$ so the sensitivity to {\it CP} violation is significantly reduced in comparison to that from $B^+\to DK^+$ decays. Given the almost identical kinematic properties between $B^+\to DK^+$ and $B^+\to D\pi^+$, the $F_{i}$ parameters are common for the two sets of yields within a single experiment. We adopt a parameterisation \cite{garra1,garra2} that utilises the common dependence on $\phi_3$ of the $B^+\to DK^+$ and $B^+\to D\pi^+$  yields by introducing the single complex variable, 
\begin{equation}
    \xi^{D\pi} = \left(\frac{r_B^{D\pi}}{r_B^{DK}}\right)e^{i\left(\delta_B^{D\pi}-\delta_{B}^{DK}\right)} \;.
\end{equation}
Defining $x_{\xi}^{D\pi}\equiv\mathrm{Re}\left(\xi^{D\pi}\right)$ and $y_{\xi}^{D\pi}\equiv\mathrm{Im}\left(\xi^{D\pi}\right)$ we can write
\begin{equation}
    x_{\pm}^{D\pi} = x_{\xi}^{D\pi}x_{\pm}^{DK}-y_{\xi}^{D\pi}y_{\pm}^{DK}\;,\;\;\; y_{\pm}^{D\pi} = x_{\xi}^{D\pi}y_{\pm}^{DK}+y_{\xi}^{D\pi}x_{\pm}^{DK} \;.
\end{equation}
The values of $x_{\pm}^{DK}$, $y_{\pm}^{DK}$, $x_{\xi}^{D\pi}$, $y_{\xi}^{D\pi}$ and $F_i$ are determined simultaneously from a fit to the $B^{+}\to D h^+$ candidates. 
The advantages of this parameterisation are the inclusion of the $\phi_3$ sensitivity from $B^{+}\to D\pi^+$ in the determination of $x_{\pm}^{DK}$ and $y_{\pm}^{DK}$ as well as much improved fit stability \cite{lhcb_gamma}. Further, the determination of $F_i$ by simultaneously fitting $B^+\to Dh^+$ removes a source of systematic uncertainty in this analysis compared to that reported in Ref.~\cite{anton}. The previous Belle analysis \cite{anton} determined the values of $F_i$ from a sample of $D^{*+}\to D^0\pi^+$ decays. The differing kinematic properties of the $B^+\to D\pi^+$ and $D^{*+}\to D^0\pi^+$ decays resulted in different $\eta\left(m_-^2,m_+^2\right)$ acceptance functions for the two samples, which was a source of systematic uncertainty.  

\begin{figure}[!t]
\centering
	\begin{tabular}{c c}
		\includegraphics[scale=0.18]{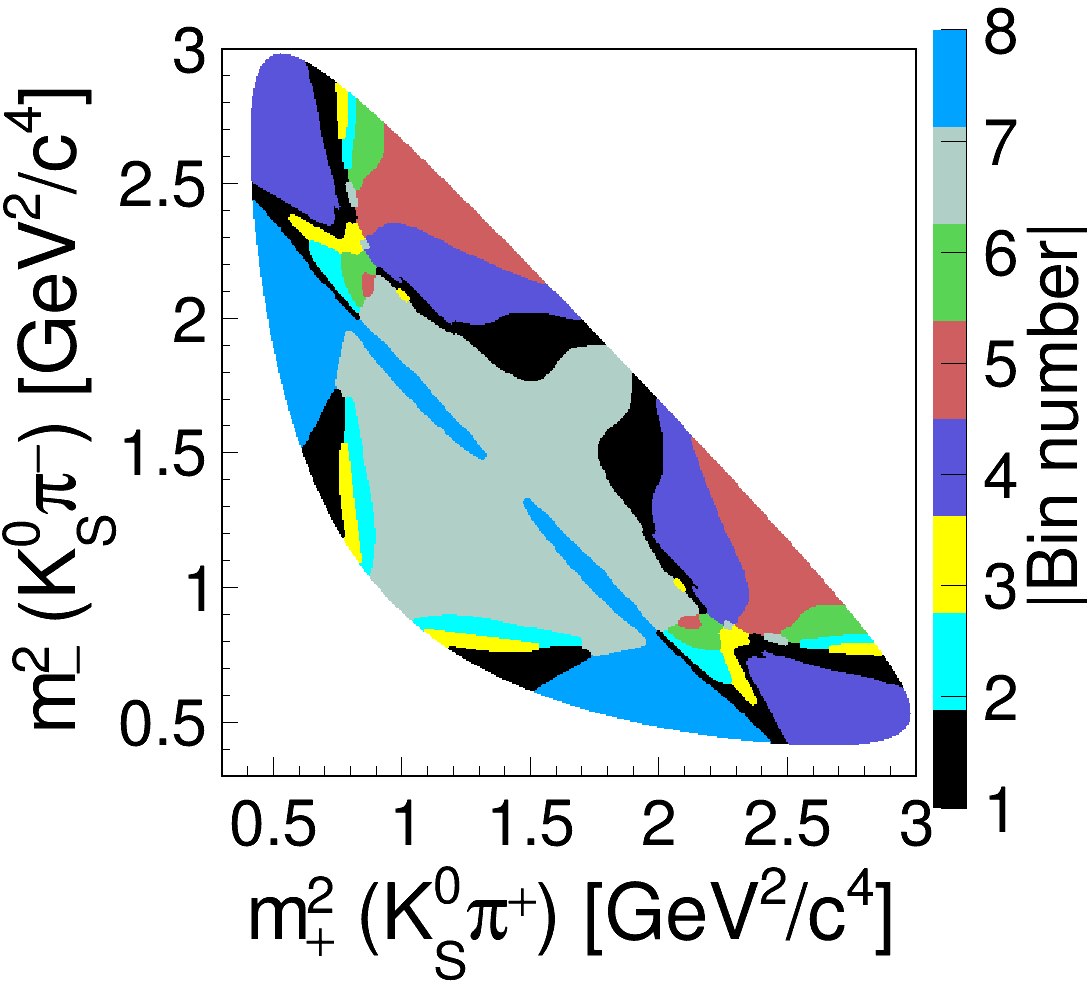} &
		\includegraphics[scale=0.18]{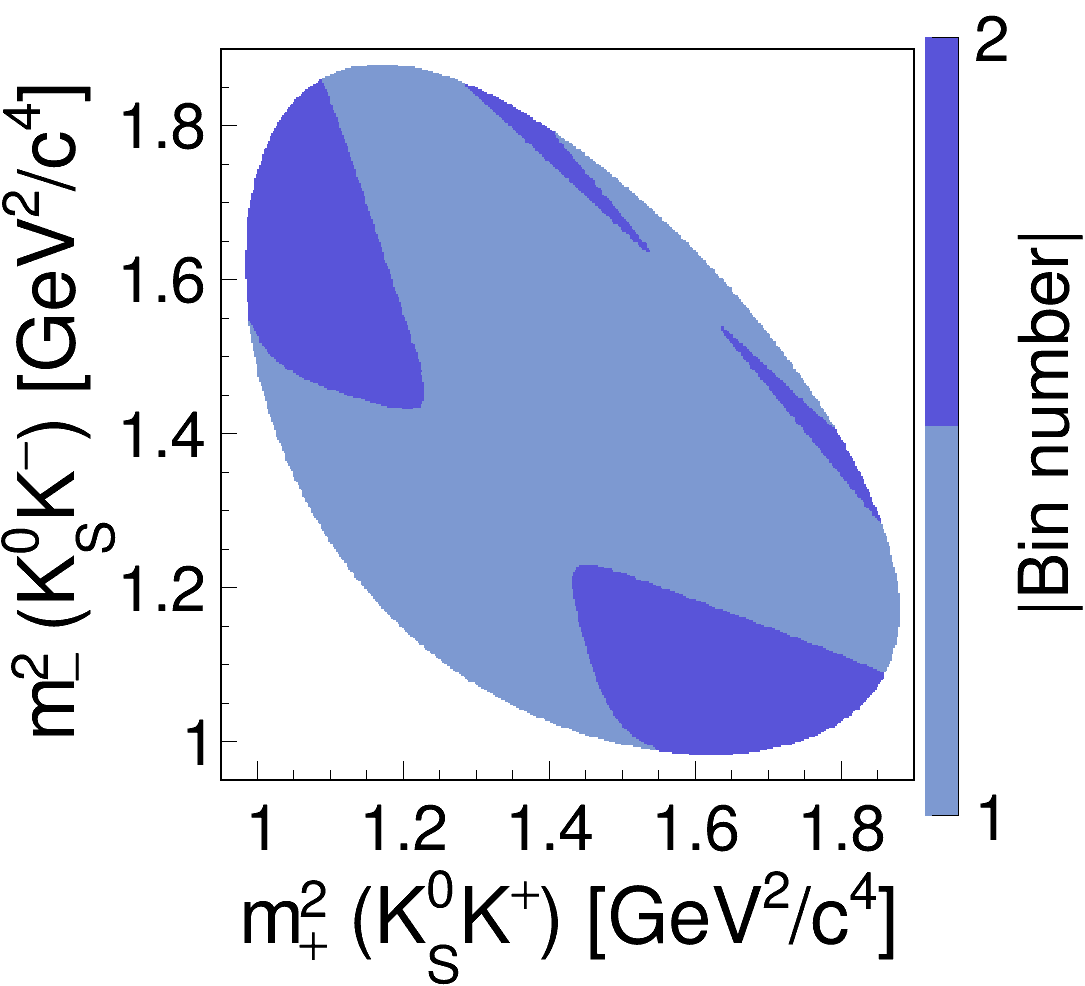}
	\end{tabular}
\caption{Binning schemes used for (left) $B^+\to D\left(K^0_{\rm S}\pi^+\pi^-\right)K^+$ decays and (right) $B^+\to D\left(K^0_{\rm S}K^+K^-\right)K^+$ decays.}
\label{fig:binning}
\end{figure}

There are three binning schemes, for both $D^0\to K^{0}_{\rm S}\pi^+\pi^-$ and $D^0\to K^{0}_{\rm S}K^+ K^-$ decays, for which $c_i$ and $s_i$ have been measured \cite{kspipi_cisi,kskk_cisi}. We adopt the $\mathcal{N}=8$ optimal binning for $B^{+}\to D\left(K^{0}_{\rm S}\pi^+\pi^-\right)h^+$ decays, which has been shown to have approximately 90\% of the statistical sensitivity of an unbinned analysis \cite{antonbinning,libby}. We adopt the $\mathcal{N}=2$ equal-strong-phase binning for $B^{+}\to D\left(K^{0}_{\rm S}K^+K^-\right)h^+$ decays, which has better fit stability than the $\mathcal{N}=3$ and $4$ schemes \cite{kskk_cisi} given the limited size of $B^+\to D\left(K_{\rm S}^{0}K^+K^-\right)K^+$ event sample. Figure~\ref{fig:binning} shows the two binning schemes used. The measurements of $c_i$ and $s_i$ ignore the effects of $D$-mixing and assume {\it CP}-conservation in $D$ decay. Ignoring both these effects in the strong-phase and model-independent $B^+\to D\left(K^0_{\rm S}h^+h^-\right)h^+$ analyses, as in this paper, results in negligible bias \cite{d-mixing}. The potential bias of ignoring $K^0$ {\it CP}-violation and regeneration has also been extensively studied \cite{mikkel} and a bias of $
\left(0.4\pm 0.1\right)^{\circ}$ on $\phi_3$ is reported. This bias is negligible in comparison to the current statistical precision and is not considered further.

%%%%%%%%%%%%%%%%%%%%%%%%%%%%%%%%%%%%%%%%%%%%%%%%%%%%%%%%%%%%%%%%%%%%%%%%%%%
\section{Belle and Belle~II detectors}
\label{sec:detector}
The Belle detector \cite{belle1,belle2} was located at the interaction point (IP) of the KEKB asymmetric-energy $e^+e^-$ collider \cite{kekb1,kekb2}. The energies of the electron and positron beams were \SI{8.0}{GeV} and \SI{3.5}{GeV}, respectively. The detector subsystems most relevant for this study are the following: the silicon vertex detector and central drift chamber (CDC), for charged particle
tracking and measurement of energy loss due to ionisation, and the aerogel threshold Cherenkov counters and time-of-flight scintillation counters, for particle identification (PID). These subsystems were situated in a magnetic field of \SI{1.5}{T}. More detailed descriptions of the Belle detector can be found in Refs.~\cite{belle1,belle2}.

The Belle~II detector~\cite{belle2_TDR} is located at the IP of the SuperKEKB asymmetric-energy $e^+e^-$ collider~\cite{superkekb}. The energies of the electron and positron beams are \SI{7.0}{GeV} and \SI{4.0}{GeV}, respectively. The target instantaneous luminosity of SuperKEKB is a factor of 30 greater than KEKB, which will lead to significantly larger beam-related backgrounds. Therefore, the Belle~II detector is an upgraded version of the Belle detector that is designed to cope with increased level of background. The detector contains several completely new subsystems, as well as substantial upgrades to others. The innermost subdetector is the vertex detector (VXD), which uses position-sensitive silicon sensors to precisely sample the trajectories of charged particles (tracks) in the vicinity of the IP. The VXD includes two inner layers of pixel sensors and four outer layers of double-sided silicon microstrip sensors. The second pixel layer is currently incomplete covering only  one sixth of the azimuthal angle. Charged-particle momenta and charges are measured by a new large-radius, helium-ethane, small-cell CDC, which also offers charged-particle-identification information through a measurement of specific ionisation. The Belle PID system has been replaced. A Cherenkov-light angle and time-of-propagation (TOP) detector surrounding the CDC provides charged-particle identification in the central detector volume, supplemented by proximity-focusing, aerogel, ring-imaging Cherenkov (ARICH) detectors in the forward  region with respect to the electron beam. The Belle CsI(Tl) crystal electromagnetic calorimeter, the Belle solenoid and iron flux return are reused in the Belle~II detector. The electromagnetic calorimeter readout electronics have been upgraded and the instrumentation in the flux return to identify $K^0_{\rm L}$ mesons and muons has been replaced. 
%%%%%%%%%%%%%%%%%%%%%%%%%%%%%%%%%%%%%%%%%%%%%%%%%%%%%%%%%%%%%%%%%%%%%%%%%%
\section{Data sets}
\label{sec:data}
The analysis uses $e^+e^-$ collision data collected at a c.m.\ energy corresponding to the mass of the $\Upsilon(4S)$ resonance. The integrated luminosities of the samples collected by Belle and Belle~II are \SI{711}{fb^{-1}}
and \SI{128}{fb^{-1}}, respectively. 
%A total of $xx \times 10^6 $ of $B\bar{B}$ pairs are used for this analysis. 

Simulation samples are used to optimise the selection criteria, estimate signal efficiencies, train multivariate discriminants, identify various sources of background and develop a model to fit data. The signal and $ e^{+}e^{-} \rightarrow \Upsilon(4S) \rightarrow B\bar{B} $ simulation samples are generated using the \textsc{EvtGen} software package~\cite{evtgen}. 
Samples of signal events are generated with the $D$-decay products following both resonant and non-resonant distributions.
The Belle simulation samples of continuum background events $ e^{+}e^{-} \rightarrow q\bar{q}$, where $q = u, d, s, c$, are generated by \textsc{Pythia}~\cite{pythia}. The Belle~II $e^{+}e^{-} \rightarrow  q\bar{q}$ simulation sample is generated using the \textsc{KKMC}~\cite{kkmc} generator interfaced with  \textsc{Pythia}. The \textsc{EvtGen} package also simulates the decay of short-lived particles.
The Belle (Belle~II) simulation samples use a \textsc{Geant3}-based simulation package~\cite{geant3} (\textsc{Geant4}~\cite{geant4}) to model the detector response to the final-state particles. Final-state radiation effects are taken into account by including the \textsc{Photos} model~\cite{photos}. Belle simulation includes the effect of beam background by overlaying data taken that is unrelated to $e^+e^-$ collisions (random triggers). Belle~II simulation samples include the effect of simulated beam-induced background caused by the Touschek effect (scattering and loss of beam particles) and by beam-gas scattering, as well as luminosity-dependent backgrounds caused by Bhabha scattering and two-photon quantum electrodynamic processes \cite{beast}.

%%%%%%%%%%%%%%%%%%%%%%%%%%%%%%%%%%%%%%%%%%%%%%%%%%%%%%%%%%%%%%%%%%%%%%%%%%
\section{Reconstruction and event selection}
\label{sec:selection}
We use the Belle~II analysis software framework (basf2)~\cite{basf2} for decay-chain reconstruction. The Belle~II data are processed using this framework, whereas the tracks and clusters in the processed Belle data are converted to basf2 format using the \textsc{B2BII} software package~\cite{b2bii}. Hence, the reconstruction software is identical for both the data samples.

The selections are similar to those described in Ref.~\cite{anton}. We apply nearly identical event selection criteria, with some slight differences due to the different detector configurations.

An overview of the selection procedure is as follows. We reconstruct the decay mode $B^+ \to D\left(K_{\rm S}^0h^{+}h^{-}\right)h^+$.  We select the $\pi^+$ and $K^+$ candidate tracks, as well as $K_{\rm S}^0$ candidates, with selections designed to maximise the product of efficiency and purity. From these samples, $D\to K^0_{\rm S}h^+h^-$ candidates are reconstructed, which are further combined with an $h^+$ track to form a $B^+$ candidate. Vertex and kinematic fits are performed, which constrain the $B$-decay products to a common vertex. Additional criteria are developed to suppress background events. The remainder of this section describes the motivation for the various selection requirements. 

Charged particles, $\pi^+$ and $K^+$, consistent with originating from $e^+e^-$ collisions are selected by requiring the distance of closest approach to the IP to be less than \SI{0.2}{cm} in the plane transverse to and \SI{1.0}{cm} along the $z$ direction; in both the Belle and Belle~II coordinate system the $z$-axis is defined to lie along the axis of symmetry of the solenoid approximately in the direction of the $e^-$ beam. These charged particles are then identified as either kaons or pions by using the information from the CDC, the aerogel threshold Cherenkov counters and time-of-flight scintillation counters of Belle, and all subdetectors of Belle~II, though that from CDC, TOP and ARICH is most significant. To identify kaon and pion candidates, we use the ratio $\mathcal{R}_{K/\pi}=\mathcal{L}(K)/\left[\mathcal{L}(K)+\mathcal{L}(\pi)\right]$, where $\mathcal{L}(h)$ is the likelihood for the particle $h$ to produce the observed particle-identification signal associated with the track. 
We use a requirement of $\mathcal{L}(K/\pi)>0.6$ to separate the kaons from pions coming directly from the $B^+\to Dh^+$ decays. 
The kaon-identification efficiency is 84\% (79\%) and the probability of a pion being misidentified as a kaon is 8\% (7\%) in Belle~\cite{pid} (Belle~II) data. A less restrictive requirement of $\mathcal{L}(K/\pi)>0.2$ is applied to the kaons used to reconstruct the $D\to K^0_{\rm S}K^+K^-$ candidates, which suppresses the $K$-$\pi$ misidentification rate while maintaining high efficiency. For Belle~II data, there is an additional requirement of $\cos\theta>-0.6$, where $\theta$ is the polar angle in the laboratory frame, applied to the $\pi$ or $K$ candidates that come directly from the $B$ meson decays. This criterion removes the tracks outside the acceptance region of the TOP and ARICH, which reduces the $K$-$\pi$ misidentification rate. Furthermore, this requirement increases the separation of the contribution from misidentified $B^+ \to D\pi^+$ decays from that of correctly reconstructed $B^+ \to DK^+$ decays~\cite{moriond2021}. We do not apply this selection to the Belle data because the contribution to the misidentified $B^+ \to D\pi^+$ decays is much smaller in this region, due to fewer tracks having $\cos{\theta}< -0.6$ as a result of the higher c.m.\ boost of the KEKB collider

Candidate $K_{\rm S}^0$ mesons are reconstructed from pairs of oppositely charged particles, selected assigning the pion-mass hypothesis, that originate from a common vertex. The $K_{\rm S}^0$ candidates are required to have an invariant mass in the interval $0.487-$\SI{0.508}{GeV/{\it c}^2}, which corresponds to $\pm 3\sigma$ around the known $K_{\rm S}^0$ mass~\cite{pdg}. Here $\sigma$ is the invariant mass resolution. A multivariate technique is used to improve the purity of the $K_{\rm S}^0$ candidate sample by rejecting combinatorial background. Algorithms based on a neural network (NN)~\cite{nisks} and a fast boosted decision tree (FastBDT) \cite{fbdt}  are used to identify the background in Belle and Belle~II data, respectively. Five input variables are common to the Belle and Belle~II algorithms: the angle between the momentum vector and the vector between the IP and the decay vertex of the $K_{\rm S}^0$ candidate, the longer and shorter distances of closest approach between the extrapolated tracks of the pion candidates and the IP, the flight distance of the $K_{\rm S}^0$ candidate projected on to the plane transverse to the $z$ axis, and the difference between the observed and known $K^{0}_{\rm S}$ mass divided by the uncertainty in the observed mass. The following seven additional variables are input to the Belle NN discriminator: $K_{\rm S}^0$ momentum in the lab frame, shortest distance between the two track helices projected along the $z$ axis, the angle between pion momentum in the $K_{\rm S}^0$ rest frame and the boost direction between the laboratory frame and the  $K_{\rm S}^0$ rest frame, and the number (presence) of CDC (silicon vertex detector) hits for each pion track. The efficiency and purity of the $K_{\rm S}^0$ selection are 87\% (91\%) and 99\% (97\%), respectively, for Belle (Belle~II). Here we define efficiency as the fraction of correctly reconstructed $K_{\rm S}^0$ candidates retained by the classifier and the purity as the fraction of correctly reconstructed $K_{\rm S}^0$ candidates in the selected sample. The values of efficiency and purity are determined from simulation samples.  Achieving a similar performance in the product of efficiency and purity in Belle~II and Belle motivates the reduced set of input variables; fewer input variables makes the algorithm more robust against any disagreements between data and simulation distributions. A kinematic-constrained fit is performed on selected candidates by restricting the reconstructed $\pi^+\pi^-$ mass to the known $K^0_{\rm S}$ mass \cite{pdg} to improve the four-momentum resolution.

    The $D\to K^0_{\rm S}\pi^+\pi^-$ $\left(D\to K^0_{\rm S}K^+K^-\right)$ meson candidates are reconstructed from a pair of oppositely charged pion (kaon) tracks and a $K_{\rm S}^0$ candidate. The invariant mass of the $D$-decay products $m\left(K_{\rm S}^0h^-h^+\right)$  is required to be in the range $1.85-$\SI{1.88}{GeV/{\it c}^2}. This interval corresponds to $\pm 3\sigma$ around the nominal $D$ mass, where $\sigma$ is the $m\left(K_{\rm S}^0h^-h^+\right)$ resolution. A kinematic-constrained fit is performed by restricting the reconstructed mass to the known $D^0$ mass \cite{pdg} to improve the four-momentum resolution. After the kinematic constraint some candidates have values of $\left(m_{+}^2,m_{-}^2\right)$ that lie outside the kinematic boundary of the $D\to K_{\rm S}^{0}h^+h^-$ Dalitz plot. These candidates are not considered further because they cannot be unambiguously associated with a Dalitz plot bin. Studies of simulated signal samples indicate this requirement reduces the signal selection efficiency by 0.3\%.%Further, half of the rejected signal events are misreconstructed as they contain a hadron that is not from the signal $D$ or $K_{\rm S}^0$ meson.}

A charged particle is combined with these $D$ candidates to form a $B^+ \to Dh^+$ candidate. The kinematic variables used to separate signal $B$ candidates are the beam-energy-constrained mass
\begin{equation}
m_{\rm bc} = c^{-2}\sqrt{E_{\rm beam}^{\ast 2} - \left| \smash[b]{\sum_i} \vec{\mathbf{p}}_{i}^{\ast}\right|^{2}c^2},
\end{equation}
and the beam-energy difference,
\begin{equation}
 \Delta E = \sum_i E_{i}^{\ast} - E_{\rm beam}^{\ast} ,
\end{equation}
where $ E_{\rm beam}^{\ast}$ is the beam energy in the c.m.\ frame and $E_{i}^{\ast}$ $\left(\vec{\mathbf{p}}_{i}^{\ast}\right)$ is the energy (momentum) of the $i$-th $B$-decay product in the c.m.\ frame. The candidates are selected in the range $m_{\rm bc} >\:$\SI{5.27}{GeV/{\it c}^2} and $-0.13 < \Delta E <\:$\SI{0.18}{GeV}. In subsequent analysis, $\Delta E$ is used as a fit variable. Generally, partially reconstructed $B^+ \to D^{(\ast)} K^{(\ast )+}$ decays peak at lower $\Delta E$ values and are difficult to model. An asymmetric $\Delta E$ window is chosen to exclude these peaking structures from the analysis. 

The dominant background comes from $e^{+}e^{-} \rightarrow q\bar{q} $ events, which are suppressed by using the difference in topology with respect to $ e^{+}e^{-} \rightarrow \Upsilon(4S) \rightarrow B\bar{B} $ decays. The $q\bar{q}$ events have particle momentum spatially correlated in two directions, forming jet-like structures. In contrast, the particles from a $B\bar{B}$ event are distributed uniformly over the $4\pi$ solid angle in the c.m.\ frame because the $B$ mesons do not have significant momentum.  Separate FastBDT classifiers are applied to Belle and Belle~II data because of the differing c.m. boost. The classifiers are trained using simulated samples with several discriminating observables related to both the whole event and signal-only angular configurations as input variables. We use five variables:
\begin{itemize}
    \item a likelihood ratio obtained from a Fisher discriminant formed from the modified Fox-Wolfram moments~\cite{ksfw1, ksfw2};
    \item the absolute value of the cosine of the angle between the signal-$B$-candidate momentum and the $z$ direction in the $e^{+}e^{-}$ c.m.\ frame;
    \item the cosine of the angle between the thrust axis of the signal $B$ candidate and the thrust axis of the rest-of-the-event (ROE);\footnote{The thrust axis for a collection of particle momenta is the direction along which the sum of the momenta is maximised. The rest-of-the-event refers to all measured tracks and clusters that are not used to reconstruct the signal.}
    \item the distance between the position of the signal $B$-decay vertex and the vertex of the ROE along the $z$ direction; and 
    \item the $B$ meson flavour-tagger output~\cite{BelleFT,FT}.\footnote{Reference~\cite{FT} describes two Belle~II flavour-tagging algorithms that have similar performance; we use the output of the category-based algorithm.} 
    \end{itemize}
These variables do not have any significant correlation with the signal extraction variable $\Delta E$. The FastBDT classifier output ($C$) distribution ranges from zero, where background events peak, to one, where the signal events peak. The signal extraction fit is to the two-dimensional distributions of $\Delta E$ and a transformed variable derived from $C$ $(C')$, which is defined in Sec.~\ref{sec:xy_parameters}. We require $C>0.15$ $(C>0.2)$, which rejects 67\% of the $q\bar{q}$ background with only 4\% signal loss when applied to the Belle (Belle~II) data set; this requirement allows a simpler modelling of the $C'$ distribution in the fit with negligible loss of statistical precision. The $e^+e^-\to c\bar{c}$ background is further suppressed by vetoing candidates arising from $D^{\ast +} \to D^0\pi^+$ decays. We veto events in the range $0.143 < \Delta m <\:$\SI{0.148} {GeV/{\it c}^2}, where $\Delta m$  is the mass difference between $D^*$ and $D$ candidates. This requirement rejects approximately 9\% of the remaining background after all other criteria are applied.

The possibility of additional peaking background from charmless $B$ decays is studied by performing the analysis on the sidebands of the $m\left(K^0_{\rm S}h^+h^-\right)$ distributions. Candidates are selected from the sideband without any kinematic constraint to the $D$ mass applied; the value of $m\left(K_{\rm S}^{0}h^+h^-\right)$ of selected sideband candidates is constrained to that of the midpoint of the sideband for subsequent analysis, such that they are treated in an identical manner to those in the $m\left(K_{\rm S}^{0}h^+h^-\right)$ signal region. Studies of the simulated samples indicate that contributions arise from $ B^+ \to K^{\ast +}\pi^-\pi^+ $, $ B^+ \to K^{\ast +}\rho^{0} $ and $ B^+ \to K^{\ast +}J/\psi$ decays. The yields are obtained from fits to sideband data following the procedure described in Section~\ref{sec:xy_parameters}. The expected yields of charmless background are found to be negligible compared to that of the signal. Therefore, the charmless background is not considered further in subsequent analysis, apart from as a source of systematic uncertainty. 

 The average $B$-candidate multiplicity of events that contain at least one candidate is approximately 1.02; events with three or more candidates are very rare. In events with more than one candidate, we retain the candidate with the minimum $\chi^2$ value calculated from the measured and known values of $m\left(K_{\rm S}^0 h^+h^-\right)$ and $m_{\rm bc}$ as well as the respective experimental resolutions; studies using simulated samples of signal decays show that this criterion selects the correctly reconstructed candidate approximately 65\% of the time for all the decay modes. The selection efficiencies are summarised in Table~\ref{table:eff}. The  $B^+\to D\left(K_{\rm S}^0\pi^+\pi^-\right)h^+$ selection efficiency is marginally improved at Belle II compared to Belle, whereas there is a significant improvement in the $B^+\to D\left(K_{\rm S}^{0}K^+ K^-\right)h^+$ selection efficiency. This increase is due to the $\mathcal{R}_{K/\pi}>0.2$ requirement on the $D$-decay products being more efficient when applied to Belle~II data compared to Belle data.  

\begin{table}[!htb]
\centering
\begin{tabular}{|l c c|}
\hline
Decay mode & \multicolumn{2}{c|}{Efficiency (\%)} \\
\hline
& Belle & Belle~II  \\ 
\hline 
$  B^+ \to D(K_{S}^0\pi\pi)\pi^+$ & 21.77 $\pm$ 0.03 & 22.13 $ \pm $ 0.05\\
$  B^+ \to D(K_{S}^0\pi\pi)K^+$ & 19.11 $\pm$ 0.03 & 19.79 $ \pm $ 0.05\\
$  B^+ \to D(K_{S}^0KK)\pi^+$ & 16.26 $\pm$ 0.03 & 18.16 $ \pm $ 0.05\\
$  B^+ \to D(K_{S}^0KK)K^+$ & 14.29 $\pm$ 0.03 & 16.73 $ \pm $ 0.05\\
\hline
\end{tabular}
\caption{$B^+\to D\left(K^0_{\rm S}h^+h^-\right)h^+$ signal efficiencies estimated from simulated non-resonant signal samples. The uncertainties are statistical only.} 
\label{table:eff}
\end{table}
%%%%%%%%%%%%%%%%%%%%%%%%%%%%%%%%%%%%%%%%%%%%%%%%%%%%%%%%%%%%%%%%%%%%%%%%%%%
\boldmath \section{$x_{\pm}^{DK}$ and $y_{\pm}^{DK}$ determination from $B^+ \to Dh^+$ decays} \unboldmath
\label{sec:xy_parameters}

We perform a simultaneous analysis of $B^+ \to DK^+$ and $B^+ \to D\pi^+$ decays. The $B^+ \to DK^+$ sample provides almost all of the sensitivity to {\it CP} violation, whereas $B^+ \to D\pi^+$ is primarily used to constrain the $F_i$ fractions. Furthermore, the $B^+\to D\pi^+$ background in the $B^+\to DK^+$ sample due to $K$-$\pi$ misidentification can be directly determined from the simultaneous analysis of these two decay channels. The whole $B^+\to Dh^+$ sample is divided into pion- and kaon-enhanced categories by applying the requirements $\mathcal{R}_{K/\pi}<0.6$ and $\mathcal{R}_{K/\pi}>0.6$ to the $h^+$, respectively.

The signal extraction involves a two-dimensional fit of the $\Delta E$ and $C'$ distributions. The latter variable is related to $C$, the output of the FastBDT, which is difficult to model analytically.\ We transform $C$ to $C'$ using an ordered list of $C$ values from the signal simulation sample such that $C'$ is the fraction of signal events present below a given value of $C$ in the list ~\cite{mutrans}. Therefore, the signal distribution of $C'$ is uniform between zero and one. In practice, due to small differences in the FastBDT-input-variable distributions between data and simulation, the signal is described by a straight line in data. The background $C^{\prime}$ distribution peaks at zero and is modelled by an exponential.

Along with the signal component, we have the following three background components in our fit:
\begin{itemize}
    \item $q\bar{q}$ background events;
    \item $B\bar{B}$ background events, coming from misreconstructed candidates from $ e^{+}e^{-} \rightarrow \Upsilon(4S) \rightarrow B\bar{B}$ decays; and
    \item peaking~background from $B^+\to Dh^+,~(h = \pi$ or $ K)$ decays due to  $K$-$\pi$ misidentification.
\end{itemize}
The two-dimensional probability density function (PDF) for each component is the product of the one-dimensional PDFs for $\Delta E$ and $C'$. Negligible correlations between $\Delta E$ and $C'$ in simulation samples support the validity of this assumption. To test whether there is any non-linear correlation, the fits are performed on simulated samples and no significant bias between the measured and generated parameters is observed, which also indicates there is negligible correlation.

The $B^+\to D\left(K^{0}_{\rm S}\pi^+\pi^-\right)h^+$ signal components are modelled with a sum of two Gaussian functions and an asymmetric Gaussian function for $\Delta E$, and a straight line for $C'$. These PDFs are common to both kaon- and pion-enhanced samples. The common mean of the $\Delta E$ functions and the slope of the straight line are extracted directly from the data, along with a scaling factor to the narrowest signal Gaussian to account for any difference in $\Delta E$ resolution between simulated and data samples; other parameters are fixed to those obtained from a fit to a large simulated sample of signal events. 

The $q\bar{q}$ background is modelled with a straight line for $\Delta E$ and a sum of two exponentials for $C'$. The slope of the straight line and the steeper exponential function's exponent are determined from the fit to data; other parameters are fixed to those obtained from a fit to the corresponding simulated sample. These PDFs are common to both pion- and kaon-enhanced samples. 

The misreconstructed $B\bar{B}$ background distribution is slightly different for pion- and kaon-enhanced samples so they are modelled separately. The $\Delta E$ ($C'$) PDFs are an exponential (straight line) and a sum of an exponential and straight line (second-order polynomial), for the pion- and kaon-enhanced samples, respectively. Only the slope of the kaon-enhanced $\Delta E$ PDF is determined from the fit. The PDF descriptions of the peaking backgrounds are the same as those of the signal components but with independent parameters.

For the $B^+\to D\left(K_{\rm S}^0K^+K^-\right)h^+$ final state, some component PDFs are modified. The $B\bar{B}$ background PDFs that describes $\Delta E$ and $C'$ for the kaon-enhanced sample are an exponential and a straight line, respectively; the slope of the exponential is a free parameter. The $q\bar{q}$ background of both samples is modelled with just a single exponential function. All other component PDFs are the same as those for the $B^+\to D\left(K_{\rm S}^0\pi^+\pi^-\right)h^+$ final state.

\begin{figure}[!t]
	\begin{tabular}{c c}
		\includegraphics[scale=0.37]{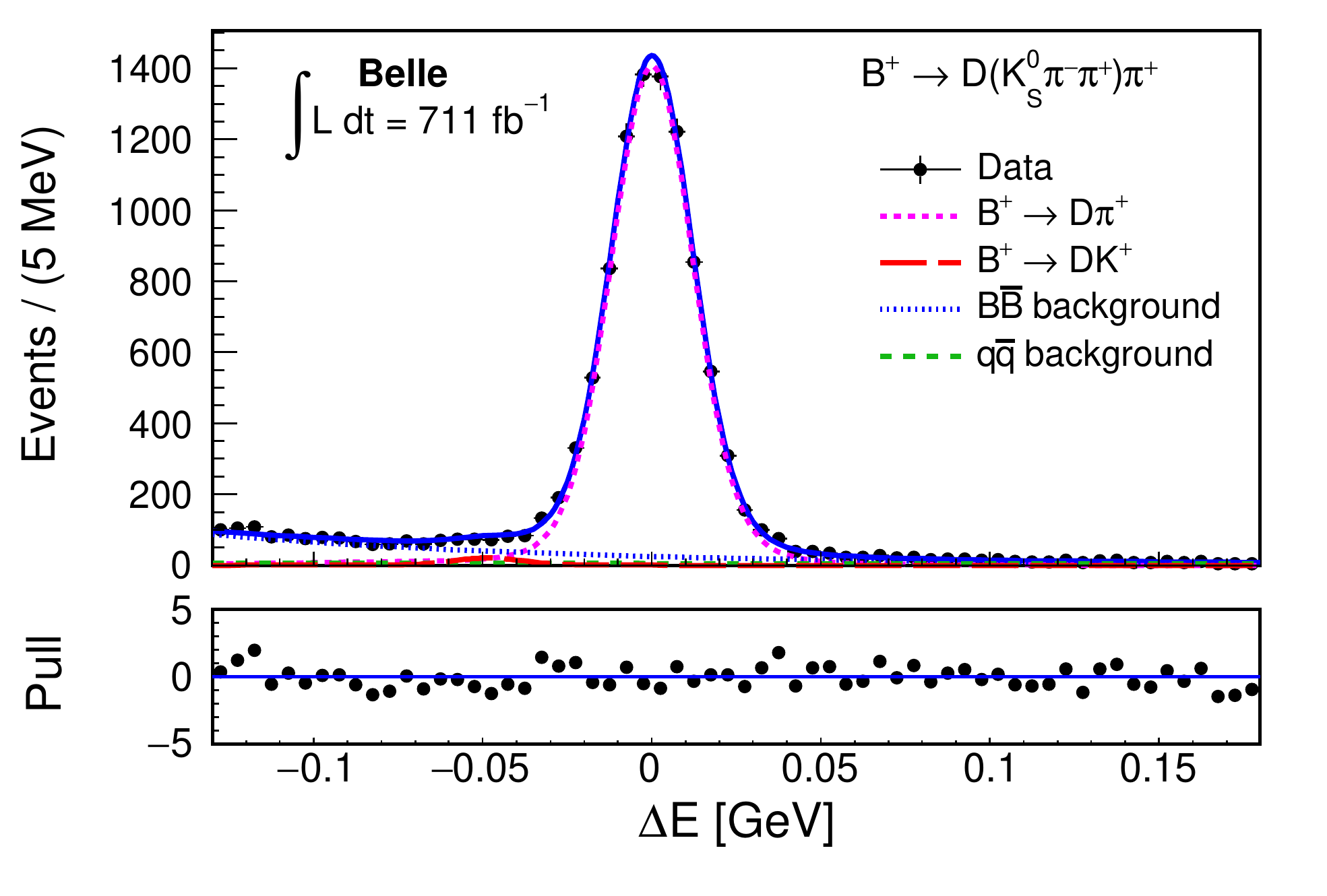} &
		\includegraphics[scale=0.37]{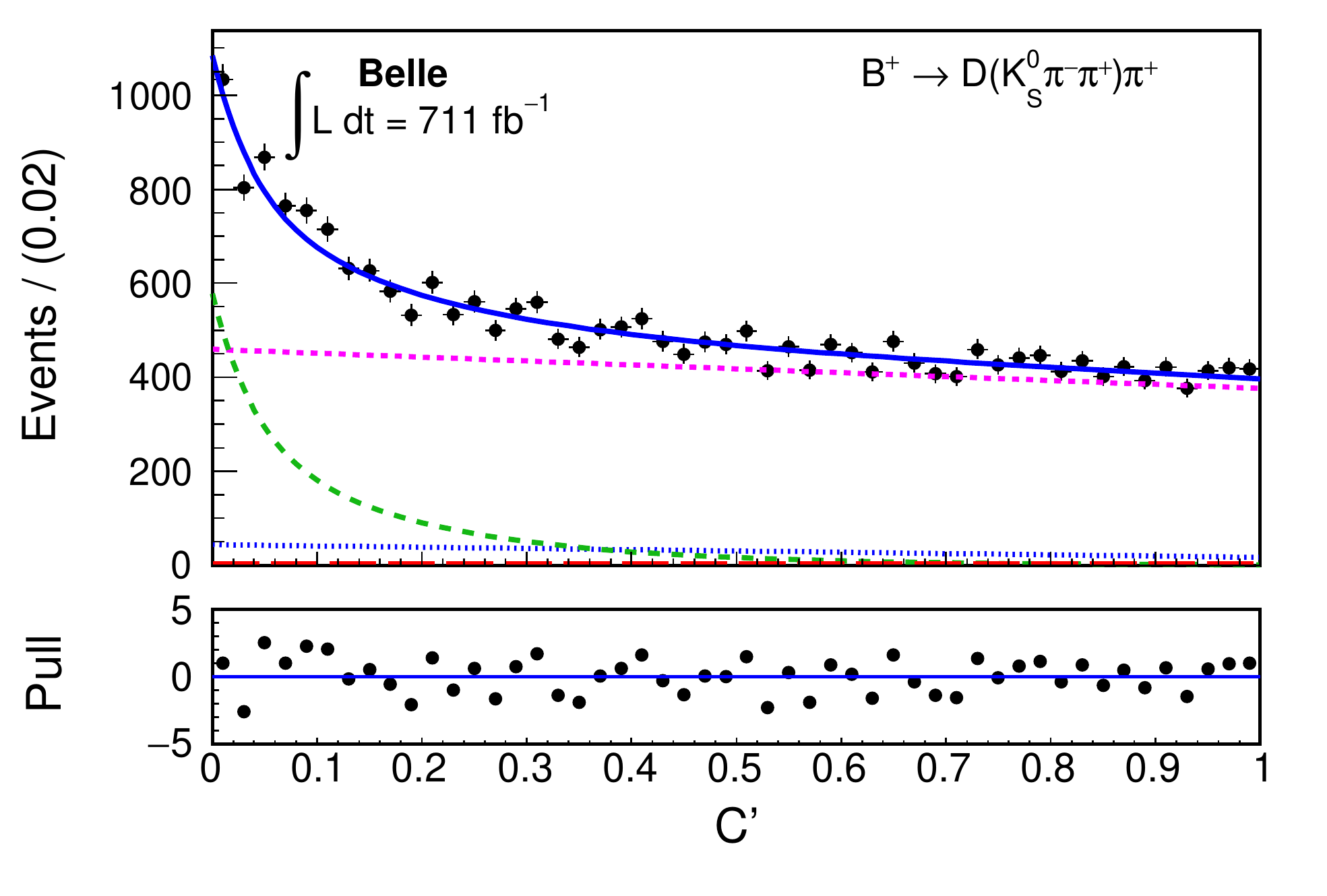} \\
		\includegraphics[scale=0.37]{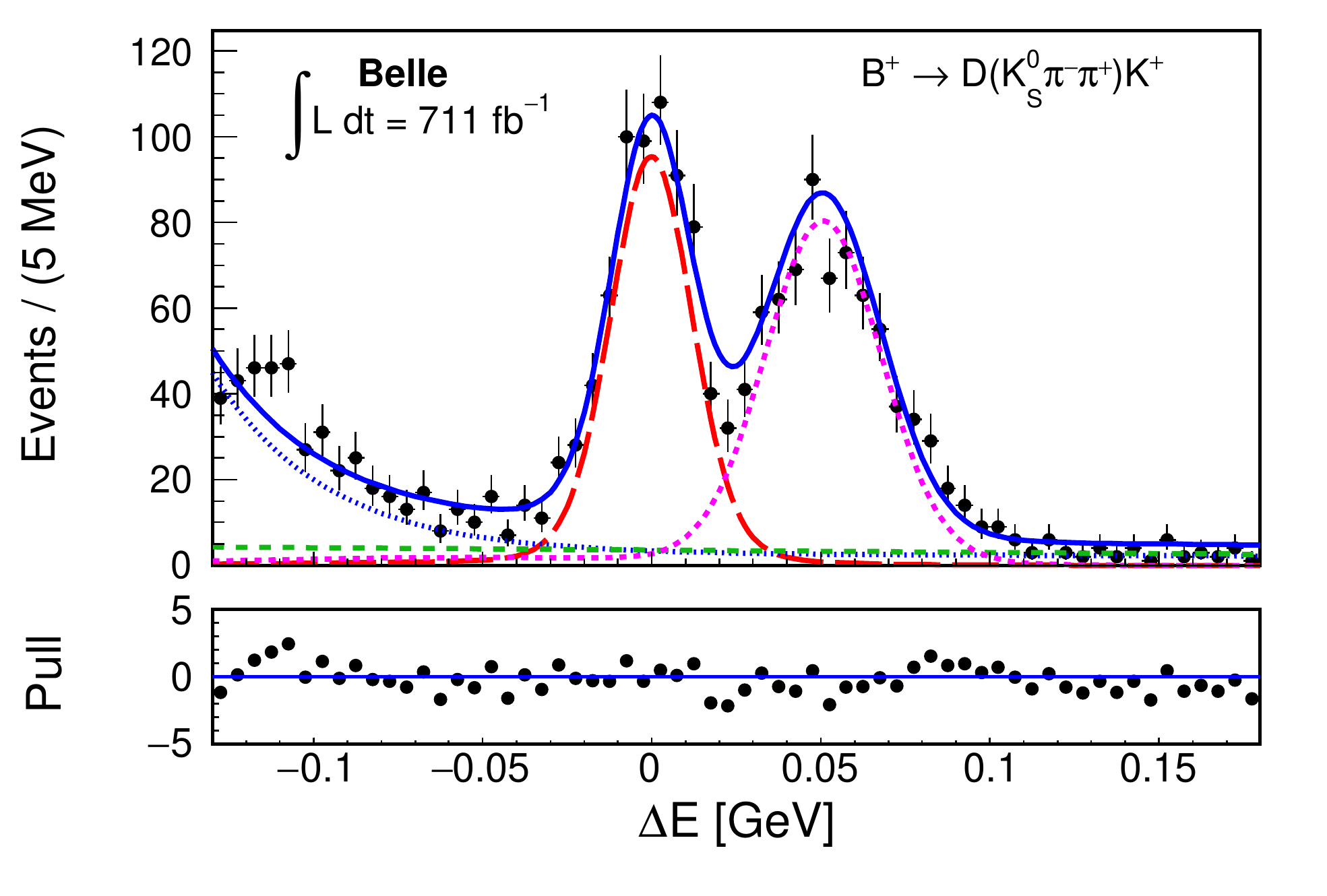} &
		\includegraphics[scale=0.37]{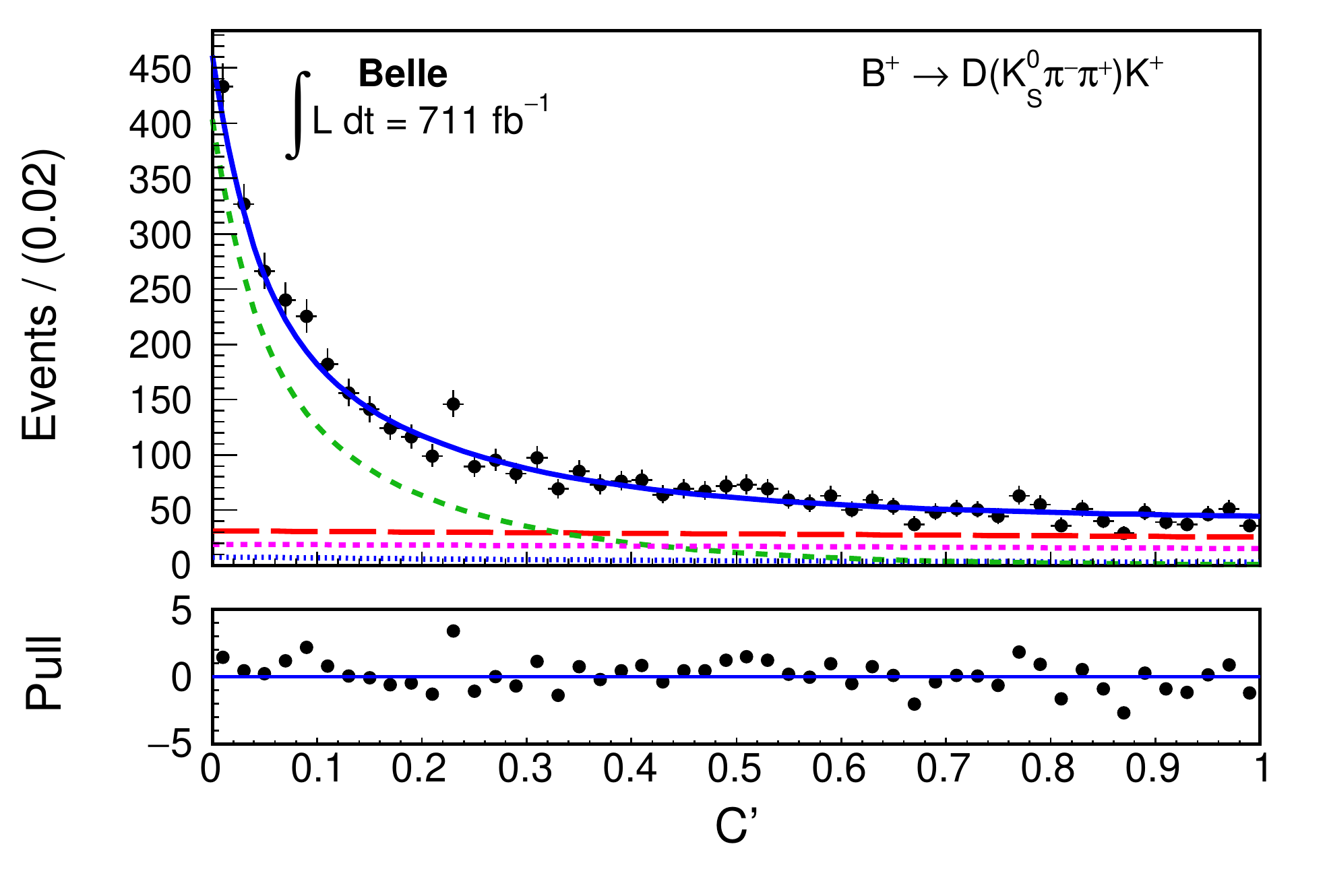} \\
	\end{tabular}
\caption{Distributions of (left) $\Delta E$ and (right) $C’$ for (top) $B^+ \to D(K_{\rm S}^0\pi^-\pi^+)\pi^+$ and (bottom) $B^+ \to D(K_{\rm S}^0\pi^-\pi^+)K^+$ candidates restricted to the signal-enhanced region in the Belle data set with fit projections overlaid. The black points with error bars represent data and the solid blue curve is the total fit. The large-dotted magenta, long-dashed red, small-dotted blue and short-dashed green curves represent $B^+ \to D\pi^+$, $B^+ \to DK^+$, $q\bar{q}$ and combinatorial $B\bar{B}$ background components, respectively. Differences between the fit function and data normalised by the uncertainty in data (pull) are shown under each panel.}
\label{fig:belle_kspipi}
\end{figure}

\begin{figure}[!t]
	\begin{tabular}{c c}
		\includegraphics[scale=0.37]{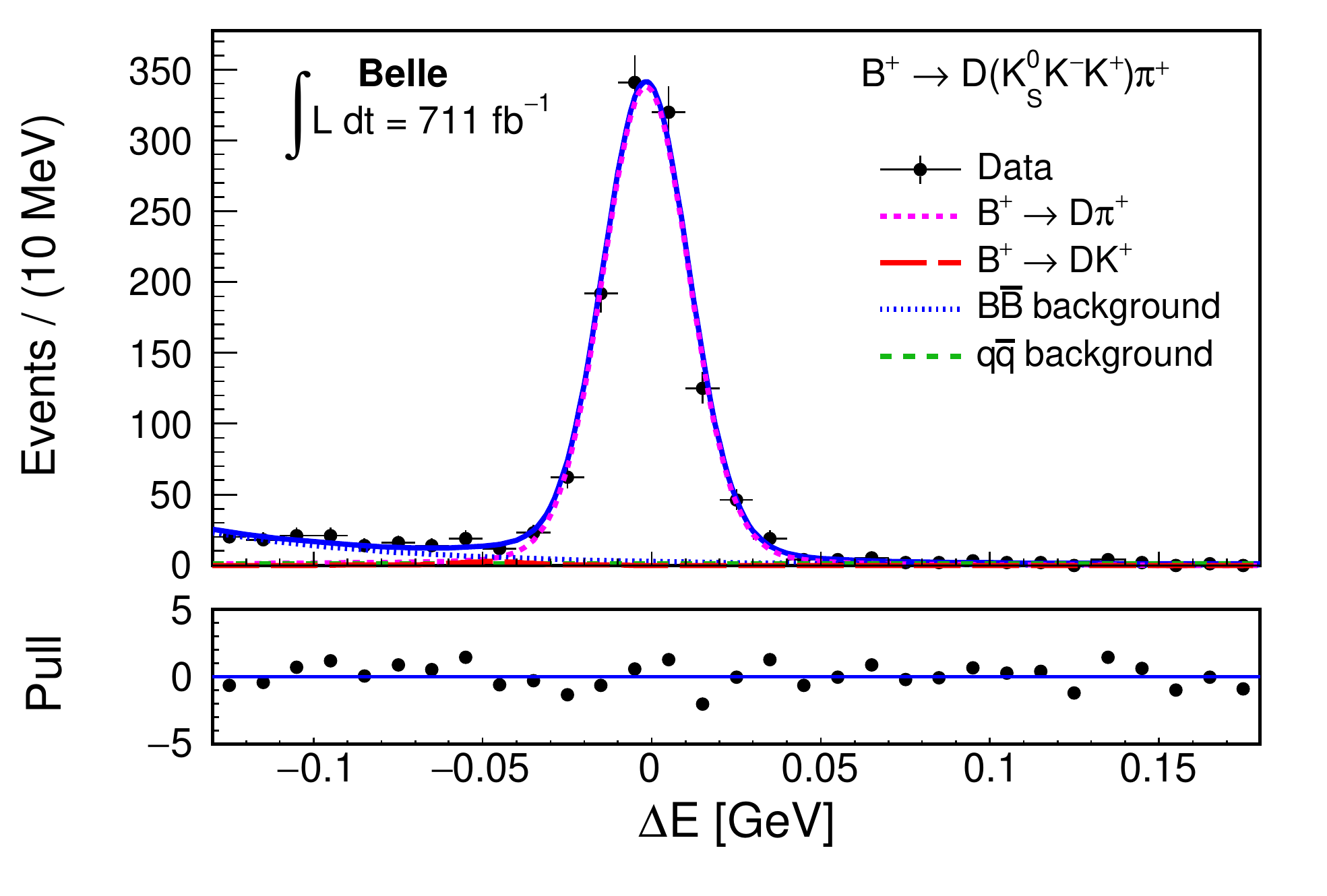} &
		\includegraphics[scale=0.37]{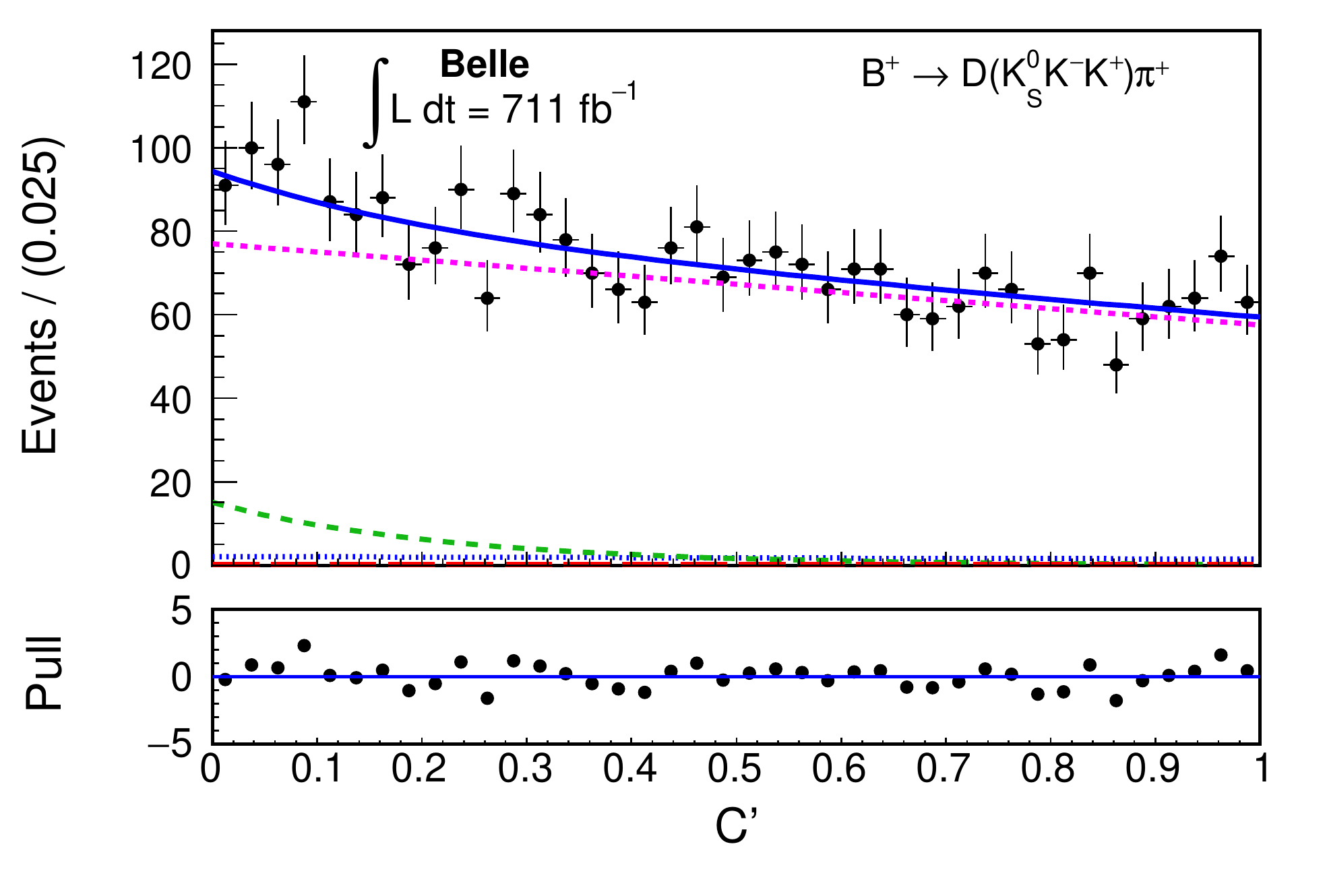} \\
		\includegraphics[scale=0.37]{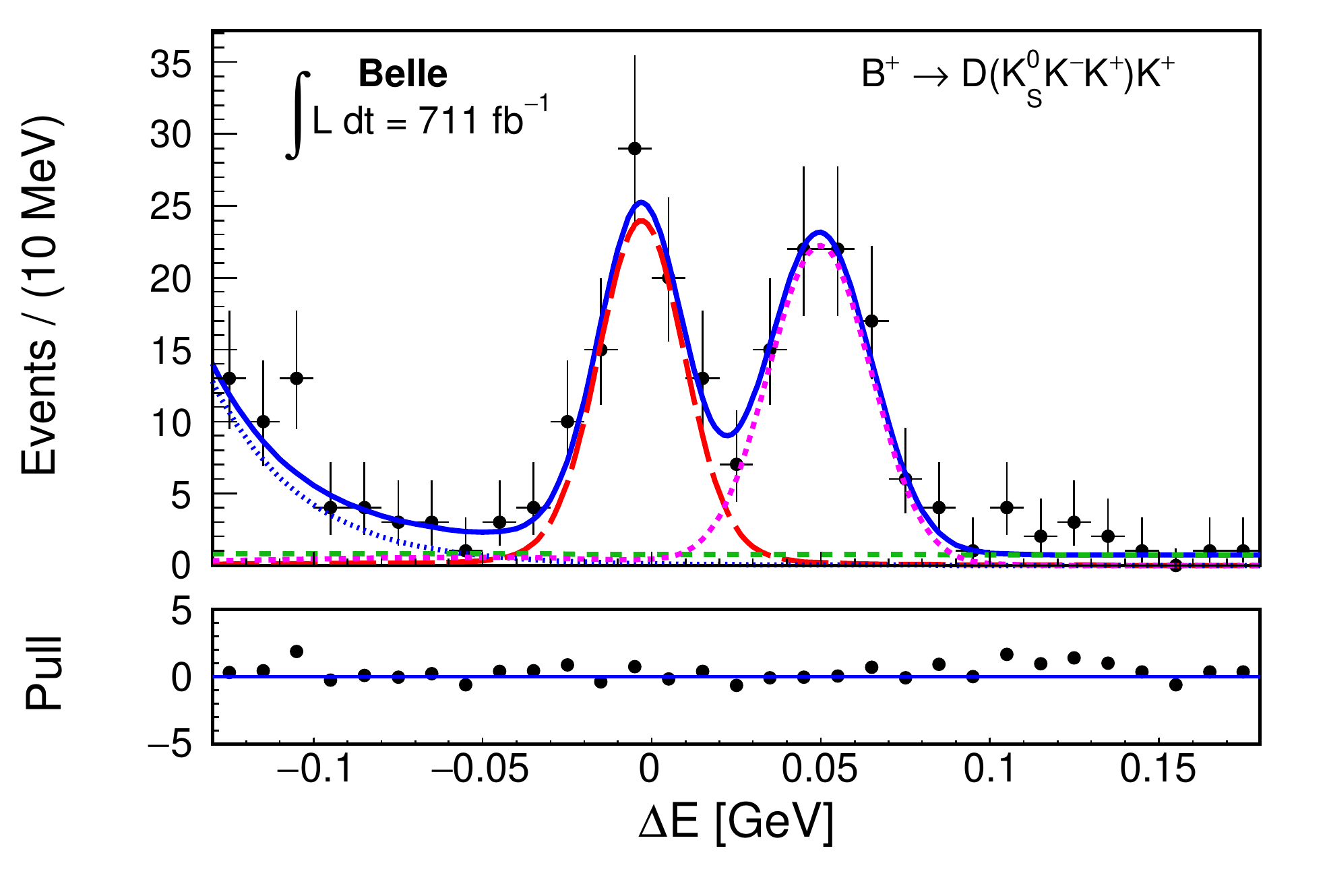} &
		\includegraphics[scale=0.37]{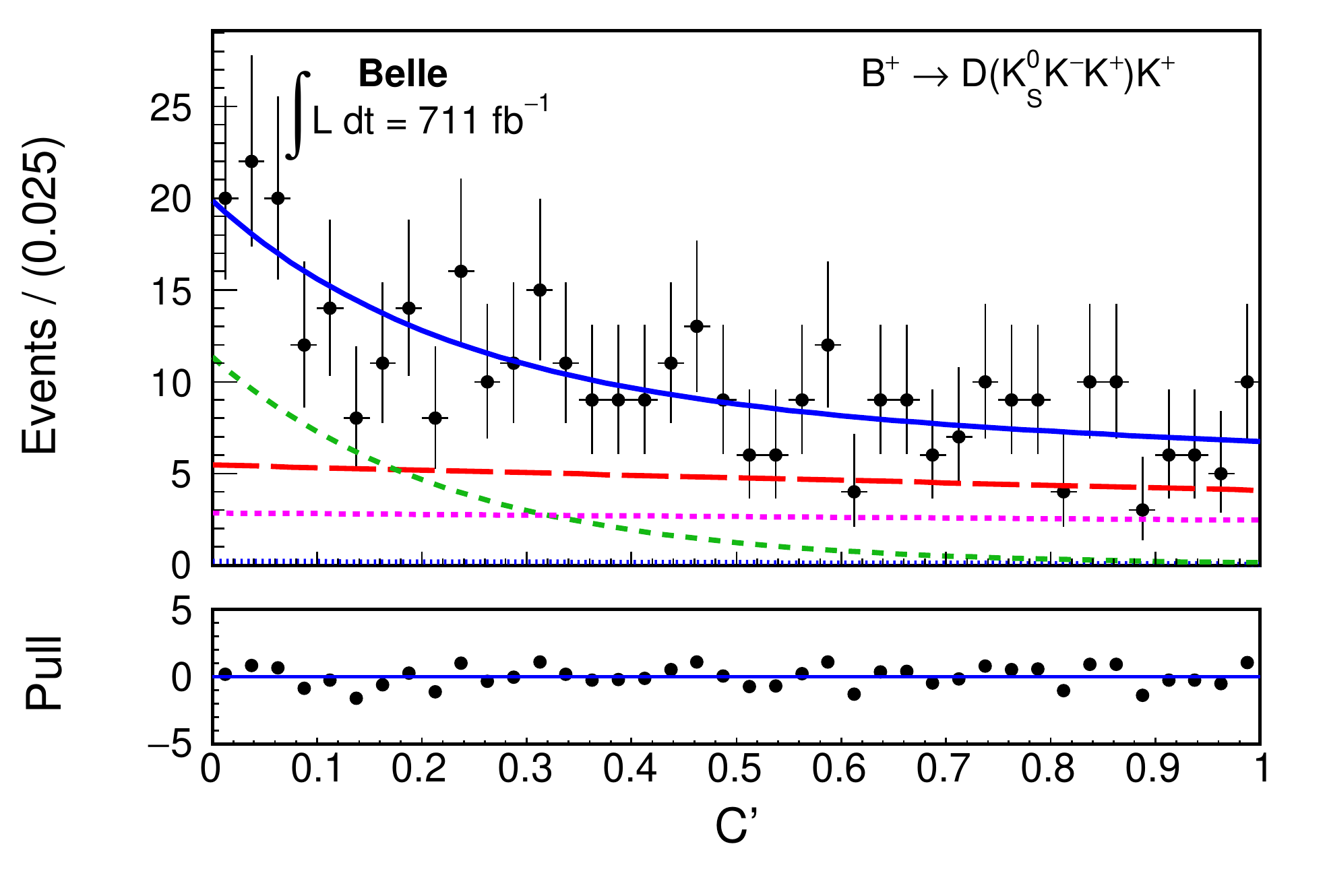} \\
	\end{tabular}
\caption{Distributions of (left) $\Delta E$ and (right) $C’$ for (top) $B^+ \to D(K_{\rm S}^0K^-K^+)\pi^+$ and (bottom) $B^+ \to D(K_{\rm S}^0K^-K^+)K^+$ candidates restricted to the signal-enhanced region in the Belle data set with fit projections overlaid. The black points with error bars represent data and the solid blue curve is the total fit. The large-dotted magenta, long-dashed red, small-dotted blue and short-dashed green curves represent $B^+ \to D\pi^+$, $B^+ \to DK^+$, $q\bar{q}$ and combinatorial $B\bar{B}$ background components, respectively. Differences between the fit function and data normalised by the uncertainty in data (pull) are shown under each panel.}
\label{fig:belle_kskk}
\end{figure}

\begin{figure}[!t]
	\begin{tabular}{c c}
		\includegraphics[scale=0.37]{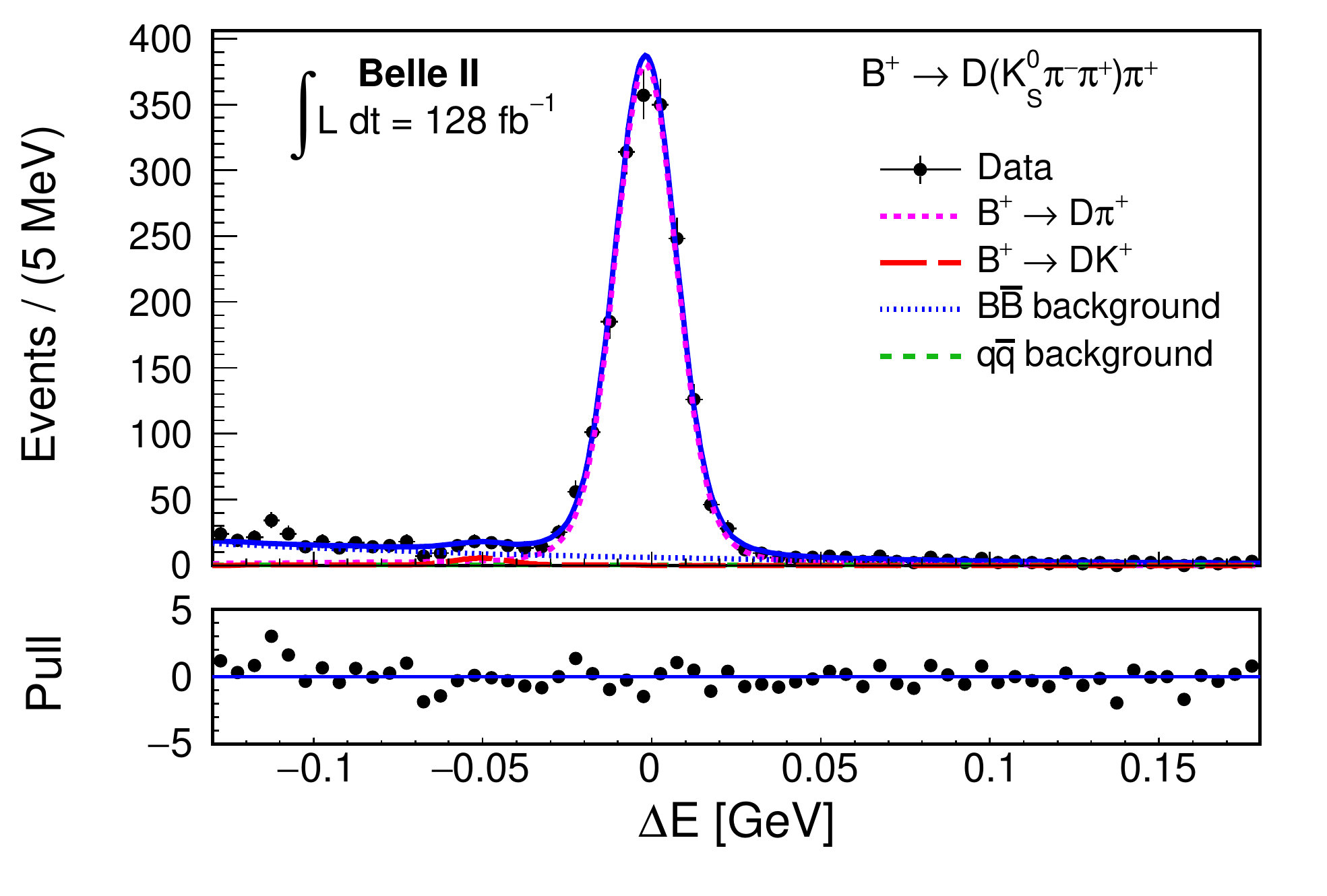} &
		\includegraphics[scale=0.37]{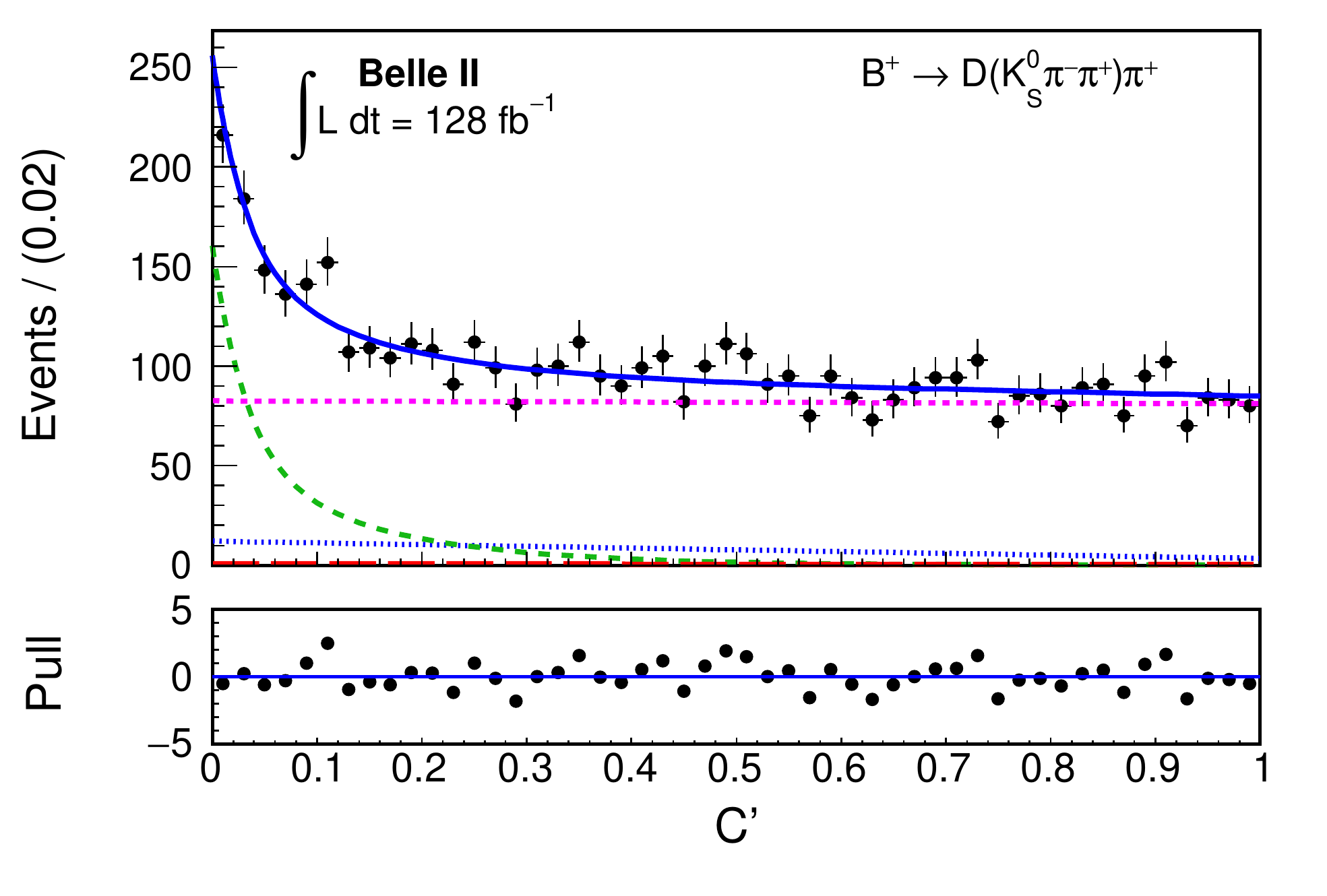} \\
		\includegraphics[scale=0.37]{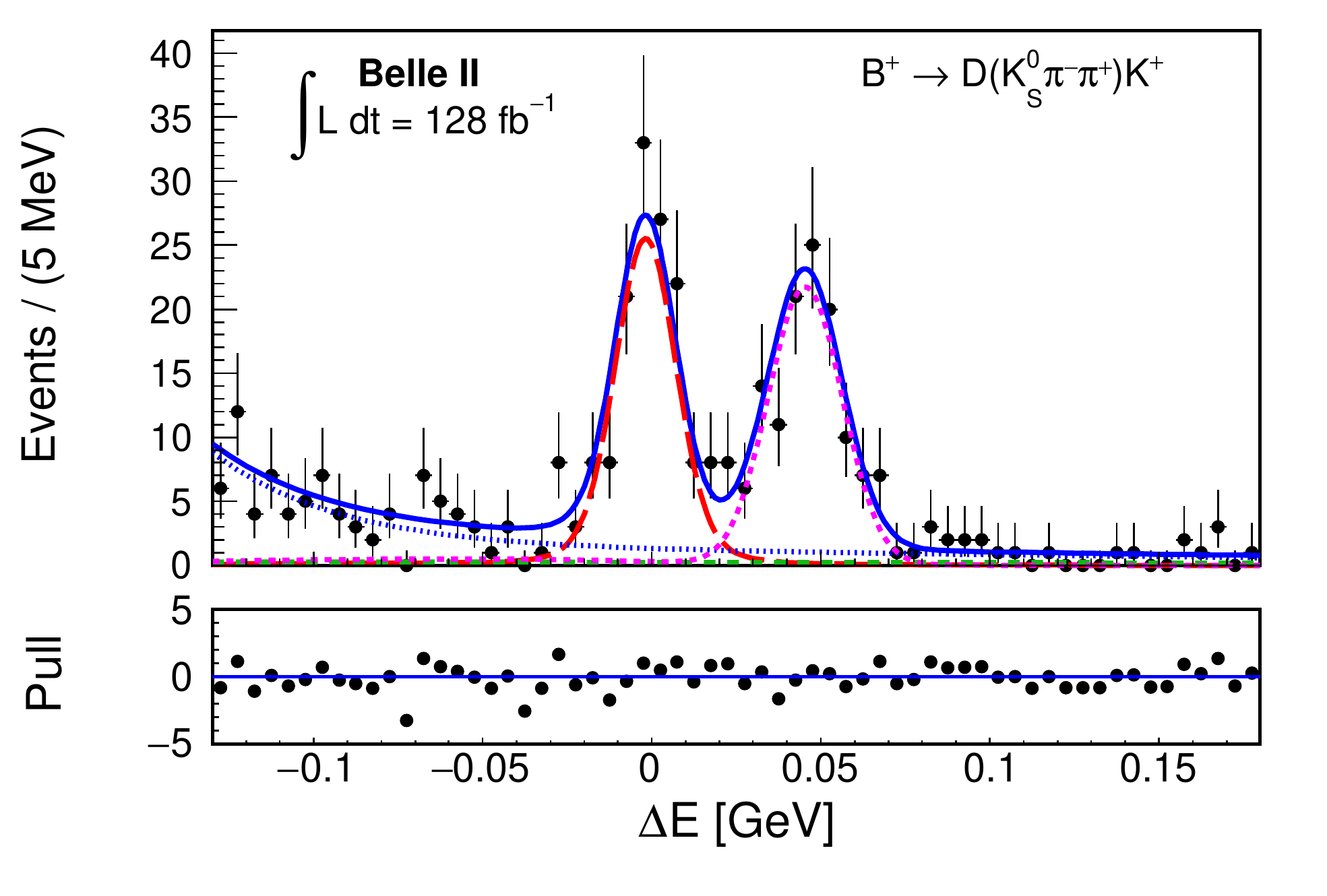} &
		\includegraphics[scale=0.37]{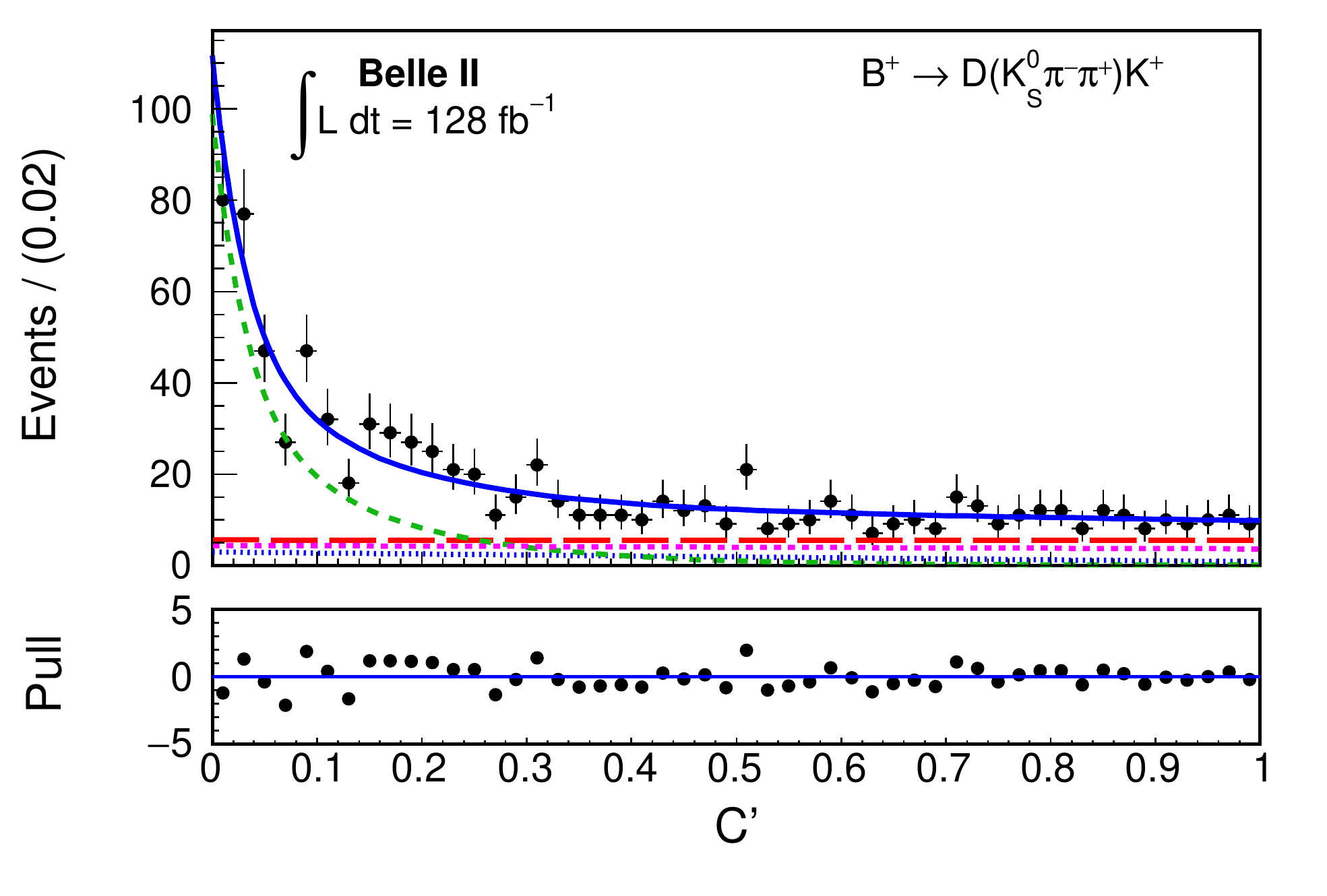} \\
	\end{tabular}
\caption{Distributions of (left) $\Delta E$ and (right) $C’$ for (top) $B^+ \to D(K_{\rm S}^0\pi^-\pi^+)\pi^+$ and (bottom) $B^+ \to D(K_{\rm S}^0\pi^-\pi^+)K^+$ candidates restricted to the signal-enhanced region in the Belle~II data set with fit projections overlaid. The black points with error bars represent data and the solid blue curve is the total fit. The large-dotted magenta, long-dashed red, small-dotted blue and short-dashed green curves represent $B^+ \to D\pi^+$, $B^+ \to DK^+$, $q\bar{q}$ and combinatorial $B\bar{B}$ background components, respectively. Differences between the fit function and data normalised by the uncertainty in data (pull) are shown under each panel.}
\label{fig:belle2_kspipi}
\end{figure}

\begin{figure}[!t]
	\begin{tabular}{c c}
		\includegraphics[scale=0.37]{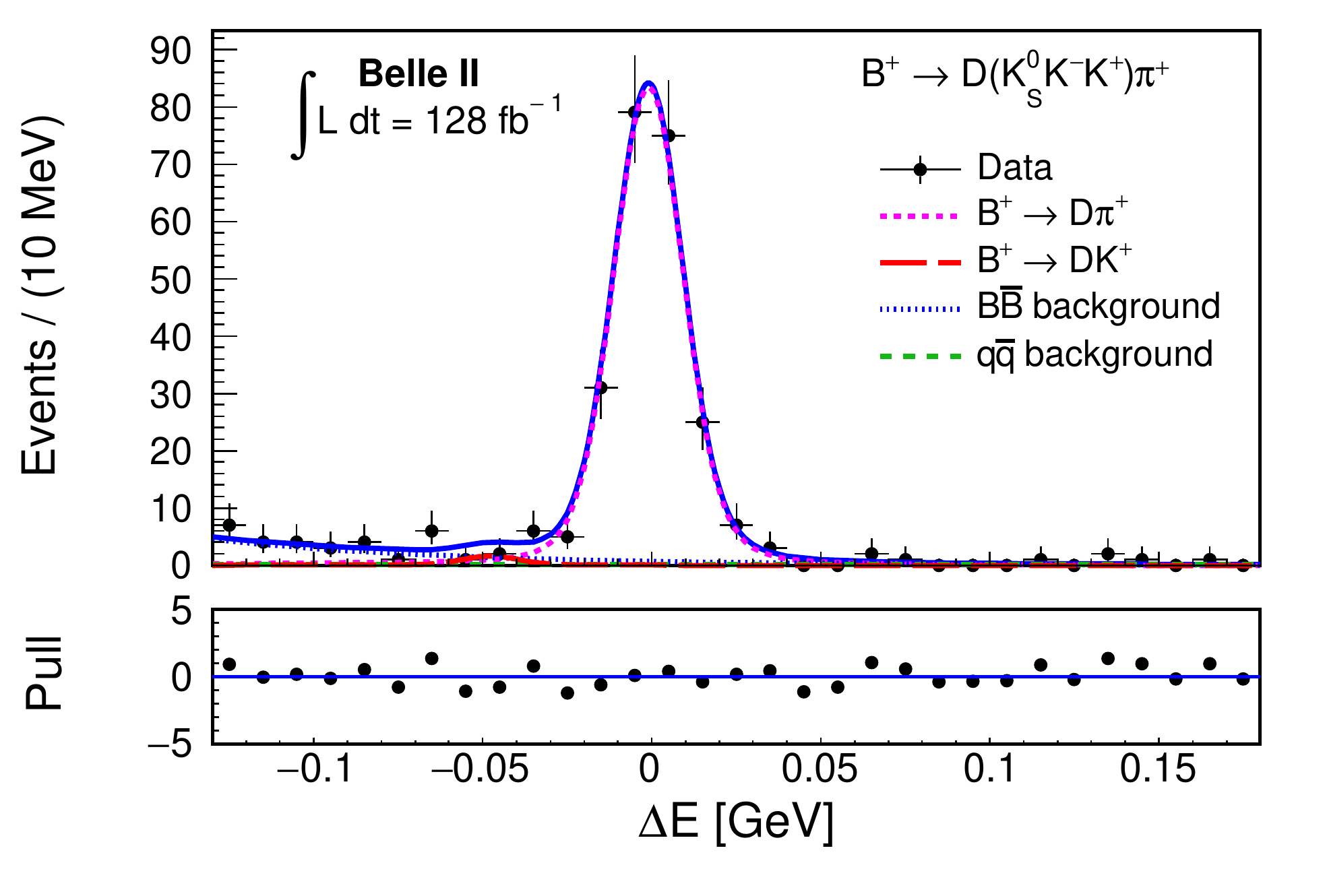} &
		\includegraphics[scale=0.37]{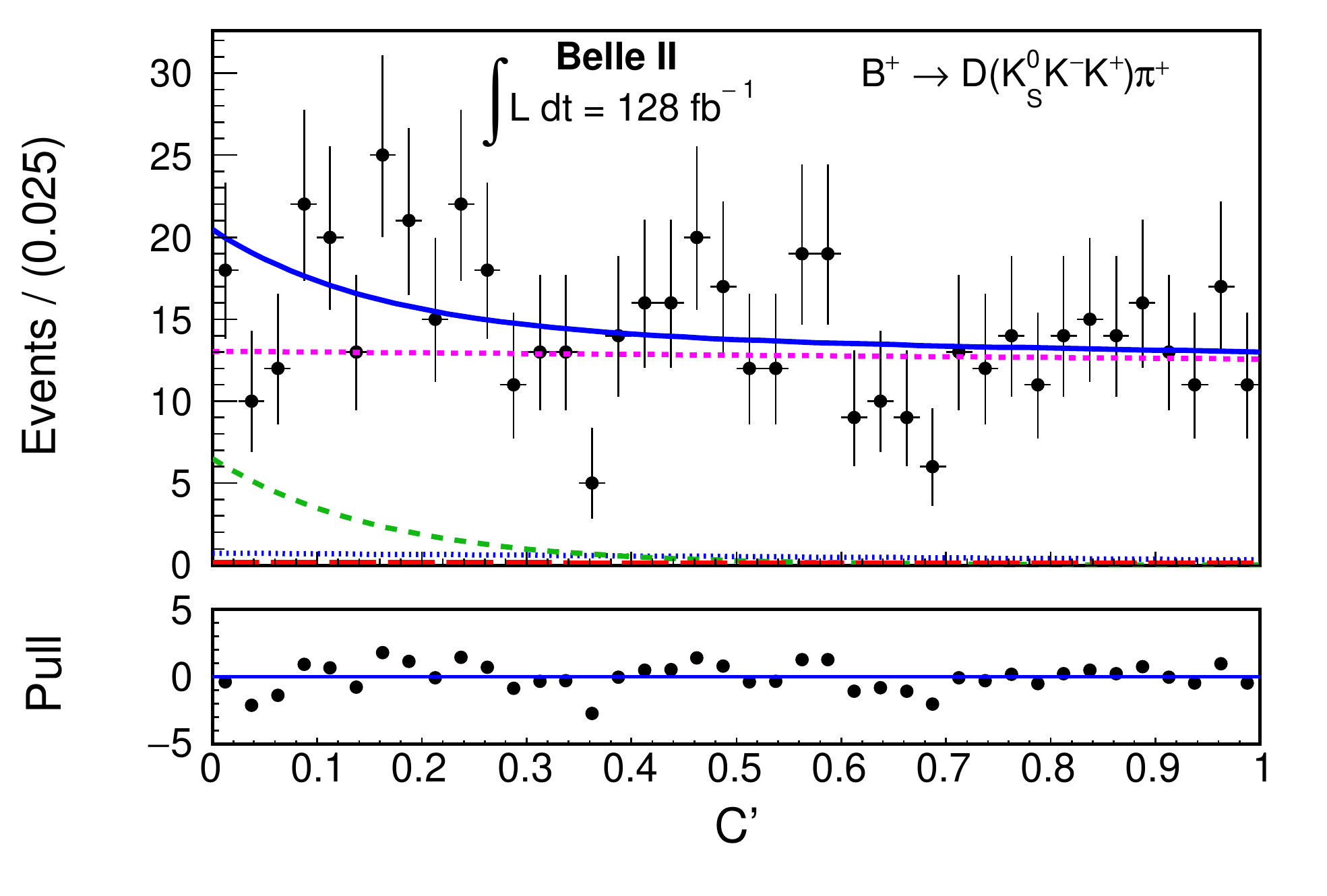} \\
		\includegraphics[scale=0.37]{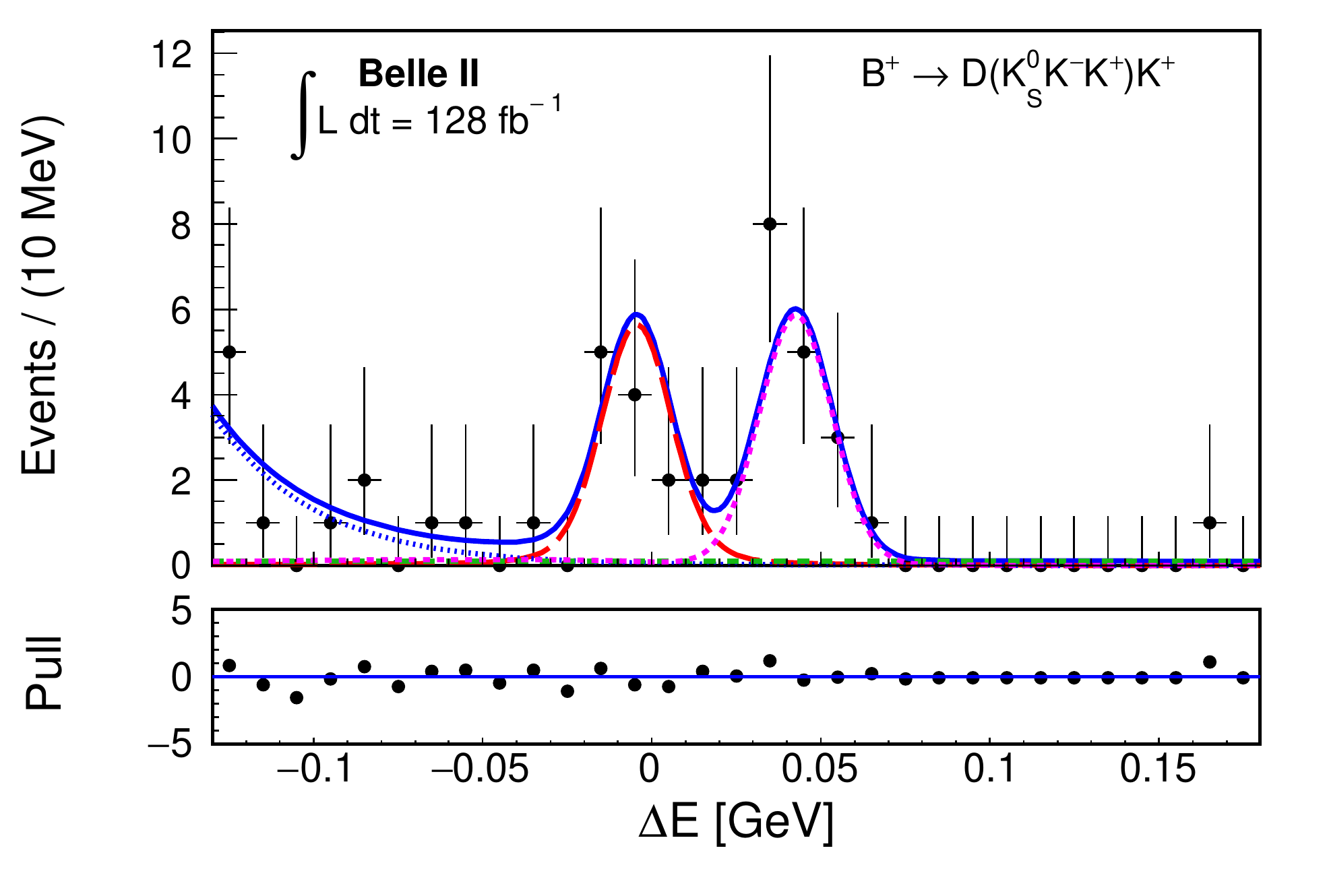} &
		\includegraphics[scale=0.37]{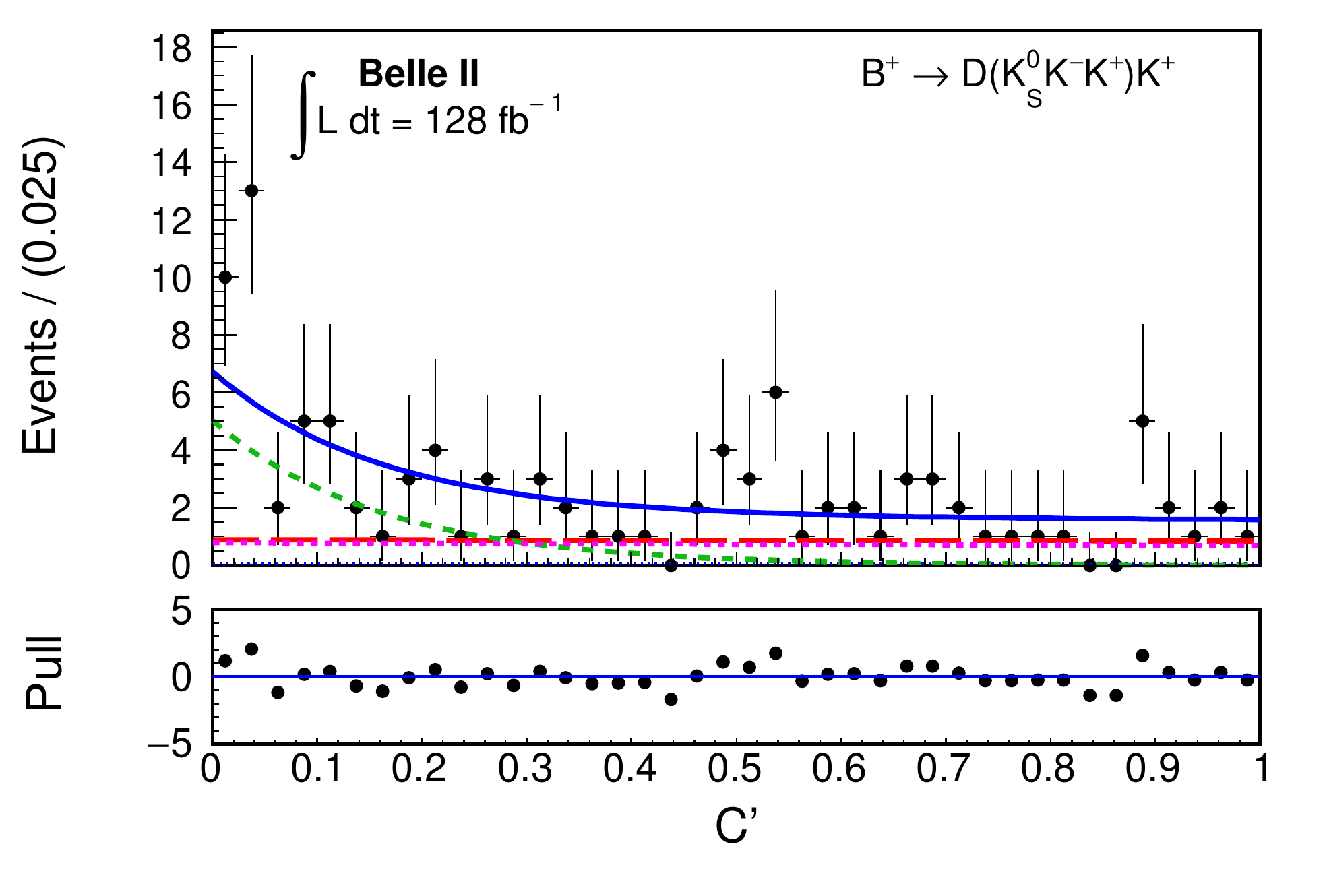} \\
	\end{tabular}
\caption{Distributions of (left) $\Delta E$ and (right) $C’$ for (top) $B^+ \to D(K_{\rm S}^0K^-K^+)\pi^+$ and (bottom) $B^+ \to D(K_{\rm S}^0K^-K^+)K^+$ candidates restricted to the signal-enhanced region in the Belle~II data set with fit projections overlaid. The black points with error bars represent data and the solid blue curve is the total fit. The large-dotted magenta, long-dashed red, small-dotted blue and short-dashed green curves represent $B^+ \to D\pi^+$, $B^+ \to DK^+$, $q\bar{q}$ and combinatorial $B\bar{B}$ background components, respectively. Differences between the fit function and data normalised by the uncertainty in data (pull) are shown under each panel.}
\label{fig:belle2_kskk}
\end{figure}

The PDF parameterisation is identical for the fits to the Belle and Belle~II data samples; only the fixed parameter values are estimated separately from the simulation samples corresponding to the respective experiments. First, we perform separate fits to the Belle and Belle~II data samples, which are subdivided based on the reconstructed $D$-decay final state. These fits are to data integrated over the $D$-decay Dalitz plot. We refer to this fit as the ``combined fit''. The signal and peaking background yields $N^{Dh}_{h^{\prime}}$, where $h^{\prime}=\pi$ or $K$ represents the respective enhancement of the sample, are parameterised in terms of kaon-identification efficiency ($\epsilon$) and $K$-$\pi$ misidentification rate ($\kappa$) using the relations 
\begin{eqnarray*}
N^{D\pi}_{\pi} & = & (1- \kappa) N_{\rm tot}^{D\pi}, \nonumber \\ 
N^{DK}_{\pi} & = & (1 - \epsilon) N_{\rm tot}^{DK}, \nonumber\\
N^{DK}_{K}  & = & \epsilon N_{\rm tot}^{DK},~ {\rm and}\nonumber\\
N^{D\pi}_{K}  & = & \kappa N_{\rm tot}^{D\pi},
\end{eqnarray*}
where $N_{\rm tot}^{D\pi}$ and $N_{\rm tot}^{DK}$ represent the total $D\pi$ and $DK$ yields without any PID selection, respectively. In this way, $\kappa$ is directly extracted from data. The value of $\epsilon$ is fixed to the value determined from data control samples of $D^{*+}\to D^{0}\left(K^-\pi^+\right)\pi^+$ decays. The total signal and background yields obtained from the combined fit are summarised in Table~\ref{tab:yields}. Distributions restricted to candidates with $|\Delta E| <\:$\SI{0.05}{GeV} and $0.65 < C' < 1$ (signal-enhanced) are shown in Figs.~\ref{fig:belle_kspipi}$-$\ref{fig:belle2_kskk} with fit projections overlaid.

\begin{table}[!t]
    \centering
    \begin{tabular}{| c | c | c c | c c |}
    \hline
     & Sample & \multicolumn{2}{c|}{Pion-enhanced} & \multicolumn{2}{c |}{Kaon-enhanced}\\
    \hline
    $D$ decay & Component & Belle & Belle~II & Belle & Belle~II \\
    \hline
    $D \to K_{\rm S}^0\pi^+\pi^-$ & $B^+ \to D\pi^+$ & $21325 \pm 162$ & $4193 \pm 70$ & $1764 \pm 64\phantom{0}$ & $\phantom{0}308 \pm 23$  \\
                                  & $B^+ \to DK^+$  & $\phantom{00}140 \pm 29\phantom{0}$ & $\phantom{00}62 \pm 11$ & $1467 \pm 53\phantom{0}$ & $\phantom{0}280 \pm 21$  \\
                                  & $B\bar{B}$ background & $\phantom{0}5040 \pm 155$ & $1223 \pm 68$ & $1309 \pm 85\phantom{0}$ & $\phantom{0}387 \pm 42$  \\
                                  & $q\bar{q}$ background & $\phantom{0}9022 \pm 172$ & $1657 \pm 69$ & $6295 \pm 122$ & $1021 \pm 47$  \\
    \hline
    $D \to K_{\rm S}^0K^+K^-$ & $B^+ \to D\pi^+$ & $ 2740 \pm 56$ & $\phantom{0}519 \pm 21$ & $\phantom{0}211 \pm 18\phantom{0}$ & $\phantom{00}50 \pm 10$ \\
                                  & $B^+ \to DK^+$ & $\phantom{000}17 \pm 4\phantom{00}$ & $\phantom{00}2.1 \pm 0.2$ & $194 \pm 17$ & $\phantom{00}34 \pm 7\phantom{0}$ \\
                                  & $B\bar{B}$ background & $\phantom{0}333 \pm 31$ & $\phantom{00}77 \pm 12$ & $110 \pm 18$ & $\phantom{00}22 \pm 7\phantom{0}$  \\
                                  & $q\bar{q}$ background & $\phantom{0}409 \pm 37$ & $\phantom{0}124 \pm 14$ & $309 \pm 28$ & $\phantom{00}92 \pm 11$  \\
    \hline
    \end{tabular}
    \caption{Signal and background yields obtained from the two-dimensional combined fit.}
    \label{tab:yields}
\end{table}
%%%%%%%%%%%%%%%%%%%%%%%%%%%%%%%%%%%%%%%%%%%%%%%%%%%%%%%%%%%%%%%%%%%%%

\begin{figure}[!t]
	\begin{tabular}{c c}
		\includegraphics[scale=0.36]{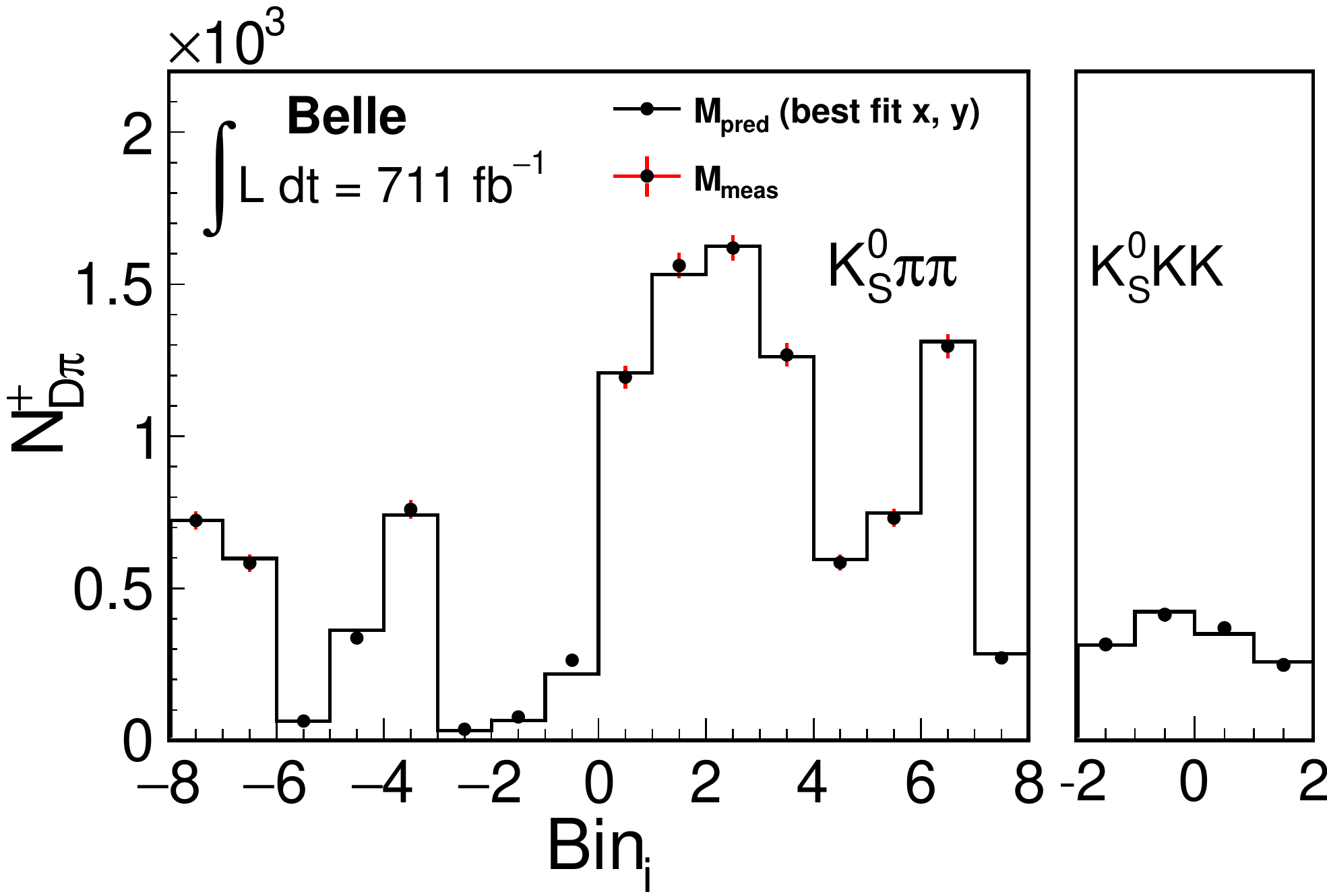} &
		\includegraphics[scale=0.36]{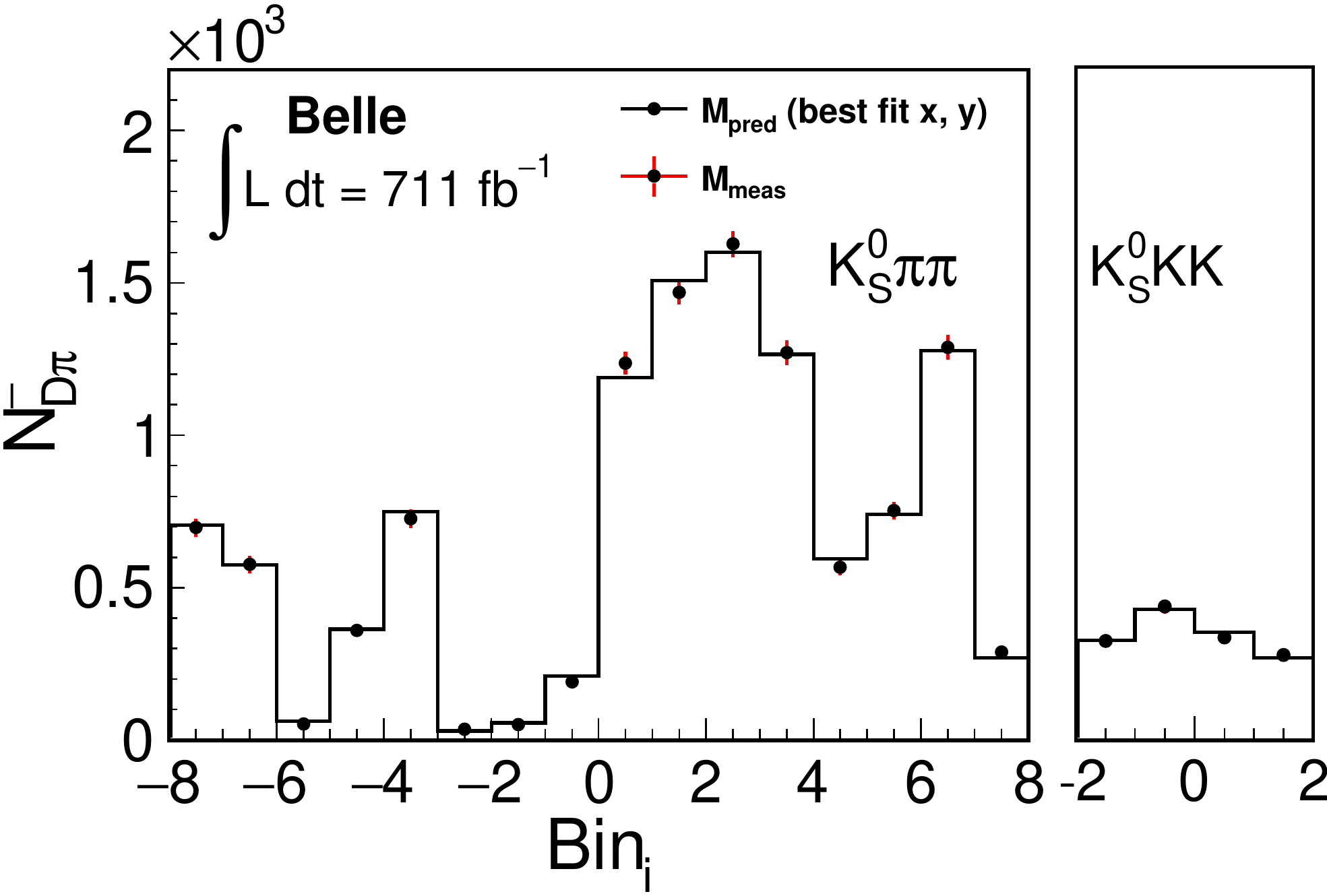} \\
		\includegraphics[scale=0.36]{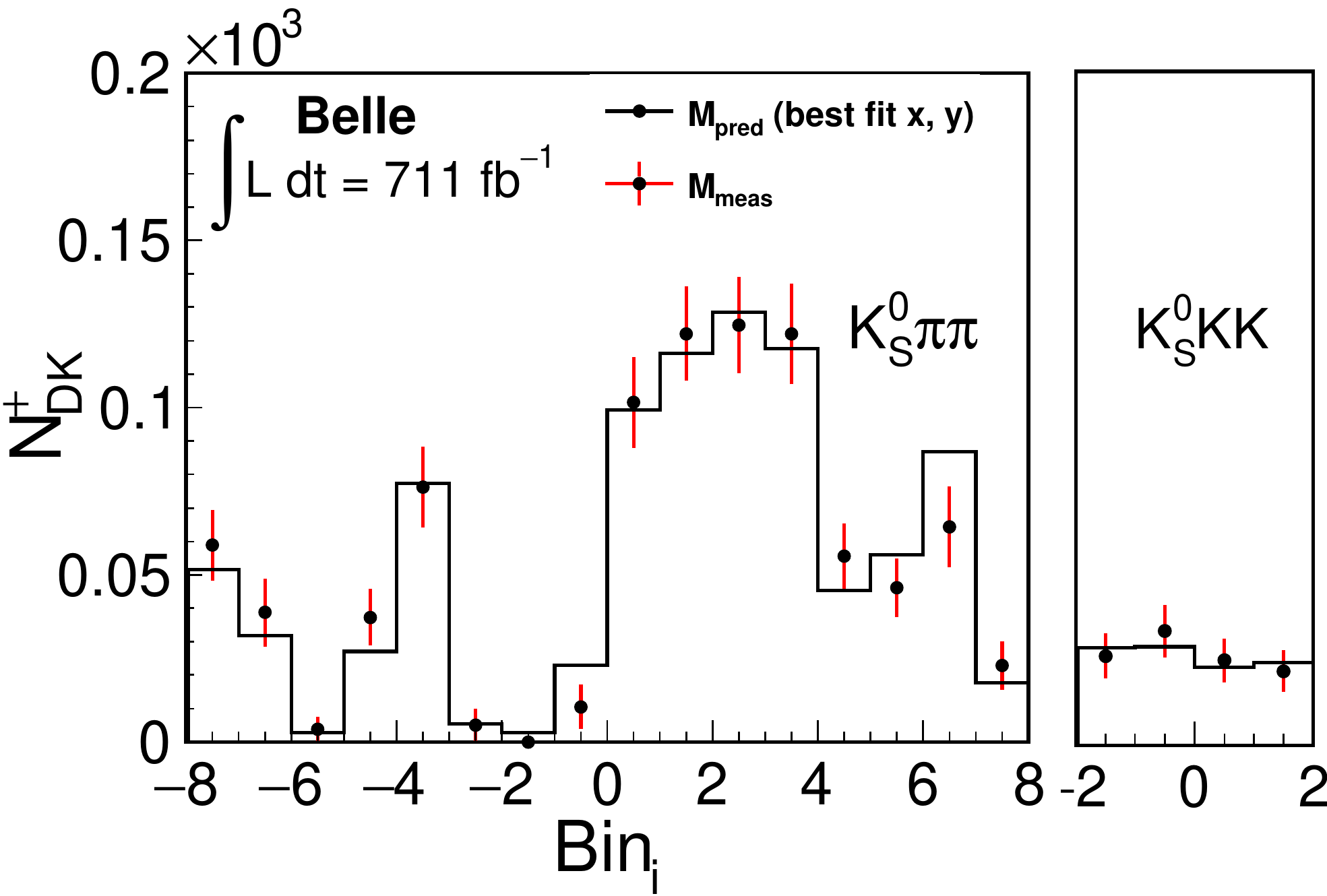} &
		\includegraphics[scale=0.36]{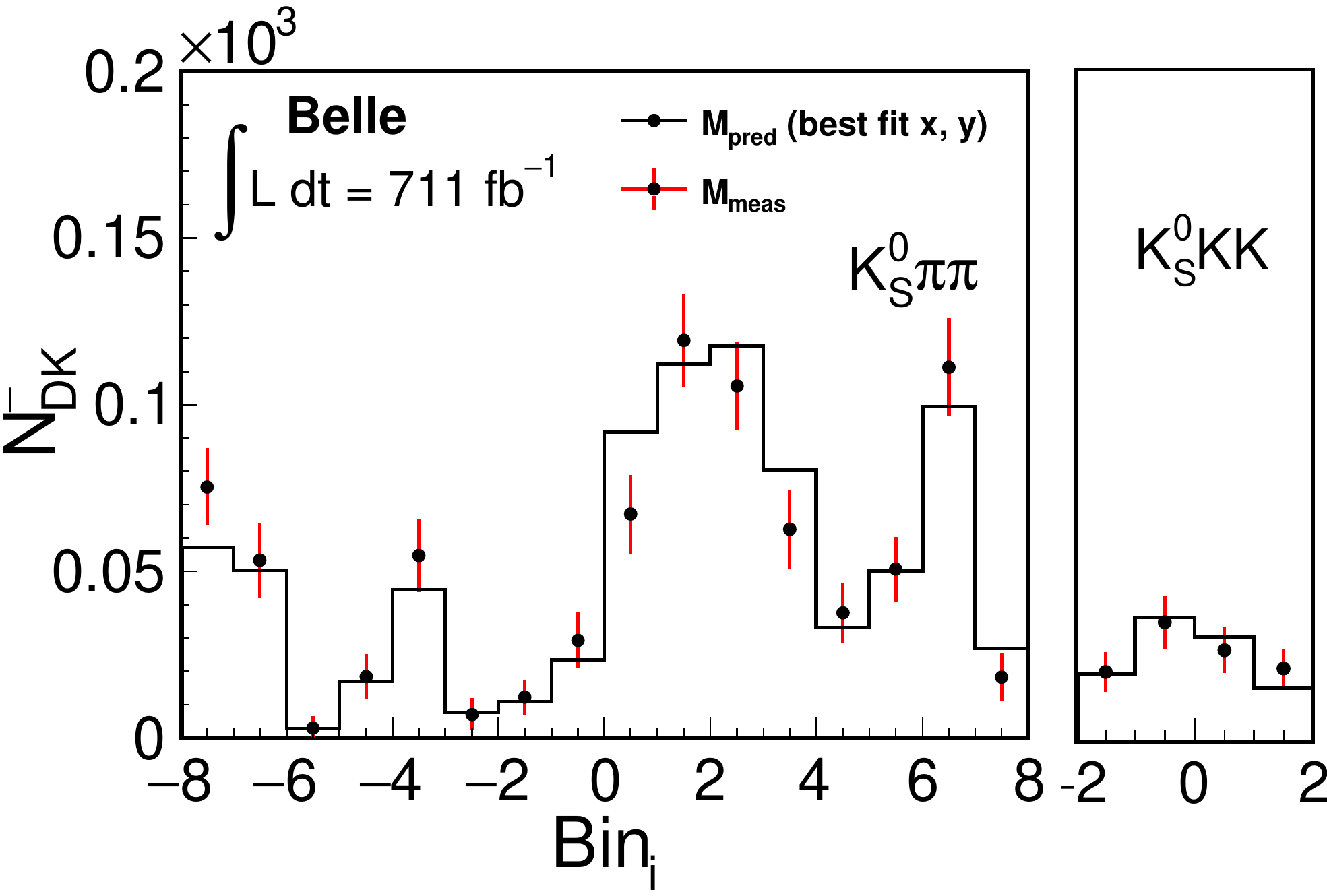} \\
	\end{tabular}
	\caption{Yields in bins for $ B^+ \to D\pi^+ $ (top left), $ B^- \to D\pi^- $ (top right), $ B^+ \to DK^+ $ (bottom left) and $ B^- \to DK^- $ (bottom right) decays in the Belle data set. The data points with error bars are the measured yields with their statistical uncertainty and the histogram is the expected yield from the best fit $ (x_\pm, y_\pm) $ parameter values.}
	\label{fig:bin_yield_belle}
\end{figure}

\begin{figure}[!t]
	\begin{tabular}{c c}
		\includegraphics[scale=0.36]{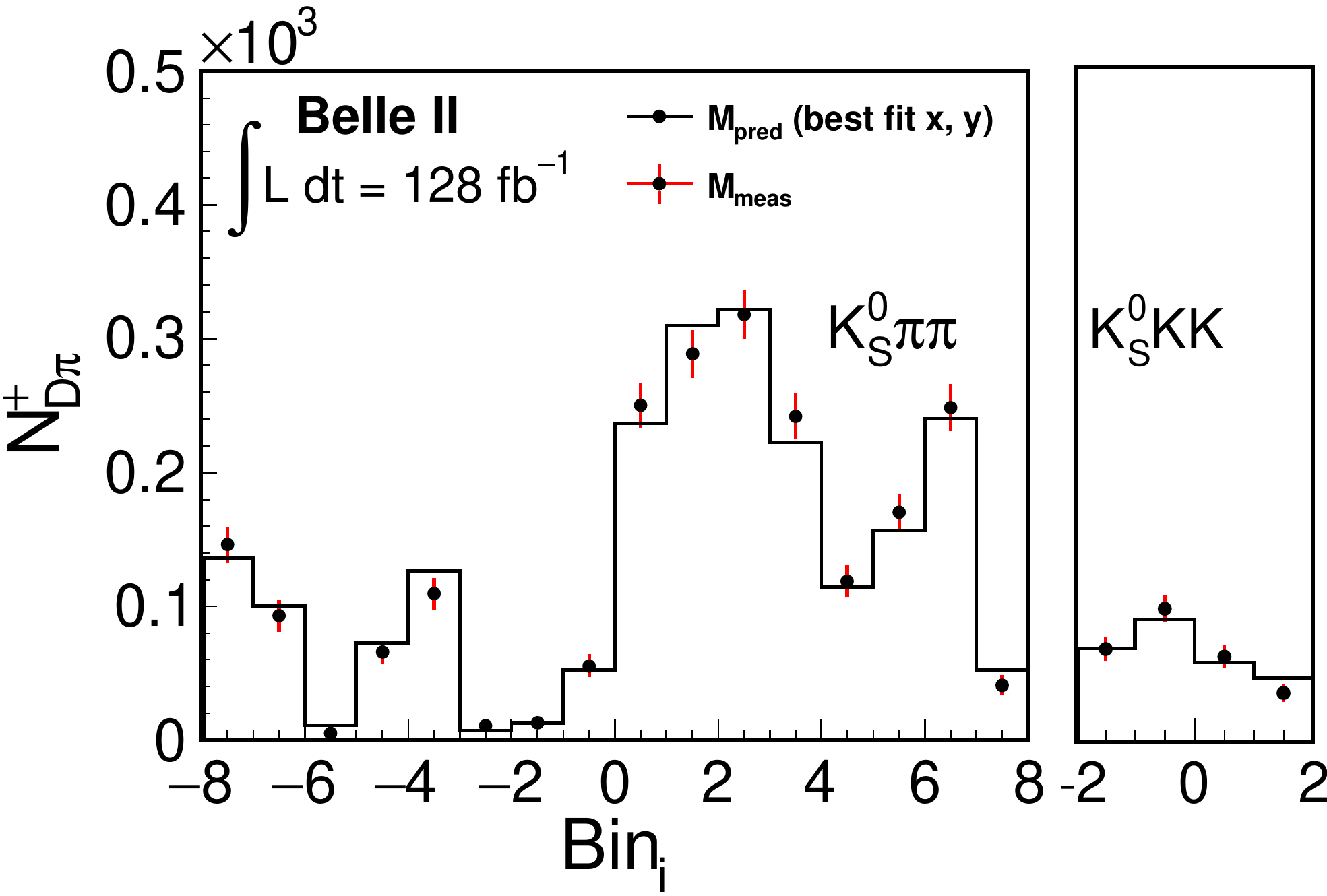} &
		\includegraphics[scale=0.36]{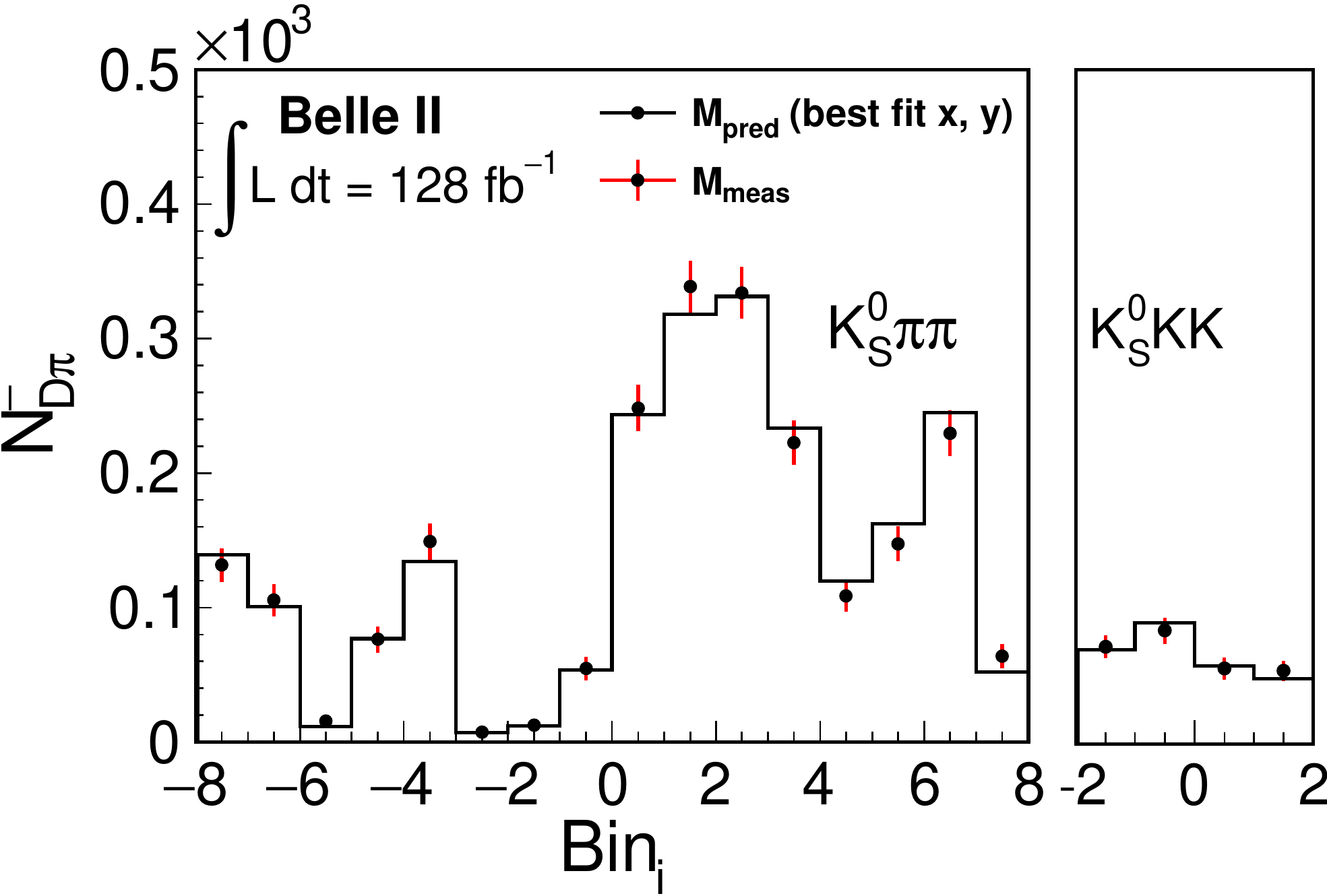} \\
		\includegraphics[scale=0.36]{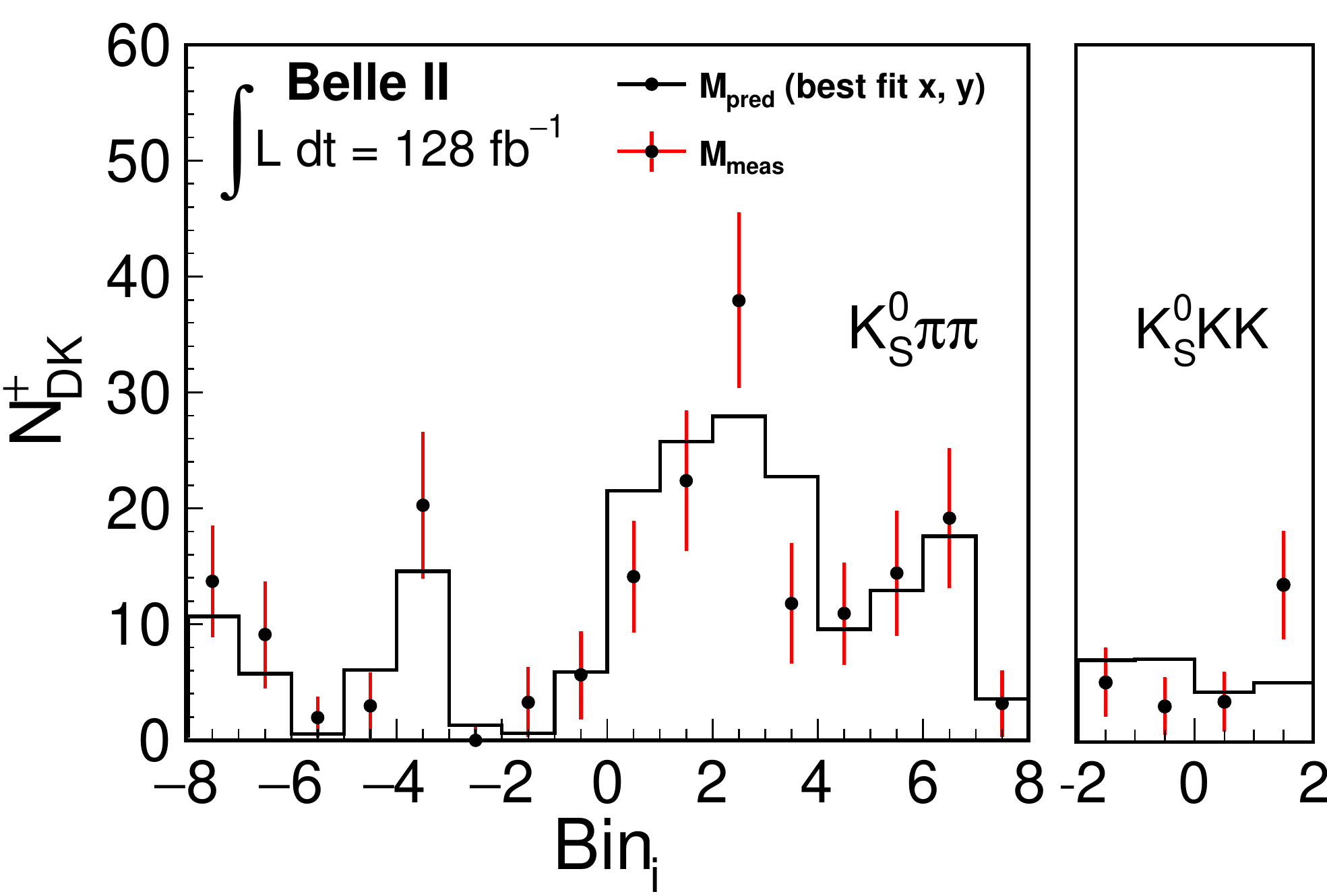} &
		\includegraphics[scale=0.36]{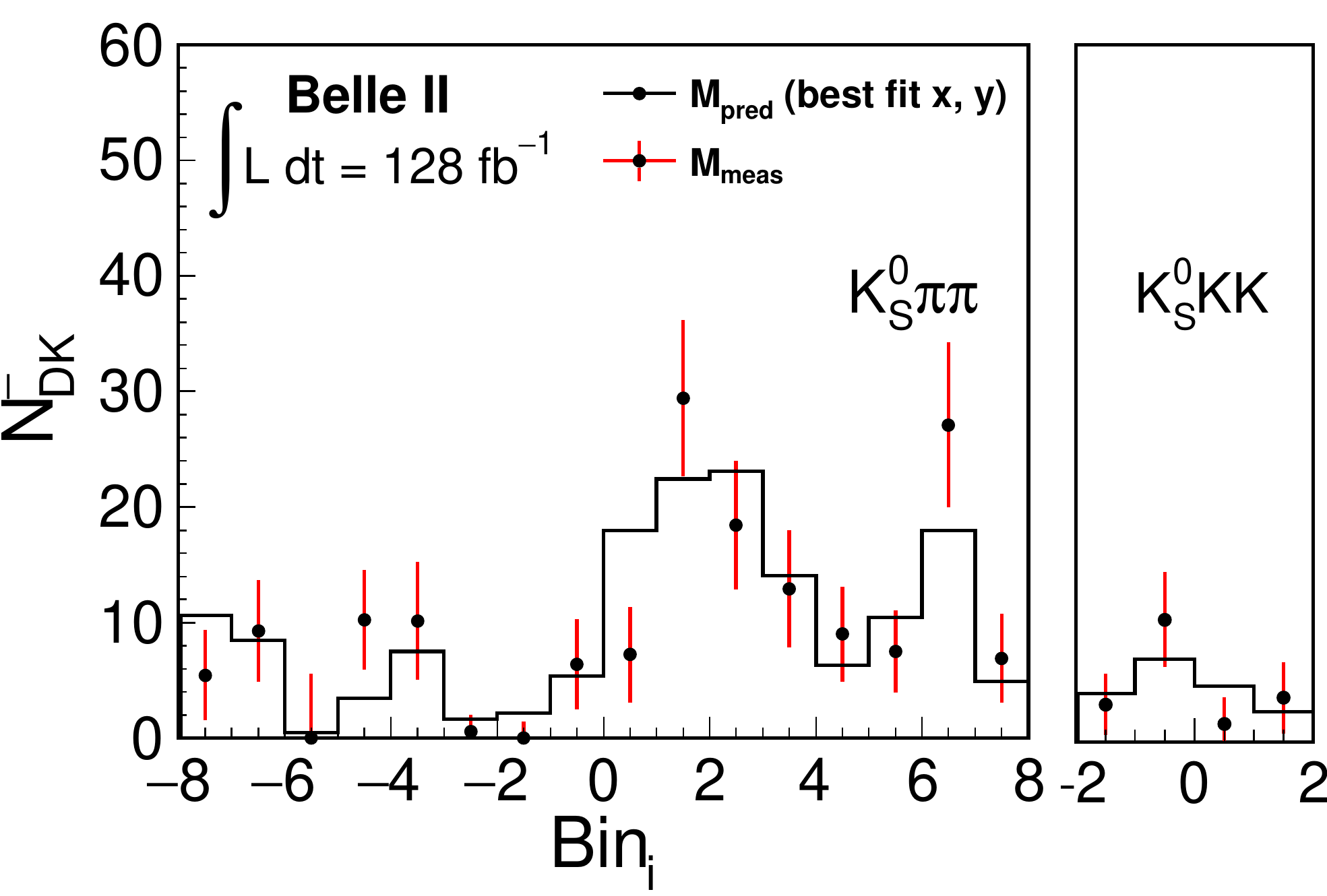} \\
	\end{tabular}
	\caption{Yields in bins for $ B^+ \to D\pi^+ $ (top left), $ B^- \to D\pi^- $ (top right), $ B^+ \to DK^+ $ (bottom left) and $ B^- \to DK^- $ (bottom right) decays in the Belle~II data set. The data points with error bars are the measured yields with their statistical uncertainty and the histogram is the expected yield from the best fit $ (x_\pm, y_\pm) $ parameter values.}
	\label{fig:bin_yield_belle2}
\end{figure}

%%%%%%%%%%%%%%%%%%%%%%%%%%%%%%%%%%%%%%%%%%%%%%%%%%%%%%%%%%%%%%%%%%%%%%%%%%%
To determine the {\it CP}-violating observables, we further divide the fit categories according to the charge of $B$ candidates and the $D$-decay Dalitz bins described in Sec.~\ref{sec:bpggsz}. This fit is performed simultaneously for the Belle and Belle~II data and is referred to as the ``binned fit''. The component PDFs for the individual bins are the same as in the fits to the unbinned data, along with the same sets of free and fixed parameters. The value of $\kappa$ is fixed in the binned fit to that obtained from the combined fit. The signal yield in each bin is parameterised according to the expressions in Eq.~(\ref{eq:byields}), which depend on common $x_{\pm}^{DK}, y_{\pm}^{DK}, x_{\xi}^{D\pi}$ and $y_{\xi}^{D\pi}$ parameters, as well as the external input values of $c_i$ and $s_i$~\citep{kspipi_cisi,kskk_cisi}. The background yields are fit independently in each bin for both $B^+$ and $B^-$ samples, which accounts for any $CP$-violation in the background. The $F_i$ and $F_{-i}$ fractions are extracted directly from the fit. As these fractions must satisfy $ \sum F_i = 1, F_i \in [0, 1] $, a fit instability can be induced due to large correlations between the $F_i$ parameters \cite{lhcb_gamma}. Hence, following Ref.~\cite{lhcb_gamma}, we reparameterise $F_i$ as a series of $2\mathcal{N}-1$ recursive fractions $ \mathcal{R}_\mathnormal{i} $ that are extracted from the fit. The $\mathcal{R}_i$ fractions are defined as
\begin{equation}
F_i =
\begin{cases}
\mathcal{R}_i & ,~i = -\mathcal{N} \\
\mathcal{R}_i\prod_{j < i}(1-\mathcal{R}_j) & ,~ -\mathcal{N} < i < +\mathcal{N} \\
\prod_{j < i}(1-\mathcal{R}_j) & ,~i = +\mathcal{N} .\\
\end{cases}
\label{eq:ki}
\end{equation}
The values of $\mathcal{R}_i$ are independent for Belle and Belle~II such that any difference in the acceptance profile is accounted for. The values of $\mathcal{R}_i$ are found to be compatible between Belle and Belle~II. However, if common $\mathcal{R}_i$ parameters are used, there is little statistical advantage in determining the {\it CP}-violating observables and an additional systematic uncertainty would be introduced related to the assumption. Figure~\ref{fig:bin_yield_belle} (\ref{fig:bin_yield_belle2}) shows the measured $B^+\to Dh^+$ yields in each bin for the Belle (Belle~II) data sample.

%%%%%%%%%%%%%%%%%%%%%%%%%%%%%%%%%%%%%%%%%%%%%%%%%%%%%%%%%%%%%%%%%%%%%%%%%

\begin{figure}[!t]
	\begin{tabular}{c c}
		\includegraphics[scale=0.36]{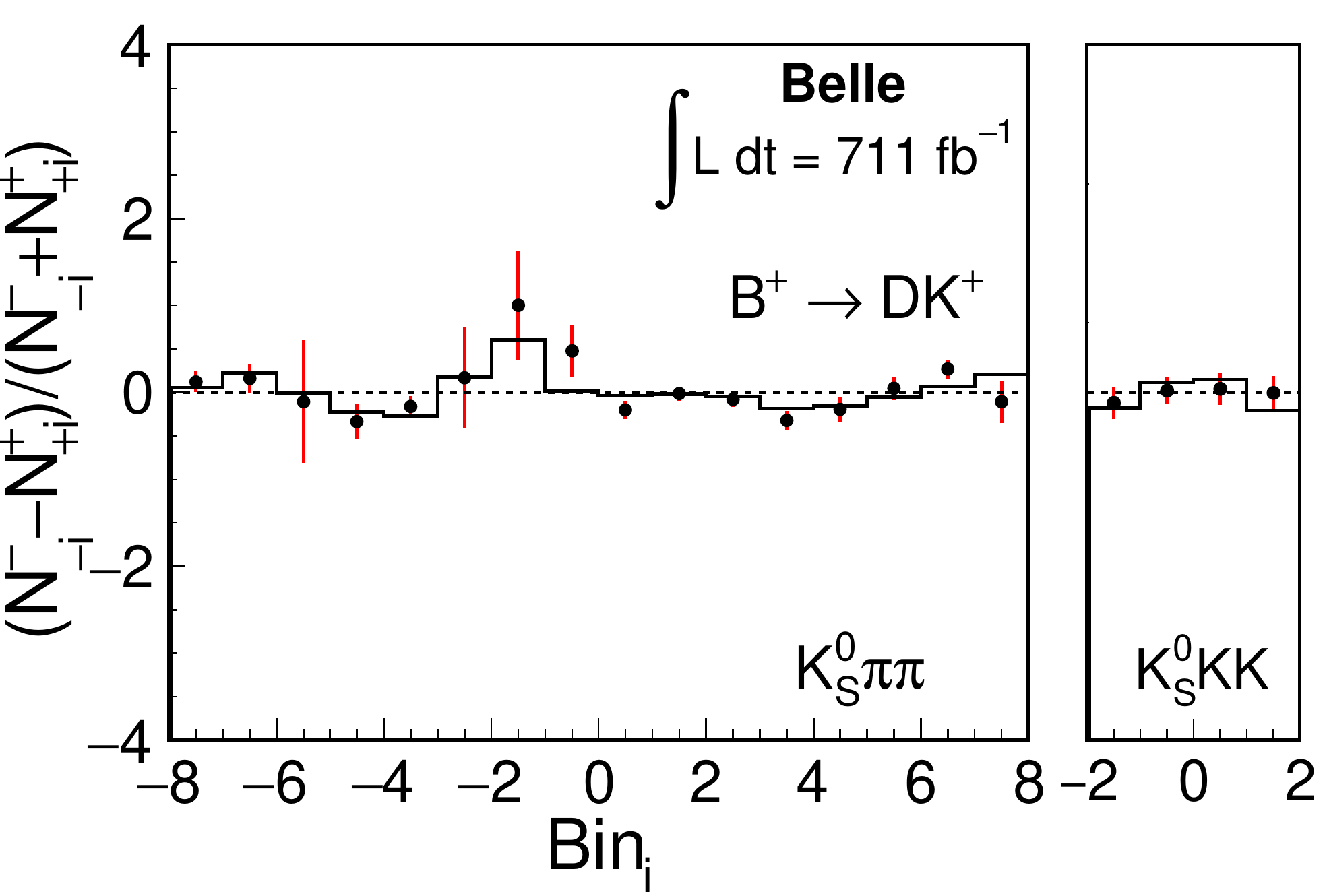} &
		\includegraphics[scale=0.36]{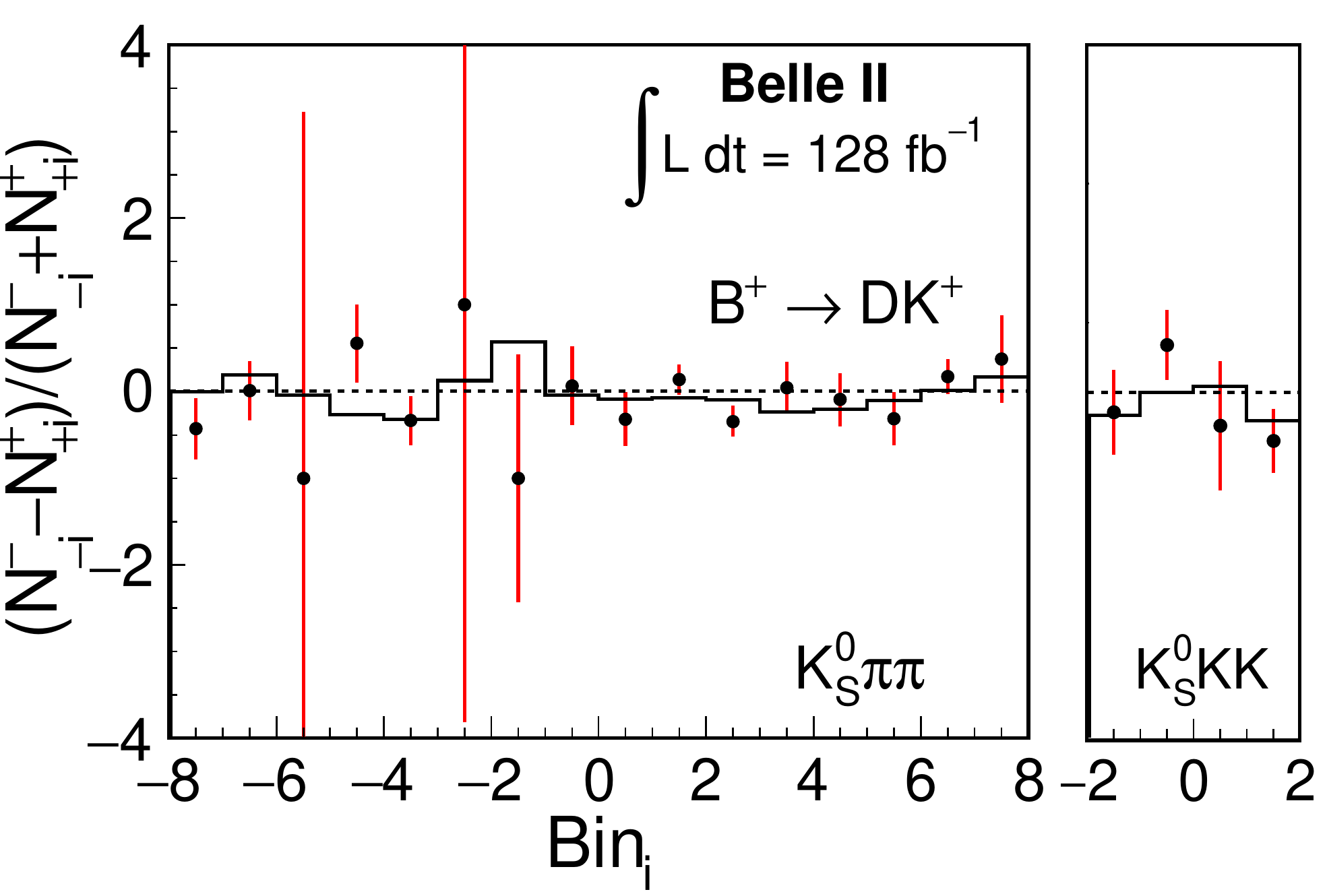} \\
		\includegraphics[scale=0.36]{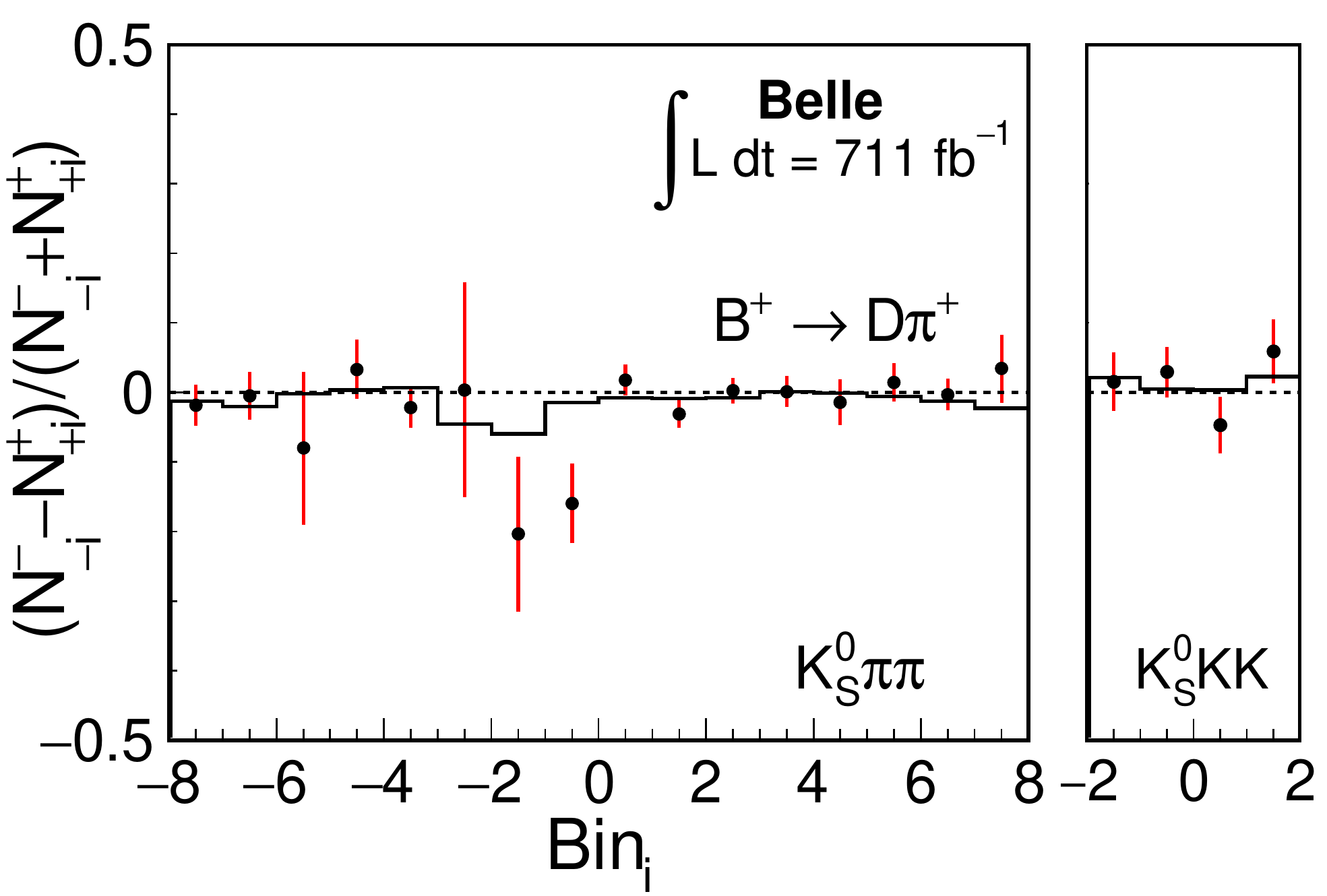} &
		\includegraphics[scale=0.36]{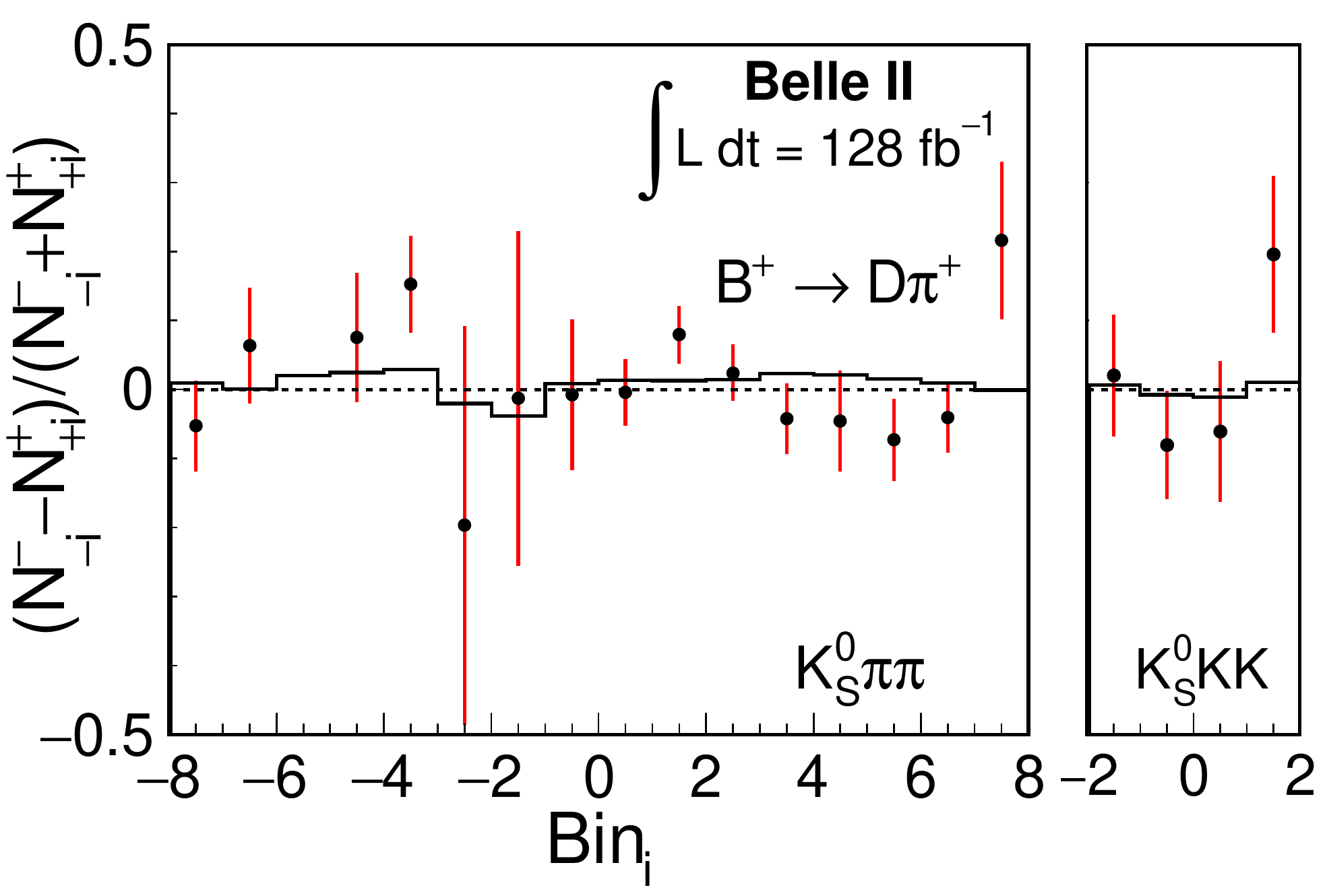} \\
	\end{tabular}
	\caption{Per-bin yield asymmetries $\left(N_{-i}^{-} - N_{+i}^{+}\right)/\left(N_{-i}^{-} + N_{+i}^{+}\right)$ in each Dalitz plot bin $i$ for $B^+ \to DK^+$ (top) and $B^+ \to D\pi^+$ (bottom) for the Belle (left) and Belle~II (right) data sets. The asymmetries produced in fits with independent bin yields are given with statistical error bars, and the prediction from the best-combined-fit values of the $(x,y)$ parameters is displayed with a solid line. The dotted line is the expectation without {\it CP} violation.}
\label{fig:asymmetry}
\end{figure}

The fit results along with their statistical and systematic uncertainties are summarised in Sec.~\ref{sec:phi3}, and the likelihood contours are shown in Fig.~\ref{fig:x_vs_y}. The correlations between the parameters are given in Appendix~\ref{app:corr}. The bin-by-bin asymmetries \linebreak $\left(N_{-i}^{-} - N_{+i}^{+}\right)/\left(N_{-i}^{-} + N_{+i}^{+}\right)$ in each Dalitz plot bin $i$ are shown in Fig.~\ref{fig:asymmetry}. Clear evidence for {\it CP} violation is seen in the Belle kaon-enhanced sample as in the earlier Belle analysis \cite{anton}. We assess the significance of the observed {\it CP} violation by comparing the likelihood to that from a fit under the no {\it CP}-violation hypothesis of $x^{DK}_+=x^{DK}_-$ and $y^{DK}_+=y^{DK}_-$. Considering only the statistical uncertainties we find the significance is $5.8$~standard deviations.

\begin{figure}[!t]
    \centering
    \includegraphics[scale=0.45]{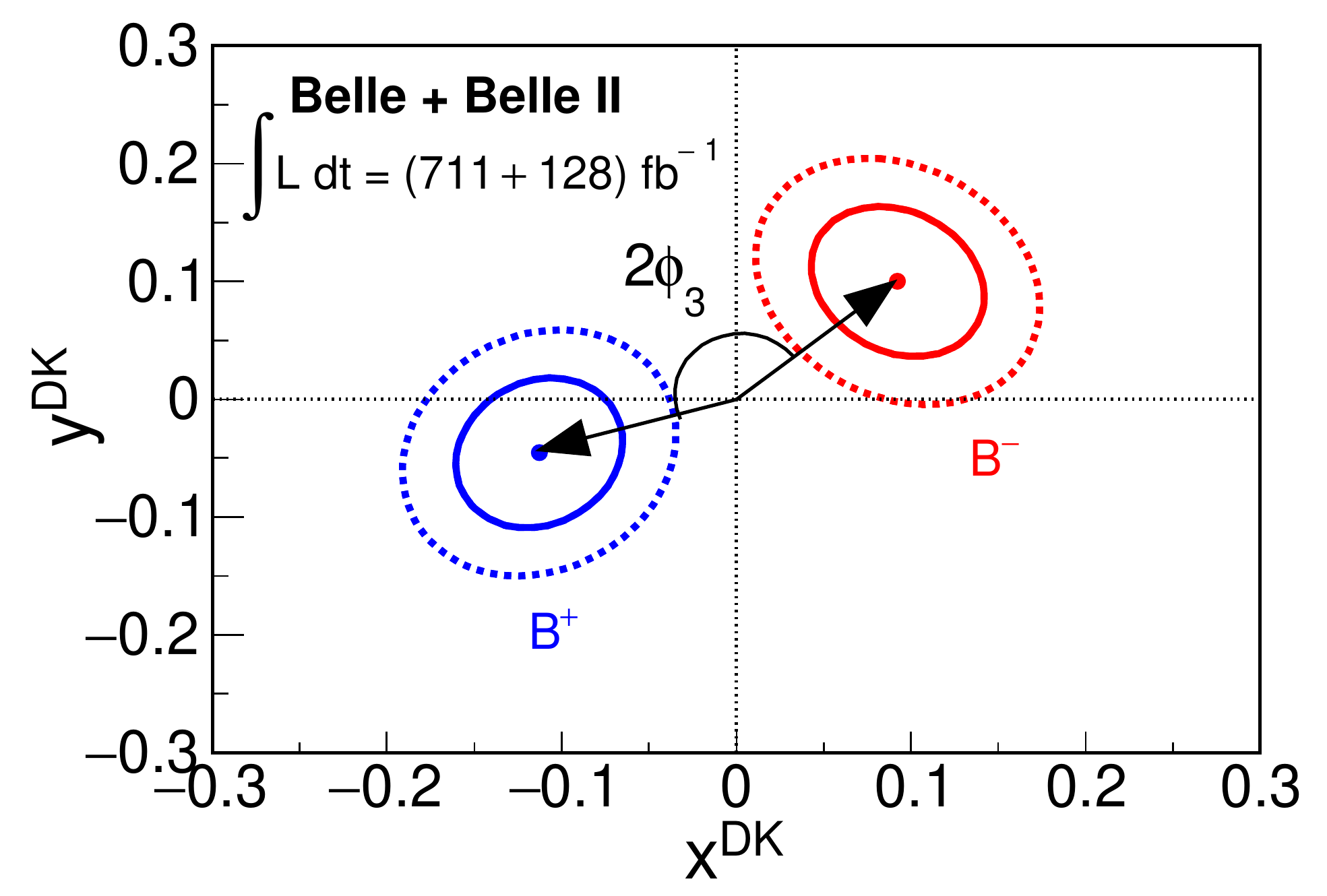}
    \caption{Two-dimensional confidence regions at (inner curve) 68.3\% and (outer curve) 95.5\% probability  for $\left(x_{+}^{DK}, y_{+}^{DK} \right)$ (blue) and $\left(x_{-}^{DK}, y_{-}^{DK} \right)$ (red) as measured in $B^+ \to DK^+$ decays from a profile likelihood scan. The dots represent central values.}
    \label{fig:x_vs_y}
\end{figure}

%%%%%%%%%%%%%%%%%%%%%%%%%%%%%%%%%%%%%%%%%%%%%%%%%%%%%%%%%%%%%%%%%%%%%%%%%%
\section{Systematic uncertainties}
\label{sec:systematics}

Several possible sources of systematic uncertainties are considered, which are listed in Table~\ref{tab:systematics}. This section explains each source and the methodology adopted to compute the systematic uncertainties. The only correlated sources of systematic uncertainty between Belle and Belle~II are the input $c_{i}$ and $s_{i} $ values, as well as the fit bias. All other systematic uncertainties are assessed independently for Belle and Belle~II, and are summed in quadrature.

In general, we smear the input nominal values by their uncertainties and then perform the fit to assess the associated systematic uncertainty. The smearing procedure is repeated 1000 times and the resulting width of the fit parameter distribution from this ensemble of fits is considered as the corresponding systematic uncertainty. If the input values are correlated, we use the Cholesky decomposition~\cite{cholesky} of the covariance matrix to smear the uncertainties, which takes the correlation into account. This approach is used to compute the contributions of the external inputs $ c_{i} $ and $ s_{i} $, and any correlated fixed parameters used to describe the PDFs. The correlations of the external inputs are taken from Refs.~\cite{kspipi_cisi,kskk_cisi}. The correlations between the fixed fit parameters are taken from the results of fits to simulated samples of the signal and background components.\footnote{We provide the detailed output of this study as supplementary material to this paper at the publisher’s website, as noted in Ref.~\cite{lhcb_gamma} this will provide sufficient information to determine the correlation between this uncertainty and the corresponding uncertainties of other $\phi_3$ measurements that also rely on the same strong-phase measurements.} The results are listed in  Table~\ref{tab:systematics}. The corresponding uncertainties are less than 20\% and 5\% of the total statistical uncertainty for the external inputs and the fixed fit parameters, respectively. The systematic uncertainties related to PID are also calculated using this same strategy, where the efficiency and misidentification rates are varied within their uncertainty, independently for Belle and Belle~II. To estimate the effect of ignoring the charmless peaking background we repeat the fits include a fixed peaking background component normalised to the yields found in the studies of the $m\left(K^0_{\rm S}h^+h^-\right)$ sideband. The background is modelled using the same PDF distributions as the signal. The resulting bias in the central values of the physics parameters, which is two orders of magnitude smaller than the statistical uncertainty, is considered as a systematic uncertainty.

Fit biases are investigated using linearity tests. Many simulated data sets of bin yields are generated for five different values of $ x^{DK}_{\pm}$, $y^{DK}_{\pm}$, $x_{\xi}^{D\pi}$ and $y_{\xi}^{D\pi}$: $ 0, \pm 0.05$ and $ \pm 0.1$. Each sample is then fit to determine the parameters. For an unbiased sample a graph of generated versus fit values should be linear with slope one and intercept zero. No significant bias is observed. The slopes agree with unity except for $ x_{\xi}^{D\pi}$, which differs by three standard deviations. As this is a nuisance parameter in the fit, it has no impact on the final $\phi_{3}$ extraction. The associated systematic uncertainty is assessed using the uncertainty in the slope and the data central values.
\begin{table}[!htb]
\centering
\begin{tabular}{| l c c c c c c |}
\hline
 Source & $ \sigma_{x_{+}^{DK}} $ & $ \sigma_{y_{+}^{DK}} $ & $ \sigma_{x_{-}^{DK}} $ & $ \sigma_{y_{-}^{DK}} $ & $ \sigma_{x_{\xi}^{D\pi} }$ & $ \sigma_{y_{\xi}^{D\pi}} $\\
\hline
  Input $ c_\mathnormal{i} $, $ s_\mathnormal{i} $ & $\phantom{<}0.22$  & $\phantom{<}0.55$ & $\phantom{<}0.23$ & 0.67 & $\phantom{<}0.73$ & $\phantom{<}0.82$\\
\hline
 PDF parametrisation & $\phantom{<}0.07$ & $\phantom{<}0.08$ & $\phantom{<}0.12$ & 0.16 & $\phantom{<}0.12$ & $\phantom{<}0.12$\\
 PID & $<0.01$ & $<0.01$ & $<0.01$ & 0.01 & $<0.01$ & $<0.01$\\
 Peaking background & $\phantom{<}0.03$ & $\phantom{<}0.05$ & $\phantom{<}0.03$ & 0.04 & $\phantom{<}0.02$ & $\phantom{<}0.10$\\
 Fit bias & $\phantom{<}0.16$ & $\phantom{<}0.06$ & $\phantom{<}0.12$ & 0.16 & $\phantom{<}0.49$ & $\phantom{<}0.10$ \\
Bin migration & $<0.01$ & $<0.01$ & $<0.01$ & $<0.01$ & $<0.01$ & $\phantom{<}0.03$\\
\hline
 Total & $\phantom{<}0.18$ & $\phantom{<}0.11$ & $\phantom{<}0.17$ & 0.23 & $\phantom{<}0.51$ & $\phantom{<}0.19$\\
 \hline
 
 Statistical & $\phantom{<}3.15$ & $\phantom{<}4.20$ & $\phantom{<}3.27$ & 4.20 & $\phantom{<}4.75$ & $\phantom{<}5.44$ \\
\hline
\end{tabular}
\caption{Systematic uncertainty summary. All values are quoted in units of $10^{-2} $.}
\label{tab:systematics}
\end{table}

We also check the contribution of migration between $D$-decay Dalitz plot bins due to the finite resolution of $m_\pm^2$. The definition of $F_i$ includes the effect of migration if it is the same in the $B^+\to D\pi^+$ and $B^+\to DK^+$ samples. However, the presence of {\it CP} violation means the Dalitz plot densities of the two samples are different, which can lead to differing levels of migration. Therefore, we generate samples of events including {\it CP} violation and fit them with and without the effect of $m_\pm^2$ resolution included. The parameter values shift less than $10^{-4}$ except for $y_{\xi}^{D\pi}$; the full bias is treated as a systematic uncertainty on $y_{\xi}^{D\pi}$.

We assume that the values of $F_i$ are the same for $B^{+}\to D\left(K^0_{\rm S}h^+ h^-\right)K^+$ and \linebreak $B^{-}\to \left(K^0_{\rm S}h^+ h^-\right)\pi^+$ decays. In principle a small difference exists due to the altered kinematics induced by the differing pion and kaon masses. We investigated the validity of our assumption in large simulated samples. No significant difference is observed in the values of $F_i$ so no related systematic uncertainty is assigned. We also consider how the Belle and Belle~II Dalitz plot acceptance might distort the effective values of $c_i$ and $s_i$, which are measured assuming a uniform acceptance. The values are calculated with and without the Belle (Belle~II) acceptance included. The deviations in the values of $c_i$ and $s_i$ are at most an order of magnitude smaller than the reported uncertainties~\cite{kspipi_cisi,kskk_cisi}, which are already considered in our measurement. Therefore, this potential source of systematic uncertainty is ignored.

As a further check of the fit performance, we generate 1000 simplified-simulated experiments with mean signal yields that correspond to our measured values of {\it CP}-violating parameters. These samples are then fit in an identical manner to the data. The results verify that the fit is stable and unbiased with the current sample size, as well as providing the appropriate statistical coverage. We also find that the uncertainties on measured {\it CP}-violating parameters in data lie within the distribution of uncertainties generated by the simplified-simulated experiments.

%%%%%%%%%%%%%%%%%%%%%%%%%%%%%%%%%%%%%%%%%%%%%%%%%%%%%%%%%%%%%%%%%%%%%%%%
\boldmath \section{Determination of $\phi_3, r_{B}^{DK}$ and $\delta_B^{DK}$} \unboldmath
\label{sec:phi3}
The parameters obtained from the fit are
\begin{equation} \label{eq:xy_param}
    \begin{split}
        x_{-}^{DK} &= \left(\phantom{-0}9.24 \pm 3.27 \pm 0.17 \pm 0.23\right) \times 10^{-2}, \\
        y_{-}^{DK} &= \left(\phantom{-}10.00 \pm 4.20 \pm 0.23 \pm 0.67\right) \times 10^{-2}, \\
        x_{+}^{DK} &= \left(-11.28 \pm 3.15 \pm 0.18 \pm 0.22\right) \times 10^{-2}, \\
        y_{+}^{DK} &= \left(\,\,\,-4.55 \pm 4.20 \pm 0.11 \pm 0.55\right) \times 10^{-2}, \\
        x_{\xi}^{D\pi} &= \left(-11.09 \pm 4.75 \pm 0.51 \pm 0.73\right) \times 10^{-2}, \\
        y_{\xi}^{D\pi} &= \left(\,\,\,-7.90 \pm 5.44 \pm 0.19 \pm 0.82\right) \times 10^{-2}, \\
    \end{split}
\end{equation}
where the first uncertainty is statistical, the second is the total experimental systematic uncertainty, and the third is the systematic uncertainty due to the external strong-phase difference inputs \cite{kspipi_cisi,kskk_cisi}.

\begin{figure}[!t]
	\begin{tabular}{c c}
		\includegraphics[scale=0.36]{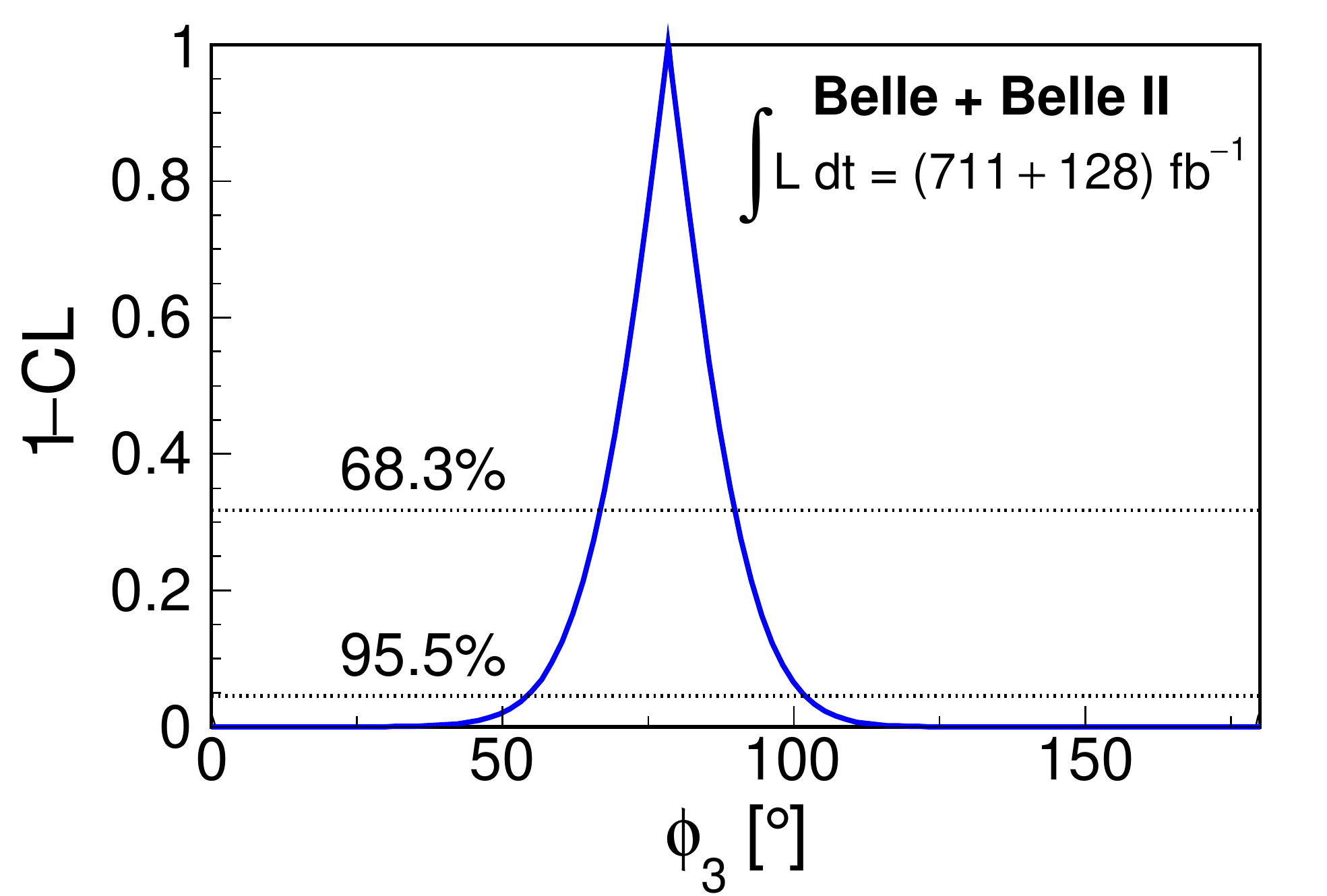} &
		\includegraphics[scale=0.36]{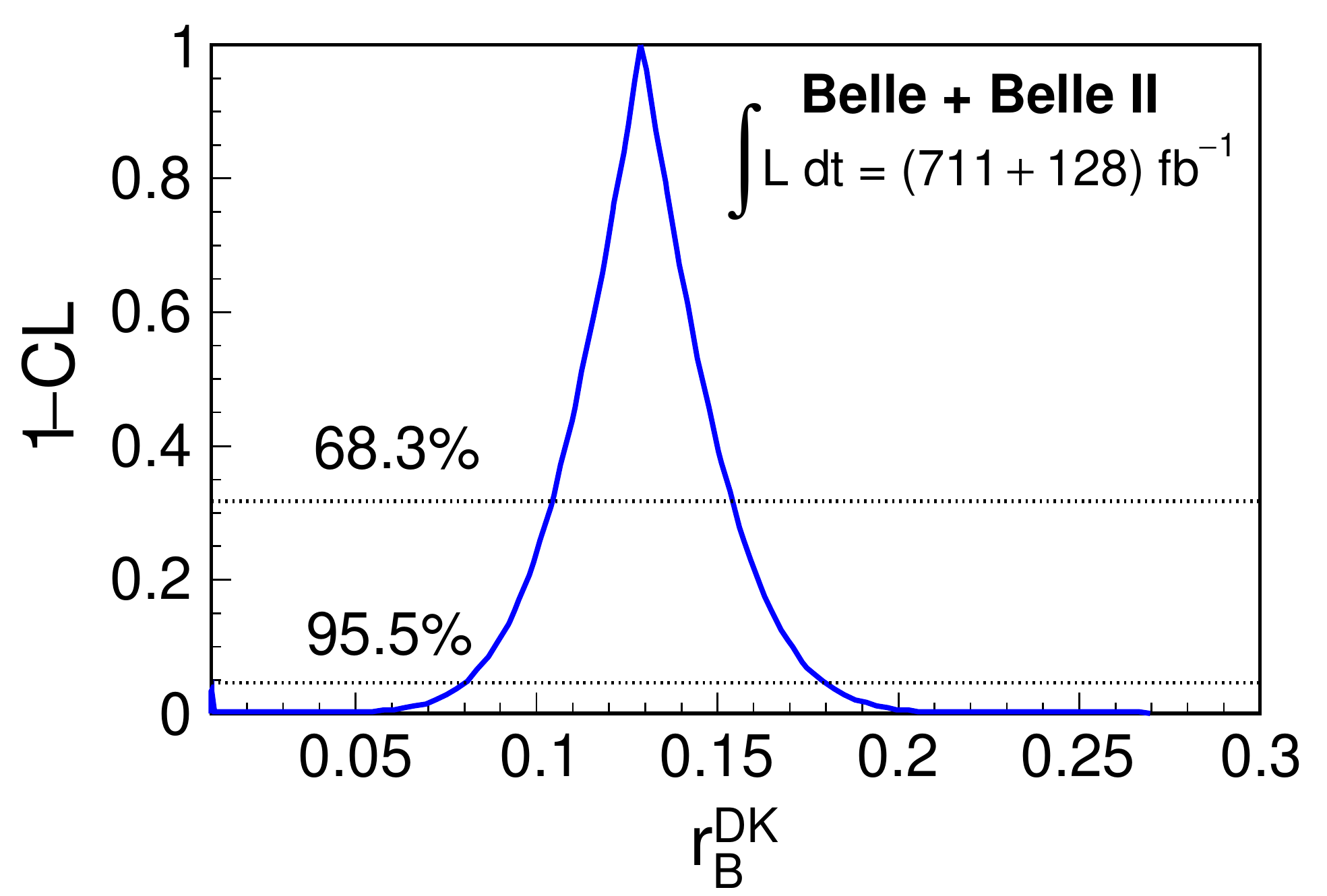}\\
	\end{tabular}
	\caption{$p$-value as a function of (left) $ \phi_{3} $ and (right) $ r_{B}^{DK} $ calculated using the methods described in Ref.~\cite{lhcb_gamma_combination}.}
	\label{fig:phi3_rB_comb}
\end{figure}

\begin{figure}[!t]
	\begin{tabular}{c c}
		\includegraphics[scale=0.36]{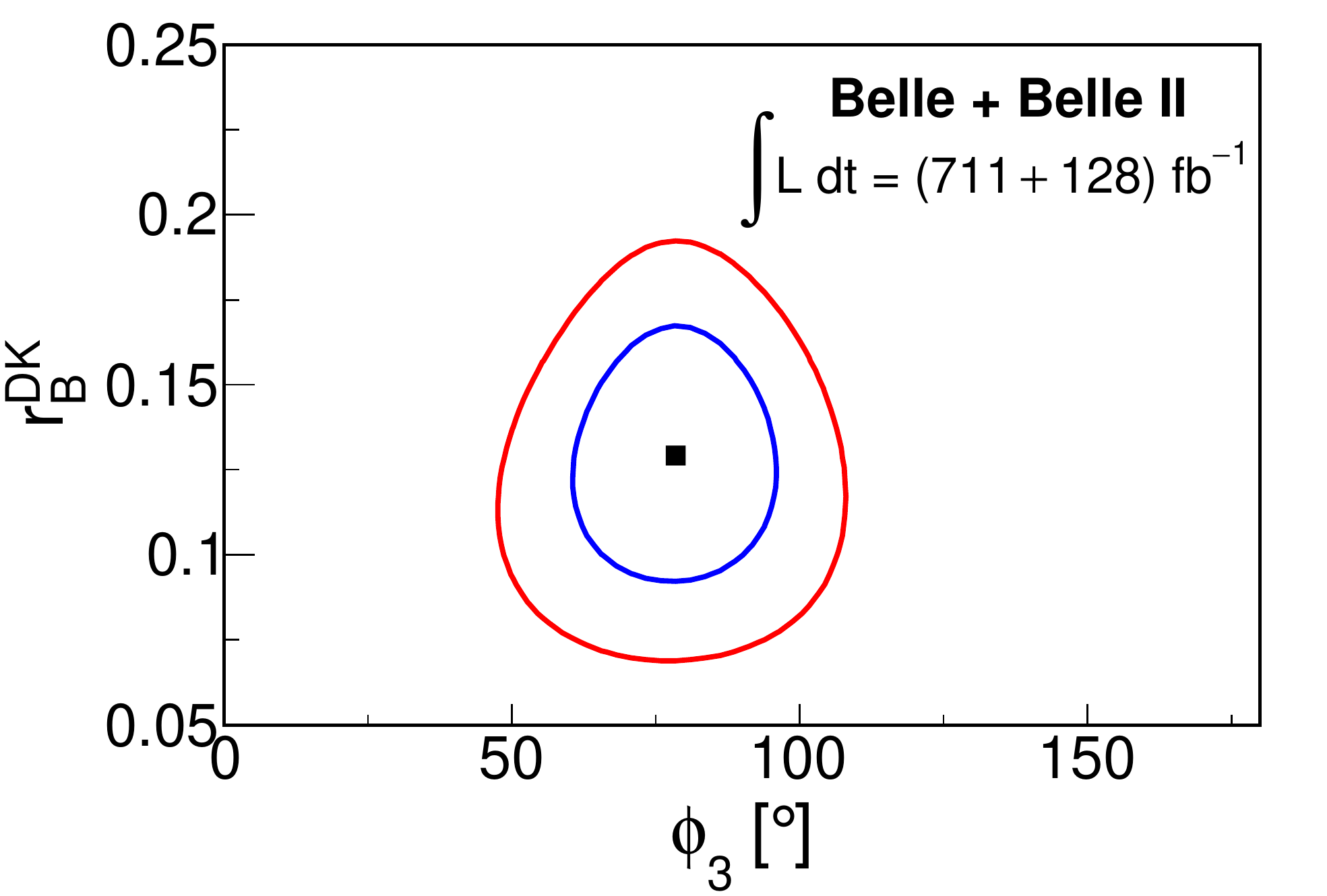} &
		\includegraphics[scale=0.36]{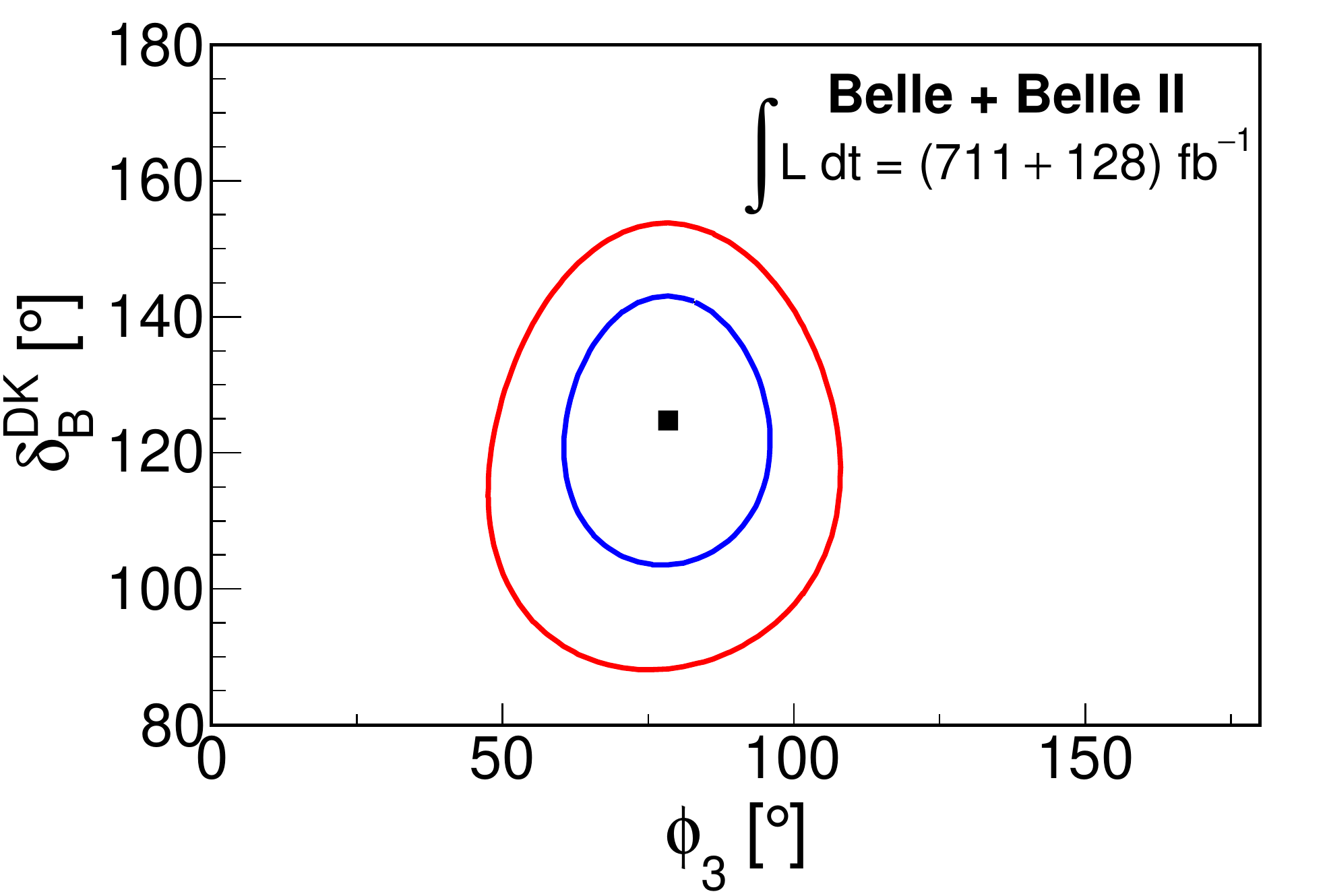} \\
	\end{tabular}
	\caption{Two-dimensional confidence regions at the (inner curve) 68\% and (outer curve) 95\%, obtained for (left) $ \phi_{3}-r_{B}^{DK} $  and (right) $ \phi_{3}-\delta_{B}^{DK}$ using the methods described in Ref.~\cite{lhcb_gamma_combination}. Note the suppressed zeroes on the vertical scales.}
	\label{fig:contour_comb}
\end{figure}

\begin{figure}[!t]
    \centering
	\includegraphics[scale=0.36]{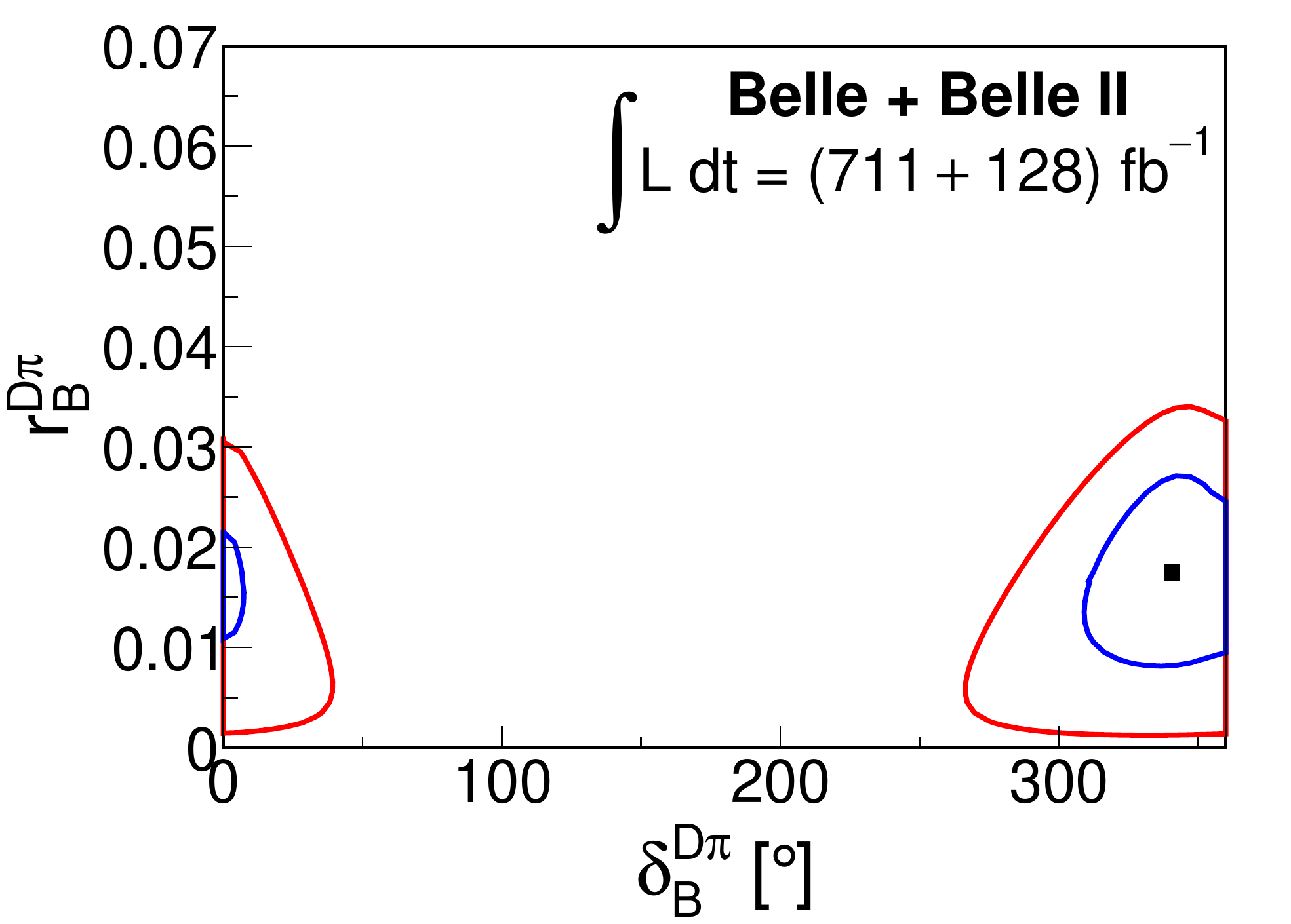}
	\caption{ Two-dimensional confidence regions at the (inner curve) 68\% and (outer curve) 95\%, obtained for $ \delta_{B}^{D\pi}-r_{B}^{D\pi} $ Ref.~\cite{lhcb_gamma_combination}.}
	\label{fig:contour_comb_dpi}
\end{figure}

The parameters $\phi_3, r_{B}^{DK}$, $\delta_B^{DK}$, $r_{B}^{D\pi}$ and $\delta_{B}^{D\pi}$ are determined from $x^{DK}_{\pm}$, $y^{DK}_{\pm}$, $x^{D\pi}_{\xi}$ and $y^{D\pi}_{\xi}$ using a frequentist approach implemented in the \textsc{GammaCombo} package~\cite{lhcb_gamma_combination}.\footnote{Of the methods available in {\sc GammaCombo}, we adopt the so-called plug-in method, which uses simulated samples and assumes the nuisance parameter values observed in data.} Generally, there is a two-fold ambiguity in the results of these physics parameters as Eqs.~(\ref{eq:byields}) are invariant under the simultaneous substitutions of $ \phi_{3} = \phi_{3} + 180^\circ $ and $ \delta_{B}^{Dh} = \delta_{B}^{Dh} + 180^\circ $. We choose the solution in the range $ 0^\circ < \phi_{3} < 180^\circ $, which is favoured by other measurements \cite{hflav}. The results are
\begin{equation}\label{eq:phi3}
\begin{split}
\phi_3 &= \left(78.4 \pm 11.4 \pm 0.5 \pm 1.0\right)^{\circ}, \\
r_{B}^{DK} &= 0.129 \pm 0.024 \pm 0.001 \pm 0.002, \\
\delta_{B}^{DK} &= \left(124.8 \pm 12.9 \pm 0.5\pm 1.7\right)^{\circ}, \\ 
r_{B}^{D\pi} &=  0.017 \pm 0.006 \pm 0.001 \pm 0.001, \\
\delta_{B}^{D\pi} &= \left(341.0 \pm 17.0 \pm 1.2 \pm 2.6\right)^{\circ}. \\ 
\end{split}
\end{equation}  
The statistical confidence intervals for $ \phi_{3} $ and $ r_{B}^{DK} $ are illustrated in Fig.~\ref{fig:phi3_rB_comb}, while Fig.~\ref{fig:contour_comb} shows the two-dimensional statistical confidence regions obtained for the ($ \phi_{3}, r_{B}^{DK} $) and ($ \phi_{3}, \delta_{B} $) parameter combinations. Fig.~\ref{fig:contour_comb_dpi} shows the two-dimensional statistical confidence region obtained for the ($\delta_{B}^{D\pi}, r_{B}^{D\pi} $)  parameter combination; the 95\% confidence region is compatible with the most precise values of these parameters reported \cite{lhcbmeta}.
The $\phi_3$ result is consistent with the previous Belle analysis \cite{anton} but the statistical precision on $\phi_3$ is improved from $15^{\circ}$ due to improved $K^0_{\rm S}$ selection and background suppression. The uncertainty related to strong-phase inputs has also decreased from $ 4^{\circ} $ because of the new measurements reported by the BESIII collaboration \cite{kspipi_cisi, kskk_cisi}. Furthermore, the experimental systematic uncertainty has decreased from $4^{\circ}$ primarily from the improved background suppression and the use of the $B^+\to D\pi^+$ sample to determine the acceptance.  
    
%%%%%%%%%%%%%%%%%%%%%%%%%%%%%%%%%%%%%%%%%%%%%%%%%%%%%%%%%%%%%%%%%%%%%%%%%%%
\section{Conclusion}
\label{sec:conclusion}
The results of the first Belle and Belle~II combined model-independent measurement of the CKM unitarity triangle angle $ \phi_{3} $ are presented. The analysis uses $B^+ \to D(K_{S}^0 h^- h^+) h^+$ decays reconstructed from a combined sample of \SI{711}{fb^{-1}} of Belle data and \SI{128}{fb^{-1}} of Belle~II data. Independently measured strong-phase difference parameters $ c_{i}$ and $s_{i} $ are used, which come from a combination of results reported by the CLEO and BESIII collaborations~\cite{kspipi_cisi,kskk_cisi}. We measure $ \phi_3 = \left(78.4 \pm 11.4 \pm 0.5 \pm 1.0 \right)^{\circ}$, where the first uncertainty is statistical, the second is the total experimental systematic uncertainty and the third is the systematic uncertainty due to the external $c_i$ and $s_i$ measurements. 

The measurement is also performed on the Belle data sample alone and the results are reported in Appendix~\ref{app:Belle_phi3}. The statistical uncertainty in $\phi_3$ is $11^{\circ}$, which is significantly improved from the $15^{\circ}$ reported in the previous Belle analysis with the same data set \cite{anton}. The improvements are primarily due to the improved background rejection and $K_{\rm S}^{0}$ selection, as well as the addition of  $B^{+}\to D(K^{0}_{\rm S} K^+ K^-)h^+$ decays. 
The inclusion of Belle~II data improves the precision of $x^{DK}_{\pm}$ and $y^{DK}_{\pm}$ parameters. However, the $\phi_3$ statistical uncertainty does not improve despite introducing 17\% more data. Belle~II data favours a much smaller value of $r_B^{DK}$, which results in a central value of 0.129 for the combined fit compared to 0.144 for the Belle data alone. The uncertainty in $\phi_3$ is inversely proportional to $r_B$, which explains the lack of improvement in $\phi_3$ sensitivity when including the Belle~II data. The world average value of $r_B$ is $0.0996\pm 0.0026$ \cite{hflav} so it is possible that the value of $r_B$ will
decrease and approach this value as additional data is included.
 
The statistical precision on $\phi_3$ is worse than the current world-average value \cite{hflav}. However, the precision is limited by the size of the data sample, so a future analysis with a Belle~II data set corresponding to \SI{10}{ab^{-1}} will provide measurements with a precision of approximately $4^\circ$ from the $B^+\to D\left( K^{0}_{\rm S}\pi^+\pi^-\right)h^+$ mode alone.\footnote{The world-average value of $r_B^{DK}$ is assumed in these extrapolations.} The use of other modes will give additional sensitivity to $\phi_3$~\cite{Belle2phybook}.
%%%%%%%%%%%%%%%%%%%%%%%%%%%%%%%%%%%%%%%%%%%%%%%%%%%%%%%%%%%%%%%%%%%%%%%%%%%%

\clearpage
\appendix
%%%%%%%%%%%%%%%%%%%%%%%%%%%%%%%%%%%%%%%%%%%%%%%%%%%%%%%%%%%%%%%%%%%%%%%%%
%%%%%%%%%%%%%%%%%%%%%%%%%%%%%%%%%%%%%%%%%%%%%%%%%%%%%%%%%%%%%%%%%%%%%%%%
\section{Correlation matrices}
\label{app:corr}
Tables~\ref{tab:corr_comb_fit}$-$\ref{tab:syst_exp_corr_comb_fit} represent the statistical, external strong phase input systematics and total experimental systematics correlation matrices for the combined data set of Belle and Belle~II.
%%%%%%%%%%%%%%%%%%%% Belle + Belle2 statistical%%%%%%%%%%%%%%%%%%%%%%
\begin{table}[!htb]
	\centering
	\begin{tabular}{|c | c c c c c c |}
		\hline 
		& $x_{-}^{DK}$ & $y_{-}^{DK}$ & $x_{+}^{DK}$ & $y_{+}^{DK}$ & $x_{\xi}^{D\pi}$ & $y_{\xi}^{D\pi}$ \\
		\hline
		$x_{-}^{DK}$ & 1 & $-0.204$ & $-0.051$ & $\phantom{-}0.063$ & $\phantom{-}0.365$ & $-0.151$ \\  
		$y_{-}^{DK}$ & & 1 &	$\phantom{-}0.014$ &	$-0.051$ &	$-0.090$ &	$\phantom{-}$0.404\\
		$x_{+}^{DK}$ & & & 1 &	$\phantom{-}0.152$ &	$-0.330$ &	$-0.057$ \\
		$y_{+}^{DK}$ & & & & 1 &	$\phantom{-}$0.026 &	$-0.391$ \\
		$x_{\xi}^{D\pi}$ & & & & & 1 &	$\phantom{-}0.080$\\
		$y_{\xi}^{D\pi}$ & & & & & & 1\\
		\hline 
	\end{tabular}
	\caption{Statistical correlation matrix obtained for the combined Belle and Belle~II data set.}
\label{tab:corr_comb_fit}
\end{table}
%%%%%%%%%%%%%%%%%% Belle + Belle2 external input %%%%%%%%%%%%%%%%%%%%%%%%
\begin{table}[!htb]
	\centering
	\begin{tabular}{|c | c c c c c c |}
		\hline 
		& $x_{-}^{DK}$ & $y_{-}^{DK}$ & $x_{+}^{DK}$ & $y_{+}^{DK}$ & $x_{\xi}^{D\pi}$ & $y_{\xi}^{D\pi}$ \\
		\hline
		$x_{-}^{DK}$ & 1 & $-0.113$ & $\phantom{-}0.069$ & $\phantom{-}0.406$ & $-0.016$ & $\phantom{-}0.114$ \\  
		$y_{-}^{DK}$ & & 1 &	$\phantom{-}0.038$ &	$-0.196$ &	$-0.692$ &	$-0.106$\\
		$x_{+}^{DK}$ & & & 1 &	$\phantom{-}0.412$ &	$-0.226$ &	$-0.469$ \\
		$y_{+}^{DK}$ & & & & 1 &	$\phantom{-}$0.180 &	$-0.069$ \\
		$x_{\xi}^{D\pi}$ & & & & & 1 &	$\phantom{-}0.622$\\
		$y_{\xi}^{D\pi}$ & & & & & & 1\\
		\hline 
	\end{tabular}
	\caption{External inputs $c_i, s_i$ systematics correlation matrix obtained for the combined Belle and Belle~II data set.}
\label{tab:syst_cisi_corr_comb_fit}
\end{table}
%%%%%%%%%%%%%%%%%% Belle + Belle2 total experimental %%%%%%%%%%%%%%%%%%%%%%%%
\begin{table}[!htb]
	\centering
	\begin{tabular}{|c | c c c c c c |}
		\hline 
		& $x_{-}^{DK}$ & $y_{-}^{DK}$ & $x_{+}^{DK}$ & $y_{+}^{DK}$ & $x_{\xi}^{D\pi}$ & $y_{\xi}^{D\pi}$ \\
		\hline
		$x_{-}^{DK}$ & 1 & $0.777$ & $0.483$ & $0.268$ & $0.839$ & $0.698$ \\  
		$y_{-}^{DK}$ & & 1 &	$0.411$ &	$0.504$ &	$0.802$ &	$0.797$\\
		$x_{+}^{DK}$ & & & 1 &	$0.680$ &	$0.766$ &	$0.377$ \\
		$y_{+}^{DK}$ & & & & 1 &	$0.480$ &	$0.303$ \\
		$x_{\xi}^{D\pi}$ & & & & & 1 &	$0.638$\\
		$y_{\xi}^{D\pi}$ & & & & & & 1\\
		\hline 
	\end{tabular}
	\caption{Experimental systematics correlation matrix obtained for the combined Belle and Belle~II data set.}
\label{tab:syst_exp_corr_comb_fit}
\end{table}

%%%%%%%%%%%%%%%%%%%%%%%%%%%%%%%%%%%%%%%%%%%%%%%%%%%%%%%%%%%%%%%%%%%%%%%%%%%%

\section{Belle data results}
\label{app:Belle_phi3}
The results obtained using only Belle data set are summarised in this section. The obtained values of the {\it CP} violating parameters are
\begin{equation}\label{eq:belle_xy}
    \begin{split}
    x_{-}^{DK}   &=  \left(\phantom{-0}9.45 \pm 3.58 \pm 0.22 \pm 0.36\right) \times 10^{-2}, \\
    y_{-}^{DK}   &=  \left(\phantom{-}12.04 \pm 4.66 \pm 0.30 \pm 0.90\right) \times 10^{-2},  \\
     x_{+}^{DK}   &= \left( -11.94 \pm 3.46 \pm 0.23 \pm 0.37\right) \times 10^{-2}, \\
 y_{+}^{DK}  &=  \left(\,\,\,-6.34 \pm 4.75 \pm 0.17 \pm 0.93\right) \times 10^{-2}, \\
 x_{\xi}^{D\pi}   &=  \left(\,\,\,-8.76 \pm 4.50 \pm 0.50 \pm 0.69\right) \times 10^{-2}, \\
 y_{\xi}^{D\pi}  &=  \left(\,\,\,-4.56 \pm 4.90 \pm 0.17 \pm 0.63\right) \times 10^{-2}. \\        
    \end{split}
\end{equation}
The obtained physics parameters values are
 \begin{equation} \label{eq:belle_phi3}
    \begin{split}
   \phi_3 &=  \left( 79.3 \pm 11.0 \pm 0.6 \pm 1.4\right)^{\circ}.  \\      
 r_{B}^{DK}  &=  0.144 \pm 0.028 \pm 0.002 \pm 0.004, \\
 \delta_{B}^{DK}  &=  \left( 130.1 \pm 12.4 \pm 0.6 \pm 2.2\right)^{\circ}.\\
 r_{B}^{D\pi} &=  0.014 \pm 0.007 \pm 0.001 \pm 0.001, \\
\delta_{B}^{D\pi} &= \left(337.5 \pm 22.0 \pm 1.3 \pm 2.9\right)^{\circ}. \\
\end{split}
\end{equation}
The correlation matrices related to statistical, external strong phase input systematics, and total experimental systematics are given in Tables~\ref{tab:correlation_belle}$-$\ref{tab:syst_exp_corr_belle}.
%%%%%%%%%%%%%%%%% Belle statistical %%%%%%%%%%%%%%%%%%%%%%%%%%%%%%%%%%%%%%%%%%
\begin{table}[!htb]
    \centering
    \begin{tabular}{|c|c c c c c c |}
    \hline
    & $x_{-}^{DK}$ & $y_{-}^{DK}$ & $x_{+}^{DK}$ & $y_{+}^{DK}$ & $x_{\xi}^{D\pi}$ & $y_{\xi}^{D\pi}$ \\
    \hline
    $x_{-}^{DK}$ & 1 & $-0.205$ & $-0.054$ & $\phantom{-}0.031$ & $\phantom{-}0.245$ & $-0.167$ \\  
    $y_{-}^{DK}$ & & 1 &	$\phantom{-}0.000$ &	$-0.045$ & $\phantom{-}0.033$ &	$\phantom{-}$0.315\\
    $x_{+}^{DK}$ & & & 1 &	$\phantom{-}0.167$ &	$-$0.298 &	$-0.017$ \\
    $y_{+}^{DK}$ & & & & 1 &	$\phantom{-}$0.103 &	$-0.323$ \\
    $x_{\xi}^{D\pi}$ & & & & & 1 &	$\phantom{-}0.184$\\
    $y_{\xi}^{D\pi}$ & & & & & & 1\\
    \hline
    \end{tabular}
    \caption{Statistical correlation matrix obtained for the Belle standalone data set.}
    \label{tab:correlation_belle}
\end{table}
%%%%%%%%%%%%%%%%%%%% Belle ci si systematics %%%%%%%%%%%%%%%%%%%%%%%%%%%%%%%%
\begin{table}[!htb]
	\centering
	\begin{tabular}{|c | c c c c c c |}
		\hline 
		& $x_{-}^{DK}$ & $y_{-}^{DK}$ & $x_{+}^{DK}$ & $y_{+}^{DK}$ & $x_{\xi}^{D\pi}$ & $y_{\xi}^{D\pi}$ \\
		\hline
		$x_{-}^{DK}$ & 1 & $\phantom{-}0.095$ & $\phantom{-}0.333$ & $\phantom{-}0.575$ & $\phantom{-}0.383$ & $\phantom{-}0.396$ \\  
		$y_{-}^{DK}$ & & 1 &	$\phantom{-}0.279$ &	$-0.054$ &	$-0.538$ &	$-0.328$\\
		$x_{+}^{DK}$ & & & 1 &	$\phantom{-}0.475$ &	$\phantom{-}0.048$ &	$-0.222$ \\
		$y_{+}^{DK}$ & & & & 1 &	$\phantom{-}$0.420 &	$\phantom{-}0.295$ \\
		$x_{\xi}^{D\pi}$ & & & & & 1 &	$\phantom{-}0.862$\\
		$y_{\xi}^{D\pi}$ & & & & & & 1\\
		\hline 
	\end{tabular}
	\caption{External inputs $c_i, s_i$ systematics correlation matrix obtained for the Belle standalone data set.}
\label{tab:syst_cisi_corr_belle}
\end{table}
%%%%%%%%%%%%%%%%%% Belle total experimental %%%%%%%%%%%%%%%%%%%%%%%%
\begin{table}[!htb]
	\centering
	\begin{tabular}{|c | c c c c c c |}
		\hline 
		& $x_{-}^{DK}$ & $y_{-}^{DK}$ & $x_{+}^{DK}$ & $y_{+}^{DK}$ & $x_{\xi}^{D\pi}$ & $y_{\xi}^{D\pi}$ \\
		\hline
		$x_{-}^{DK}$ & 1 & $0.508$ & $0.251$ & $0.074$ & $0.674$ & $0.573$ \\  
		$y_{-}^{DK}$ & & 1 &	$0.109$ &	$0.071$ &	$0.661$ &	$0.683$\\
		$x_{+}^{DK}$ & & & 1 &	$0.467$ &	$0.589$ &	$0.324$ \\
		$y_{+}^{DK}$ & & & & 1 &	$0.254$ &	$0.125$ \\
		$x_{\xi}^{D\pi}$ & & & & & 1 &	$0.724$\\
		$y_{\xi}^{D\pi}$ & & & & & & 1\\
		\hline 
	\end{tabular}
	\caption{Experimental systematics correlation matrix obtained for the Belle standalone data set.}
\label{tab:syst_exp_corr_belle}
\end{table}

\acknowledgments
We thank Matt Kenzie for help with the {\sc GammaCombo} package and Anita for calculating the effect of the Belle (II) acceptance on the values of $c_i$ and $s_i$.
We thank the SuperKEKB group for the excellent operation of the
accelerator; the KEK cryogenics group for the efficient
operation of the solenoid; the KEK computer group for
on-site computing support; and the raw-data centers at
BNL, DESY, GridKa, IN2P3, and INFN for off-site computing support.
This work was supported by the following funding sources:
%Armenia
Science Committee of the Republic of Armenia Grant No.~20TTCG-1C010;
%Australia
Australian Research Council and research Grants
No.~DP180102629,
No.~DP170102389,
No.~DP170102204,
No.~DP150103061,
No.~FT130100303,
No.~FT130100018,
and
No.~FT120100745;
%Austria
Austrian Federal Ministry of Education, Science and Research,
Austrian Science Fund No.~P 31361-N36, and
Horizon 2020 ERC Starting Grant No.~947006 ``InterLeptons'';
%Canada
Natural Sciences and Engineering Research Council of Canada, Compute Canada and CANARIE;
%China
Chinese Academy of Sciences and research Grant No.~QYZDJ-SSW-SLH011,
National Natural Science Foundation of China and research Grants
No.~11521505,
No.~11575017,
No.~11675166,
No.~11761141009,
No.~11705209,
and
No.~11975076,
LiaoNing Revitalization Talents Program under Contract No.~XLYC1807135,
Shanghai Municipal Science and Technology Committee under Contract No.~19ZR1403000,
Shanghai Pujiang Program under Grant No.~18PJ1401000,
and the CAS Center for Excellence in Particle Physics (CCEPP);
%Czech Republic
the Ministry of Education, Youth, and Sports of the Czech Republic under Contract No.~LTT17020 and
Charles University Grant No.~SVV 260448;
%EU
European Research Council, Seventh Framework PIEF-GA-2013-622527,
Horizon 2020 ERC-Advanced Grants No.~267104 and No.~884719,
Horizon 2020 ERC-Consolidator Grant No.~819127,
Horizon 2020 Marie Sklodowska-Curie Grant Agreement No.~700525 "NIOBE",
and
Horizon 2020 Marie Sklodowska-Curie RISE project JENNIFER2 Grant Agreement No.~822070 (European grants);
%France
L'Institut National de Physique Nucl\'{e}aire et de Physique des Particules (IN2P3) du CNRS (France);
%Germany
BMBF, DFG, HGF, MPG, and AvH Foundation (Germany);
%India
Department of Atomic Energy under Project Identification No.~RTI 4002 and Department of Science and Technology (India);
%Israel
Israel Science Foundation Grant No.~2476/17,
U.S.-Israel Binational Science Foundation Grant No.~2016113, and
Israel Ministry of Science Grant No.~3-16543;
%Italy
Istituto Nazionale di Fisica Nucleare and the research grants BELLE2;
%Japan
Japan Society for the Promotion of Science, Grant-in-Aid for Scientific Research Grants
No.~16H03968,
No.~16H03993,
No.~16H06492,
No.~16K05323,
No.~17H01133,
No.~17H05405,
No.~18K03621,
No.~18H03710,
No.~18H05226,
No.~19H00682, % Niigata
No.~26220706,
and
No.~26400255,
the National Institute of Informatics, and Science Information NETwork 5 (SINET5), 
and
the Ministry of Education, Culture, Sports, Science, and Technology (MEXT) of Japan;  
%Korea
National Research Foundation (NRF) of Korea Grants
No.~2016R1\-D1A1B\-01010135,
No.~2016R1\-D1A1B\-02012900,
No.~2018R1\-A2B\-3003643,
No.~2018R1\-A6A1A\-06024970,
No.~2018R1\-D1A1B\-07047294,
No.~2019K1\-A3A7A\-09033840,
and
No.~2019R1\-I1A3A\-01058933,
Radiation Science Research Institute,
Foreign Large-size Research Facility Application Supporting project,
the Global Science Experimental Data Hub Center of the Korea Institute of Science and Technology Information
and
KREONET/GLORIAD;
%Malaysia
Universiti Malaya RU grant, Akademi Sains Malaysia, and Ministry of Education Malaysia;
%Mexico
% CINVESTAV-IPN, UNAM, UAS, BUAP and CONACYT are funded under
Frontiers of Science Program Contracts
No.~FOINS-296,
No.~CB-221329,
No.~CB-236394,
No.~CB-254409,
and
No.~CB-180023, and No.~SEP-CINVESTAV research Grant No.~237 (Mexico);
%Poland
the Polish Ministry of Science and Higher Education and the National Science Center;
%Russia
the Ministry of Science and Higher Education of the Russian Federation,
Agreement No.~14.W03.31.0026, and
the HSE University Basic Research Program, Moscow;
%Saudi Arabia
University of Tabuk research Grants
No.~S-0256-1438 and No.~S-0280-1439 (Saudi Arabia);
%Slovenia
Slovenian Research Agency and research Grants
No.~J1-9124
and
No.~P1-0135;
%Spain
Agencia Estatal de Investigacion, Spain Grants
No.~FPA2014-55613-P
and
No.~FPA2017-84445-P,
and
No.~CIDEGENT/2018/020 of Generalitat Valenciana;
%Taiwan
Ministry of Science and Technology and research Grants
No.~MOST106-2112-M-002-005-MY3
and
No.~MOST107-2119-M-002-035-MY3,
and the Ministry of Education (Taiwan);
%Thailand
Thailand Center of Excellence in Physics;
%Turkey
TUBITAK ULAKBIM (Turkey);
%Ukraine
National Research Foundation of Ukraine, project No.~2020.02/0257,
and
Ministry of Education and Science of Ukraine;
%USA
the U.S. National Science Foundation and research Grants
No.~PHY-1913789 % Indiana CEEM
and
No.~PHY-2111604, % Luther
and the U.S. Department of Energy and research Awards
No.~DE-AC06-76RLO1830, % PNNL
No.~DE-SC0007983, % Wayne State
No.~DE-SC0009824, % Florida
No.~DE-SC0009973, % VPI
No.~DE-SC0010007, % Duke
No.~DE-SC0010073, % South Carolina
No.~DE-SC0010118, % Carnegie Mellon
No.~DE-SC0010504, % Hawaii
No.~DE-SC0011784, % Cincinnati
No.~DE-SC0012704, % BNL
No.~DE-SC0019230, % Duke
No.~DE-SC0021274; % Mississippi
%last group
and
%Vietnam
the Vietnam Academy of Science and Technology (VAST) under Grant No.~DL0000.05/21-23.

%%%%%%%%%%%%%%%%%%%%%%%%%%%%%%%%%%%%%%%%%%%%%%%%%%%%%%%%%%%%%%%%%%%%%%%%%%%%%%%%%

\input{files/ref}
\end{document}

%% file: files/definitions.tex
\usepackage{graphicx} % Include figure files
\usepackage{epstopdf} % Allow to include eps files
\usepackage{dcolumn} % Align table columns on decimal point
\usepackage{xcolor} % Color support
\usepackage{amssymb} % Extended symbol collection
\usepackage{hyperref} % Generate pdfs with hyperlinks
\usepackage[utf8]{inputenc} % Allow UTF8 characters
\usepackage[english]{babel} % All publications are in English
\usepackage[displaymath, mathlines]{lineno} % Line numbers for drafts
\usepackage{blindtext} % For testing

\graphicspath{{figures/}} % Use figures directory for figures

%% file: files/pub007.tex
\stepcounter{AffiliationCounter}\edef\instCPPM{\protect\theAffiliationCounter}
\stepcounter{AffiliationCounter}\edef\instYerevan{\protect\theAffiliationCounter}
\stepcounter{AffiliationCounter}\edef\instIKER{\protect\theAffiliationCounter}
\stepcounter{AffiliationCounter}\edef\instBNL{\protect\theAffiliationCounter}
\stepcounter{AffiliationCounter}\edef\instBINP{\protect\theAffiliationCounter}
\stepcounter{AffiliationCounter}\edef\instCMU{\protect\theAffiliationCounter}
\stepcounter{AffiliationCounter}\edef\instCinvestavIPN{\protect\theAffiliationCounter}
\stepcounter{AffiliationCounter}\edef\instPrague{\protect\theAffiliationCounter}
\stepcounter{AffiliationCounter}\edef\instChiangMai{\protect\theAffiliationCounter}
\stepcounter{AffiliationCounter}\edef\instChiba{\protect\theAffiliationCounter}
\stepcounter{AffiliationCounter}\edef\instCAU{\protect\theAffiliationCounter}
\stepcounter{AffiliationCounter}\edef\instConacyt{\protect\theAffiliationCounter}
\stepcounter{AffiliationCounter}\edef\instDESY{\protect\theAffiliationCounter}
\stepcounter{AffiliationCounter}\edef\instDuke{\protect\theAffiliationCounter}
\stepcounter{AffiliationCounter}\edef\instITAR{\protect\theAffiliationCounter}
\stepcounter{AffiliationCounter}\edef\instRomaENEA{\protect\theAffiliationCounter}
\stepcounter{AffiliationCounter}\edef\instFudan{\protect\theAffiliationCounter}
\stepcounter{AffiliationCounter}\edef\instSOKENDAI{\protect\theAffiliationCounter}
\stepcounter{AffiliationCounter}\edef\instGyeongsang{\protect\theAffiliationCounter}
\stepcounter{AffiliationCounter}\edef\instHanyang{\protect\theAffiliationCounter}
\stepcounter{AffiliationCounter}\edef\instKEK{\protect\theAffiliationCounter}
\stepcounter{AffiliationCounter}\edef\instJPARC{\protect\theAffiliationCounter}
\stepcounter{AffiliationCounter}\edef\instHiroshima{\protect\theAffiliationCounter}
\stepcounter{AffiliationCounter}\edef\instFrascati{\protect\theAffiliationCounter}
\stepcounter{AffiliationCounter}\edef\instNapoliINFN{\protect\theAffiliationCounter}
\stepcounter{AffiliationCounter}\edef\instPadovaINFN{\protect\theAffiliationCounter}
\stepcounter{AffiliationCounter}\edef\instPerugiaINFN{\protect\theAffiliationCounter}
\stepcounter{AffiliationCounter}\edef\instPisaINFN{\protect\theAffiliationCounter}
\stepcounter{AffiliationCounter}\edef\instRomaINFN{\protect\theAffiliationCounter}
\stepcounter{AffiliationCounter}\edef\instRomaTreINFN{\protect\theAffiliationCounter}
\stepcounter{AffiliationCounter}\edef\instTorinoINFN{\protect\theAffiliationCounter}
\stepcounter{AffiliationCounter}\edef\instTriesteINFN{\protect\theAffiliationCounter}
\stepcounter{AffiliationCounter}\edef\instIISER{\protect\theAffiliationCounter}
\stepcounter{AffiliationCounter}\edef\instIITBhubaneswar{\protect\theAffiliationCounter}
\stepcounter{AffiliationCounter}\edef\instIITGuwahati{\protect\theAffiliationCounter}
\stepcounter{AffiliationCounter}\edef\instIITHyderabad{\protect\theAffiliationCounter}
\stepcounter{AffiliationCounter}\edef\instIITMadras{\protect\theAffiliationCounter}
\stepcounter{AffiliationCounter}\edef\instIndiana{\protect\theAffiliationCounter}
\stepcounter{AffiliationCounter}\edef\instIHEPRussia{\protect\theAffiliationCounter}
\stepcounter{AffiliationCounter}\edef\instHEPHYVienna{\protect\theAffiliationCounter}
\stepcounter{AffiliationCounter}\edef\instIPP{\protect\theAffiliationCounter}
\stepcounter{AffiliationCounter}\edef\instIOP{\protect\theAffiliationCounter}
\stepcounter{AffiliationCounter}\edef\instISU{\protect\theAffiliationCounter}
\stepcounter{AffiliationCounter}\edef\instJAEA{\protect\theAffiliationCounter}
\stepcounter{AffiliationCounter}\edef\instMainz{\protect\theAffiliationCounter}
\stepcounter{AffiliationCounter}\edef\instGiessen{\protect\theAffiliationCounter}
\stepcounter{AffiliationCounter}\edef\instKarlsruhe{\protect\theAffiliationCounter}
\stepcounter{AffiliationCounter}\edef\instKAU{\protect\theAffiliationCounter}
\stepcounter{AffiliationCounter}\edef\instKISTI{\protect\theAffiliationCounter}
\stepcounter{AffiliationCounter}\edef\instKoreaUnivKU{\protect\theAffiliationCounter}
\stepcounter{AffiliationCounter}\edef\instKyotoU{\protect\theAffiliationCounter}
\stepcounter{AffiliationCounter}\edef\instKyungpook{\protect\theAffiliationCounter}
\stepcounter{AffiliationCounter}\edef\instLPI{\protect\theAffiliationCounter}
\stepcounter{AffiliationCounter}\edef\instLNNU{\protect\theAffiliationCounter}
\stepcounter{AffiliationCounter}\edef\instLMU{\protect\theAffiliationCounter}
\stepcounter{AffiliationCounter}\edef\instMNITJaipur{\protect\theAffiliationCounter}
\stepcounter{AffiliationCounter}\edef\instMPP{\protect\theAffiliationCounter}
\stepcounter{AffiliationCounter}\edef\instMcGill{\protect\theAffiliationCounter}
\stepcounter{AffiliationCounter}\edef\instNagoya{\protect\theAffiliationCounter}
\stepcounter{AffiliationCounter}\edef\instNagoyaIAR{\protect\theAffiliationCounter}
\stepcounter{AffiliationCounter}\edef\instNagoyaKMI{\protect\theAffiliationCounter}
\stepcounter{AffiliationCounter}\edef\instNaraWu{\protect\theAffiliationCounter}
\stepcounter{AffiliationCounter}\edef\instNCU{\protect\theAffiliationCounter}
\stepcounter{AffiliationCounter}\edef\instHSE{\protect\theAffiliationCounter}
\stepcounter{AffiliationCounter}\edef\instNTUTaiwan{\protect\theAffiliationCounter}
\stepcounter{AffiliationCounter}\edef\instKrakow{\protect\theAffiliationCounter}
\stepcounter{AffiliationCounter}\edef\instNiigata{\protect\theAffiliationCounter}
\stepcounter{AffiliationCounter}\edef\instNSU{\protect\theAffiliationCounter}
\stepcounter{AffiliationCounter}\edef\instOsakaCity{\protect\theAffiliationCounter}
\stepcounter{AffiliationCounter}\edef\instRCNP{\protect\theAffiliationCounter}
\stepcounter{AffiliationCounter}\edef\instPNNL{\protect\theAffiliationCounter}
\stepcounter{AffiliationCounter}\edef\instPanjab{\protect\theAffiliationCounter}
\stepcounter{AffiliationCounter}\edef\instPanjabPAU{\protect\theAffiliationCounter}
\stepcounter{AffiliationCounter}\edef\instRIKENMSL{\protect\theAffiliationCounter}
\stepcounter{AffiliationCounter}\edef\instXavier{\protect\theAffiliationCounter}
\stepcounter{AffiliationCounter}\edef\instSPU{\protect\theAffiliationCounter}
\stepcounter{AffiliationCounter}\edef\instSoongsil{\protect\theAffiliationCounter}
\stepcounter{AffiliationCounter}\edef\instLjubljanaJSI{\protect\theAffiliationCounter}
\stepcounter{AffiliationCounter}\edef\instSKKU{\protect\theAffiliationCounter}
\stepcounter{AffiliationCounter}\edef\instKyiv{\protect\theAffiliationCounter}
\stepcounter{AffiliationCounter}\edef\instTata{\protect\theAffiliationCounter}
\stepcounter{AffiliationCounter}\edef\instTUM{\protect\theAffiliationCounter}
\stepcounter{AffiliationCounter}\edef\instTelAviv{\protect\theAffiliationCounter}
\stepcounter{AffiliationCounter}\edef\instTitech{\protect\theAffiliationCounter}
\stepcounter{AffiliationCounter}\edef\instUAS{\protect\theAffiliationCounter}
\stepcounter{AffiliationCounter}\edef\instNapoliUNIV{\protect\theAffiliationCounter}
\stepcounter{AffiliationCounter}\edef\instPadovaUNIV{\protect\theAffiliationCounter}
\stepcounter{AffiliationCounter}\edef\instPerugiaUNIV{\protect\theAffiliationCounter}
\stepcounter{AffiliationCounter}\edef\instPisaUNIV{\protect\theAffiliationCounter}
\stepcounter{AffiliationCounter}\edef\instRomaTreUNIV{\protect\theAffiliationCounter}
\stepcounter{AffiliationCounter}\edef\instTorinoUNIV{\protect\theAffiliationCounter}
\stepcounter{AffiliationCounter}\edef\instTriesteUNIV{\protect\theAffiliationCounter}
\stepcounter{AffiliationCounter}\edef\instIJCLab{\protect\theAffiliationCounter}
\stepcounter{AffiliationCounter}\edef\instIPHC{\protect\theAffiliationCounter}
\stepcounter{AffiliationCounter}\edef\instBilbao{\protect\theAffiliationCounter}
\stepcounter{AffiliationCounter}\edef\instBonn{\protect\theAffiliationCounter}
\stepcounter{AffiliationCounter}\edef\instUBC{\protect\theAffiliationCounter}
\stepcounter{AffiliationCounter}\edef\instCincinnati{\protect\theAffiliationCounter}
\stepcounter{AffiliationCounter}\edef\instHawaii{\protect\theAffiliationCounter}
\stepcounter{AffiliationCounter}\edef\instLjubljanaUniLJ{\protect\theAffiliationCounter}
\stepcounter{AffiliationCounter}\edef\instLouisville{\protect\theAffiliationCounter}
\stepcounter{AffiliationCounter}\edef\instLjubljanaUM{\protect\theAffiliationCounter}
\stepcounter{AffiliationCounter}\edef\instMelbourne{\protect\theAffiliationCounter}
\stepcounter{AffiliationCounter}\edef\instMississippi{\protect\theAffiliationCounter}
\stepcounter{AffiliationCounter}\edef\instPittsburgh{\protect\theAffiliationCounter}
\stepcounter{AffiliationCounter}\edef\instUSTC{\protect\theAffiliationCounter}
\stepcounter{AffiliationCounter}\edef\instSAlabama{\protect\theAffiliationCounter}
\stepcounter{AffiliationCounter}\edef\instSydney{\protect\theAffiliationCounter}
\stepcounter{AffiliationCounter}\edef\instTabuk{\protect\theAffiliationCounter}
\stepcounter{AffiliationCounter}\edef\instUTokyo{\protect\theAffiliationCounter}
\stepcounter{AffiliationCounter}\edef\instEri{\protect\theAffiliationCounter}
\stepcounter{AffiliationCounter}\edef\instIPMU{\protect\theAffiliationCounter}
\stepcounter{AffiliationCounter}\edef\instVictoria{\protect\theAffiliationCounter}
\stepcounter{AffiliationCounter}\edef\instVPI{\protect\theAffiliationCounter}
\stepcounter{AffiliationCounter}\edef\instWayneState{\protect\theAffiliationCounter}
\stepcounter{AffiliationCounter}\edef\instYamagata{\protect\theAffiliationCounter}
\stepcounter{AffiliationCounter}\edef\instYonsei{\protect\theAffiliationCounter}
\collaboration{The Belle II Collaboration}
  \author[\instTriesteINFN]{F.~Abudin{\'e}n,} % 2250
  \author[\instPanjab]{L.~Aggarwal,} % 10083
  \author[\instXavier]{H.~Ahmed,} % 11323
  \author[\instUTokyo]{H.~Aihara,} % 2223
  \author[\instYerevan]{N.~Akopov,} % 9443
  \author[\instTabuk,\instKAU]{S.~Al~Said,} % Tabuk
  \author[\instNapoliUNIV,\instNapoliINFN]{A.~Aloisio,} % 2194
  \author[\instIOP]{N.~Anh~Ky,} % 2218
  \author[\instBNL]{D.~M.~Asner,} % 4684
  \author[\instCincinnati]{H.~Atmacan,} % 2538
% ...
\collaboration{Belle and Belle II Collaborations}